\documentclass[a4paper,11pt]{article}

\usepackage{amsmath,amsthm,amsbsy,amssymb,amsfonts,graphicx,mathrsfs,bm,cite,color}
\usepackage{hyperref}

\makeatletter
\catcode`\@=11
\@addtoreset{equation}{section}

\makeatother

\makeatletter

\@addtoreset{table}{section}
\makeatother

\usepackage[utf8]{inputenc}
\usepackage[all]{xy}

\renewcommand{\thefootnote}{\arabic{footnote}}

\newcommand{\Exp}[1]{\operatorname{e}^{#1}}

\newcommand{\tr}{\operatorname{tr}}

\newcommand{\rmd}{{\mathrm{d}}}
\newcommand{\rmT}{{\tt T}}
\newcommand{\nn}{\nonumber}
\newcommand{\Lie}{\pounds}
\newcommand{\gLie}{\hat{\pounds}}

\newcommand{\cE}{\mathcal E}\newcommand{\cF}{\mathcal F}
\newcommand{\cG}{\mathcal G}\newcommand{\cH}{\mathcal H}
\newcommand{\cI}{\mathcal I}
\newcommand{\cK}{\mathcal K}
\newcommand{\cM}{\mathcal M}
\newcommand{\cP}{\mathcal P}
\newcommand{\cR}{\mathcal R}

\newcommand{\cX}{\mathcal X}

\newcommand{\lR}{\mathbb{R}}

\newcommand{\dprime}{^{\prime\!\prime}}

\newcommand{\sfa}{\mathsf{a}}
\newcommand{\sfb}{\mathsf{b}}
\newcommand{\sfc}{\mathsf{c}}
\newcommand{\sfd}{\mathsf{d}}
\newcommand{\sfe}{\mathsf{e}}
\newcommand{\sff}{\mathsf{f}}

\newcommand{\Pa}{\tt{A}}
\newcommand{\Pb}{\tt{B}}
\newcommand{\Pc}{\tt{C}}
\newcommand{\Pd}{\tt{D}}

\newcommand{\rr}{r}

\newcommand{\SL}{\text{SL}}

\newcommand{\GL}{\text{GL}}
\newcommand{\gl}{\mathfrak{gl}}
\newcommand{\OO}{\text{O}}

\allowdisplaybreaks[3]

\begin{document}

\begin{titlepage}
\renewcommand{\thefootnote}{\fnsymbol{footnote}}

\vspace*{1cm}

\centerline{\Large\textbf{Extended Drinfel'd algebras and non-Abelian duality}}%

\vspace{1.5cm}

\centerline{\large Yuho Sakatani}

\vspace{0.2cm}

\begin{center}
{\it Department of Physics, Kyoto Prefectural University of Medicine,}\\
{\it Kyoto 606-0823, Japan}\\
{\small\texttt{yuho@koto.kpu-m.ac.jp}}
\end{center}

\vspace*{2mm}

\begin{abstract} 
A Drinfel'd algebra gives the systematic construction of generalized parallelizable spaces and this allows us to study an extended $T$-duality, known as the Poisson--Lie $T$-duality. Recently, in order to find a generalized $U$-duality, an extended Drinfel'd algebra (ExDA), called the Exceptional Drinfel'd algebra (EDA) was proposed and a natural extension of the usual $U$-duality was studied both in the context of supergravity and membrane theory. In this paper, we clarify the general structure of ExDAs and show that an ExDA always gives a generalized parallelizable space, which may be regarded as a group manifold with generalized Nambu--Lie structures. We also discuss generalized Yang--Baxter deformations that are based on coboundary ExDAs. As important examples, we consider the $E_{n(n)}$ EDA for $n\leq 8$ and study various aspects, both in terms of M-theory and type IIB theory.
\end{abstract}

\thispagestyle{empty}
\end{titlepage}

\setcounter{footnote}{0}

\newpage

\tableofcontents

\newpage

\section{Introduction}
\label{sec:introduction}

The Poisson--Lie $T$-duality \cite{hep-th:9502122,hep-th:9509095} is an extension of the well-established Abelian $T$-duality. 
This is based on a Lie group, called the Drinfel'd double. 
The Drinfel'd double contains a (maximally isotropic) subalgebra $\mathfrak{g}$ and the group manifold of $G=\exp \mathfrak{g}$ plays the role of the target space of string theory.
A Drinfel'd double generally contains various subalgebras $\mathfrak{g}$, and a different choice of the subalgebra yields a different target space. 
Since the equations of motion of supergravity or string theory are insensitive to the choice of subalgebra, arbitrariness in the choice of the subalgebra gives rise to the non-trivial target space duality.

String theory has a larger duality group called $U$-duality, and a natural question is whether there are $U$-duality analogue of the Poisson--Lie $T$-duality. 
In order to address this question, an extended Drinfel'd algebra was proposed in \cite{1911.06320,1911.07833}. 
Since the $U$-duality group is the exceptional group $E_{n(n)}$, this algebra is called the exceptional Drinfel'd algebra (EDA). 
Similar to the case of the Poisson--Lie $T$-duality, the EDA contains various maximally isotropic subalgebras and each subalgebra gives a target space. 
Whether this extended $U$-duality can be used as a solution generating technique in supergravity is still an open question, but the answer is positive as discussed in \cite{1911.06320,1911.07833,2006.12452} by using several examples. 
In addition, as was discussed in \cite{2001.09983}, the equations of motion of the (topological) membrane worldvolume theory can be expressed in a covariant way under the extended $U$-duality. 
Encouraged by these results, further details of the EDA have been studied recently (see \cite{2003.06164,2007.01213
,2007.08510} for recent works). 

In the original works \cite{1911.06320,1911.07833}, the $E_{n(n)}$ EDA was proposed for $n\leq 4$ and it was extended to $n\leq 6$ in \cite{2007.08510}. 
While the extension is straightforward, increasing the rank $n$ increases the dimension of the EDA and makes it more difficult to study various properties of the EDA. 
The main purpose of this paper is to study the properties of the $E_{n(n)}$ EDA in a unified way that is less dependent on $n$.
In fact, most of the properties of the EDA do not depend on the details of the duality group $\cG$\,, and thus we consider an extended Drinfel'd algebra (ExDA) associated with a general group $\cG$. 
When $\cG$ is the $\OO(d,d)$ $T$-duality group, the ExDA reduces to the Lie algebra of the Drinfel'd double while when $\cG$ is the $E_{n(n)}$ $U$-duality group, the ExDA reduces to the $E_{n(n)}$ EDA. 
In general, other duality groups are also possible, but in this paper, we will focus on these two cases.

We also present the $E_{n(n)}$ EDA for $n\leq 8$, as an application of the general discussion of the ExDA.
Previous studies of the EDA have adopted the M-theory picture, where the dimension of the subalgebra $\mathfrak{g}$ is $n$\,, but as discussed in a series of studies \cite{hep-th:0104081,hep-th:0107181,hep-th:0307098,hep-th:0402140}, $U$-duality covariant tensors can be decomposed into tensors in both M-theory and type IIB theory.
In this paper, in addition to the M-theory picture, we also construct the EDA in type IIB picture. 
In the type IIB picture, the dimension of $\mathfrak{g}$ is $n-1$ although the dimension of the whole EDA is the same as the M-theory picture (see \cite{1911.06320,2006.12452} for some comments on the type IIB picture). 

\subsection{A brief sketch of Non-Abelian Duality}
\label{sec:E-model}

In order to illustrate the main results, we here explain the Poisson--Lie $T$-duality in more detail. 
For this purpose, it is very useful to use the language of the Extended Field Theory (ExFT) which includes Double Field Theory (DFT) \cite{hep-th:9302036,hep-th:9305073,hep-th:9308133,0904.4664,1006.4823} and Exceptional Field Theory (EFT) \cite{1008.1763,1111.0459,1206.7045,1208.5884,1308.1673,1312.0614,1312.4542,1406.3348} as special cases, $\cG=\OO(d,d)$ and $\cG=E_{n(n)}$, respectively. 
ExFT is a manifestly $\cG$-covariant formulation of supergravity, extending the physical space of $\mathfrak{n}$ dimensions into an extended space of $D$ dimensions.
The details are reviewed in section \ref{sec:ExDA}, but of particular importance is that the infinitesimal diffeomorphisms are generated by the generalized Lie derivative $\gLie_V$, which is a modification of the usual Lie derivative $\Lie_v$\,. 
In the following, we explain that the existence of certain frame fields $E_A{}^I$ ($A,I=1,\dotsc,D$) in DFT that satisfy
\begin{align}
 \gLie_{E_A} E_B{}^I = - X_{AB}{}^C\,E_C{}^I \,,
\label{eq:E-Lie}
\end{align}
where $X_{AB}{}^C$ are the structure constants of the Drinfel'd algebra, is very important in the Poisson--Lie $T$-duality (see \cite{1810.11446} where this was pointed out). 

Let us consider a string sigma model, known as the double sigma model \cite{Tseytlin:1990nb,Tseytlin:1990va,1111.1828},
\begin{align}
 S= -\frac{1}{2} \int \rmd^2\sigma\,\bigl[\cM_{IJ}(x)\,\partial_1 x^I\,\partial_1x^J - \eta_{IJ}\,\partial_0x^I\,\partial_1x^J \bigr] \,,
\end{align}
where $2\pi\alpha'=1$ and $I,J=1,\dotsc,D\,(\equiv2\,d)$\,. 
We decompose the index as $(x^I)=(x^m,\,\tilde{x}_m)$ ($m=1,\dotsc,d$) and parameterize the generalized metric $\cM_{IJ}$ and the $\OO(d,d)$ metric $\eta_{IJ}$ as
\begin{align}
 \cM_{IJ} \equiv \begin{pmatrix} g_{mn} -B_{mp}\,g^{pq}\,B_{qn} & B_{mp}\,g^{pn} \\ -g^{mp}\,B_{pn} & g^{mn} \end{pmatrix}, \qquad
 \eta_{IJ} = \begin{pmatrix} 0 & \delta_m^n \\ \delta^m_n & 0 \end{pmatrix}.
\end{align}
By assuming that $\cM_{IJ}$ depends only on $x^m$\,, the equations of motion reproduce the standard equations of motion for $x^m(\sigma)$\,, and the double sigma model is (classically) equivalent to the (bosonic) string theory (see \cite{1111.1828} for details). 
In particular, when the target space is constant, the action is manifestly invariant under the $\OO(d,d)$ $T$-duality transformation
\begin{align}
 x^I \to (\Lambda^{-1})_J{}^I\,x^J\,,\qquad 
 \cM_{IJ} \to \Lambda_I{}^K\,\Lambda_J{}^L\,\cM_{KL}\qquad (\Lambda\in\OO(d,d))\,.
\end{align}

In the canonical formalism, we define the conjugate momenta as
\begin{align}
 P_I = \eta_{IJ} \,\partial_1 x^J\,,
\end{align}
and define the Poisson bracket as
\begin{align}
 \{x^I(\sigma)\,,P_J(\sigma')\} = 2\,\delta^I_J\,\delta(\sigma-\sigma')\,,\quad 
 \{x^I(\sigma)\,,x^J(\sigma')\} = 0 = \{P_I(\sigma)\,,P_J(\sigma')\} \,.
\end{align}
The Hamiltonian is obtained as
\begin{align}
 H = \frac{1}{2}\int\rmd\sigma\,\cM_{IJ}(x)\,\partial_1x^I\,\partial_1x^J \,.
\label{eq:H-DSM}
\end{align}
We find that there are the second-class constraints $\widetilde{\cP}_I \equiv P_I - \eta_{IJ} \,\partial_1x^J$\,, and the Dirac bracket is obtained as
\begin{align}
\begin{split}
 [x^I(\sigma),\,x^J(\sigma')] &= - \eta^{IJ}\, \epsilon (\sigma -\sigma' )\,,
\qquad
 [x^I(\sigma),\,P_J(\sigma')] = \delta^I_J\,\delta(\sigma-\sigma')\,,
\\
 [P_I(\sigma),\,P_J(\sigma')] &= \eta_{IJ}\, \delta'(\sigma -\sigma')\,.
\end{split}
\label{eq:Dirac-bracket}
\end{align}
Then the dynamics is described by this Dirac bracket and the Hamiltonian \eqref{eq:H-DSM}.

Now we consider the case where the generalized metric is of the form
\begin{align}
 \cM_{IJ}(x) = E_I{}^A(x)\,E_J{}^B(x)\,\hat{\cH}_{AB}\,,
\label{eq:PL-original}
\end{align}
where $\hat{\cH}_{AB}$ is constant and $E_I{}^A(x)$ is the inverse matrix of the generalized frame fields $E_A{}^I(x)$ obeying Eq.~\eqref{eq:E-Lie}. 
If we define the currents
\begin{align}
 j_A(\sigma) \equiv E_A{}^I(x(\sigma))\,P_I(\sigma)\,,
\end{align}
the Hamiltonian can be expressed as
\begin{align}
 H = \frac{1}{2}\int\rmd\sigma\,\hat{\cH}^{AB}\,j_A(\sigma) \,j_B(\sigma) \,.
\label{eq:H-E-model}
\end{align}
Using Eq.~\eqref{eq:E-Lie} and the Dirac bracket \eqref{eq:Dirac-bracket}, we obtain
\begin{align}
 [j_A(\sigma),\,j_B(\sigma')] = X_{AB}{}^C\, j_C(\sigma)\,\delta(\sigma-\sigma') 
 + \eta_{AB}\, \delta'(\sigma -\sigma')\,,
\end{align}
and the physical system governed by this current algebra and the Hamiltonian \eqref{eq:H-E-model} is called the $\cE$-model \cite{hep-th:9512040,hep-th:9605212,1508.05832}.
Under this setup, we can discuss the Poisson--Lie $T$-duality as follows. 

Suppose that there exists another set of generalized frame fields $E'_A{}^I$ that satisfy
\begin{align}
 \gLie_{E'_A}E'_B{}^I = -X'_{AB}{}^C\,E'_C\,,\qquad 
 X'_{AB}{}^C \equiv C_A{}^{A'}\,C_B{}^{B'}\,(C^{-1})_{C'}{}^C\,X_{A'B'}{}^{C'}\,,
\label{eq:E-Lie-prime}
\end{align}
where $C_A{}^B\in\OO(d,d)$ is a constant matrix and $E'_A{}^I\neq C_A{}^B\,E_B{}^I$\,. 
Then we can define another currents $j'_A(\sigma) \equiv E'_A{}^I(X(\sigma)) \,P_I(\sigma)$ obeying
\begin{align}
 [j'_A(\sigma),\,j'_B(\sigma')] = X'_{AB}{}^C\, j'_C(\sigma)\,\delta(\sigma-\sigma') 
 + \eta_{AB}\, \delta'(\sigma -\sigma')\,.
\end{align}
Introducing a new Hamiltonian
\begin{align}
 H' = \frac{1}{2}\int\rmd\sigma\,\hat{\cH}'^{AB}\,j'_A(\sigma) \,j'_B(\sigma)\,,\qquad 
 \hat{\cH}'_{AB}\equiv C_A{}^C\,C_B{}^D\,\hat{\cH}_{CD}\,,
\end{align}
we see that the dynamics in the primed system is the same as the original one because the role of $j_A$ is played by $C_A{}^B\,j'_B$\,. 
This shows that string theory defined on the original background \eqref{eq:PL-original} is equivalent to that defined on another curved background
\begin{align}
 \cH'_{IJ} = E'_I{}^A(X)\,E'_J{}^B(X)\,\hat{\cH}'_{AB} \ \bigl(\neq \cH_{IJ}\bigr)\,.
\end{align}
This is the Poisson--Lie $T$-duality. 
Here, whether we can actually find the generalized frame fields $E_A{}^I$ and $E'_A{}^I$ satisfying \eqref{eq:E-Lie} and \eqref{eq:E-Lie-prime} is very important.
As was found in \cite{hep-th:9502122,hep-th:9509095} (and in \cite{1810.11446} in the language of DFT), by using the Drinfel'd algebra, we can systematically construct such generalized frame fields $E_A{}^I$ and $E'_A{}^I$\,. 
The systematic construction of the generalized frame fields is the great benefit of the Drinfel'd algebra. 

The same discussion holds also for membrane theory \cite{2001.09983}, where the $T$-duality group is extended to the $U$-duality group. 
In membrane theory, the extended momenta $P_I(\sigma)$ satisfy the Poisson bracket \cite{1208.1232}
\begin{align}
 [P_I(\sigma),\,P_J(\sigma')] = \rho_{IJ}^{\sfa}(\sigma) \,\partial_{\sfa} \delta^2(\sigma -\sigma')\,,
\end{align}
where $\sigma^{\sfa}=1,2$ are spatial coordinates over the membrane and $\rho_{IJ}^{\sfa}(\sigma)$ is a certain non-constant matrix. 
The Hamiltonian can be expressed as
\begin{align}
 H = \frac{1}{2}\int\rmd^2\sigma\sqrt{h}\,\cM^{IJ}(x)\,P_I\,P_J \,,
\end{align}
similar to the double sigma model. 
Again, we suppose that the generalized metric has the form \eqref{eq:PL-original} and assume that the generalized frame fields satisfies
\begin{align}
 \gLie_{E_A}E_B{}^I = -X_{AB}{}^C\,E_C\,,
\label{eq:E-X-Lie}
\end{align}
by means of the generalized Lie derivative in EFT. 
Here, $X_{AB}{}^C$ are the structure constants of the EDA and, in general, the lower indices are not antisymmetric $X_{AB}{}^C\neq X_{[AB]}{}^C$ (namely, the EDA is a Leibniz algebra). 
Then, by defining the currents $j_A(\sigma) \equiv E_A{}^I(X(\sigma)) \,P_I(\sigma)$, the Hamiltonian and the Poisson bracket of the currents become
\begin{align}
\begin{split}
 &H = \frac{1}{2}\int\rmd^2\sigma\sqrt{h}\,\hat{\cH}^{AB}\,j_A(\sigma) \,j_B(\sigma)\,,
\\
 &[j_A(\sigma),\,j_B(\sigma')] = X_{[AB]}{}^C\,j_C(\sigma)\,\delta^2(\sigma-\sigma') + \frac{1}{2}\,\bigl[\rho_{AB}^{\sfa}(\sigma) - \rho_{AB}^{\sfa}(\sigma')\bigr] \,\partial_{\sfa} \delta^2(\sigma -\sigma')\,.
\end{split}
\end{align}
We thus obtain a natural extension of the $\cE$-model \cite{2001.09983}. 
At least when the target space is three dimensional, the membrane theory has the $E_{3(3)}$ $U$-duality covariance.\footnote{When the dimension of the target space is greater than three, the rotation $P_I\to \Lambda_I{}^J\,P_J$ $(\Lambda_I{}^J\in E_{n(n)})$ is not a symmetry because it breaks some integrability \cite{1509.02915}, and even the Abelian $U$-duality cannot be realized.} 
Then, similar to the case of $T$-duality, we can consider $E'_A{}^I$ satisfying \eqref{eq:E-Lie-prime} and we find that membrane theory has the symmetry under the non-Abelian $U$-duality \cite{2001.09983}. 

As described above, the existence of the generalized frame fields $E_A{}^I$ satisfying Eq.~\eqref{eq:E-X-Lie} is very important in realizing the non-Abelian duality.
When such generalized frame fields exist, the target space is called a generalized parallelizable space \cite{0807.4527,1401.3360,1410.8145} and that plays an important role in considering a dimensional reduction preserving supersymmetry. 
As we show in this paper, a major advantage of the ExDA is that, once an ExDA and its maximally isotropic subalgebra are specified, a generalized parallelizable space can be explicitly constructed.

\subsection{Summary of results}

Now that we have explained the importance of ExDA, we turn to the details of the ExDA and explain the main results.
To discuss the general properties of the ExDA for a general duality group $\cG$\,, we shall use a notation of the embedding tensor formalism developed in gauged supergravity \cite{hep-th:9804056,hep-th:0010076,hep-th:0212239,hep-th:0311224,hep-th:0412173,hep-th:0501243,hep-th:0507289,0705.2101,0801.1294,0809.5180}. 
We consider a $D$-dimensional Leibniz algebra
\begin{align}
 T_A\circ T_B = X_{AB}{}^C\, T_C \,,
\end{align}
which has the structure constants of the form
\begin{align}
\begin{split}
 X_{AB}{}^C &= \Theta_A{}^{\bm{\alpha}}\,(t_{\bm{\alpha}})_B{}^C
 -\bigl[\tfrac{1}{1+\beta}\,(t^{\bm{\alpha}})_A{}^D\,(t_{\bm{\alpha}})_B{}^C + \delta_A^D\,\delta_B^C\bigr]\,\vartheta_D\,,
\\
 \Theta_A{}^{\bm{\alpha}} &\equiv \mathbb{P}_A{}^{\bm{\alpha}}{}^B{}_{\bm{\beta}}\,F_B{}^{\bm{\beta}} \,,\qquad
 \vartheta_A \equiv F_A{}^0 - \beta\, F_B{}^{\bm{\alpha}} \, (t_{\bm{\alpha}})_A{}^B \,.
\end{split}
\label{eq:EDA-structure}
\end{align}
Here, $\{t_{\bm{\alpha}}\}$ are the generators of the duality group $\cG$, $\beta$ is a constant specific to $\cG$, $\mathbb{P}$ is a projection on a certain representation, and the information of the structure constants is contained in $F_A\equiv F_A{}^{\bm{\alpha}}\,t_{\bm{\alpha}} + F_A{}^0\,t_0$ ($t_0$: generator of a scale symmetry $\mathbb{R}^+$ \cite{1112.3989,1212.1586}). 

So far, this is the general structure of the gauge algebra for maximal gauged supergravities (in the presence of the trombone gauging) \cite{0809.5180}. 
The additional requirement specific to the ExDA is that this Leibniz algebra contains an $\mathfrak{n}$-dimensional subalgebra $\mathfrak{g}$\,, $[T_a,\,T_b] = f_{ab}{}^c\,T_c$\,, that is maximally isotropic with respect to some bilinear form $\langle T_a ,\,T_b \rangle = 0$\,. 
This requirement and the consistency condition of ExFT imply that $F_A$ can be parameterized as
\begin{align}
 F_A = \delta_A^b\,\bigl(F_{bc}{}^d\, K^c{}_d + F_{b}{}^{\hat{\Pa}}\, R_{\hat{\Pa}} + F_a{}^0\,t_0 \bigr) \,,
\end{align}
where $K^a{}_b$ is the generator of the $\mathfrak{gl}(\mathfrak{n})$ subalgebra and $\{K^a{}_b,R_{\hat{\Pa}}\}$ is a set of non-positive-level generators of $\cG$\,. 
Substituting this $F_A$ into Eq.~\eqref{eq:EDA-structure}, we obtain an ExDA. 

If an ExDA and its maximally isotropic subalgebra $\mathfrak{g}$ is specified, we can construct the generalized frame fields $E_A{}^I$ by following the procedure known in the Poisson--Lie $T$-duality. 
In the previous studies, it was very hard to check that the constructed $E_A{}^I$ indeed satisfy the relation \eqref{eq:E-X-Lie}.
However, thanks to the symbolic expression \eqref{eq:EDA-structure}, here we can generally show the relation for a wide class of duality groups $\cG$. 
Namely, we show that the ExDA always constructs a generalized parallelizable space for an arbitrary choice of the subalgebra $\mathfrak{g}$\,. 
This is one of the main results and plays an important role in the non-Abelian duality.

As known in the Drinfel'd algebra or the $E_{n(n)}$ EDA ($n\leq 6$), the Leibniz identities of the ExDA consist of the cocycle conditions, fundamental identities, and some additional identities. 
In this paper, we provide the general form of these identities. 
In particular, for the cocycle conditions, we obtain the coboundary ansatz that trivially satisfies the cocycle conditions. 
The coboundary ansatz is characterized by certain multi-vectors, and they are closely related to a generalized classical $r$-matrix. 
We propose the generalized classical Yang--Baxter equation by extending the results of \cite{1911.06320,1911.07833,2007.08510}. 
We also discuss the Yang--Baxter deformation, which is a specific example of the non-Abelian duality based on the classical $r$-matrices.

This paper is organized as follows. 
In section \ref{sec:ExFT}, we briefly review the necessary ingredients of ExFT. 
There, several non-standard notation for the duality algebra is also introduced. 
In section \ref{sec:ExDA}, we introduce the ExDA and explain various general properties. 
In section \ref{sec:NAD}, we explain how to generate solutions of ExFT through the non-Abelian duality. 
In sections \ref{sec:M} and \ref{sec:IIB}, we study the $E_{n(n)}$ EDA in the M-theory picture and the type IIB picture, respectively. 
Section \ref{sec:discussion} is devoted to summary and discussions. 

\section{Review of ExFT}
\label{sec:ExFT}

In this section, we set up our notation by shortly reviewing ExFT (see \cite{2006.09777} for a review). 
The well-studied ExFTs are DFT and EFT where the duality group $\cG$ is $\OO(d,d)$ and $E_{n(n)}$, respectively. 
In this paper, we particularly focus on $E_{n(n)}$ EFT with $n\leq 8$\,, but the extended Drinfel'd algebra can be defined for other duality groups, where various important properties of ExDA will be satisfied. 
We thus make a presentation in a way that does not depend on the details of the duality group $\cG$, as much as possible.

In ExFT, we introduce a $D$-dimensional extended space with the generalized coordinates $x^I$ ($I=1,\dotsc,D$). 
Diffeomorphisms in the extended space is generated by the generalized Lie derivative (see \cite{1208.5884} for details)
\begin{align}
 \gLie_V W^I \equiv V^J\,\partial_J W^I - W^J\,\partial_J V^I + Y^{IK}_{JL}\,\partial_K V^L\, W^J \,,
\end{align}
where $Y^{IJ}_{KL}$ is some invariant tensor that is explained in section \ref{sec:Y-tensor}. 
For consistency, for all fields defined on the extended space, we impose the so-called section condition
\begin{align}
 Y^{IJ}_{KL} \, \partial_I \otimes \partial_J =0\,.
\label{eq:SC}
\end{align}
According to this condition, all fields are defined on a maximally isotropic subspace (called the physical subspace) with coordinates $x^i$ ($i=1,\dotsc,\mathfrak{n}$). 
After choosing such a subspace, representations of the duality group can be decomposed into representations of the $\GL(\mathfrak{n})$ subgroup, which is associated with coordinates transformations on the physical space. 

\subsection{Decompsotions of generators}
\label{sec:generators}

The set of generators $\{t_{\bm{\alpha}}\}$ of each duality group $\cG$ can be decomposed as follows:
\begin{alignat}{2}
 \OO(d,d):& \ & & \{t_{\bm{\alpha}}\} =\{\tfrac{R_{a_1a_2}}{\sqrt{2!}},\,K^a{}_b,\,\tfrac{R^{a_1a_2}}{\sqrt{2!}}\} \qquad (a,b=1,\dotsc,d)\,,
\\
 E_{n(n)}:& \ & & \{t_{\bm{\alpha}}\} =\{\tfrac{R_{a_1\cdots a_8,a'}}{\sqrt{8!}},\,\tfrac{R_{a_1\cdots a_6}}{\sqrt{6!}},\,\tfrac{R_{a_1a_2a_3}}{\sqrt{3!}},\,K^a{}_b,\,\tfrac{R^{a_1a_2a_3}}{\sqrt{3!}},\,\tfrac{R^{a_1\cdots a_6}}{\sqrt{6!}},\tfrac{R^{a_1\cdots a_8,a'}}{\sqrt{8!}}\}
\nn\\
 \text{\small(M-theory)}\!\!\!\!\!\!&&&\qquad\qquad (a,b=1,\dotsc,\mathfrak{n}=n)\,,
\end{alignat}
where the multiple indices are totally antisymmetric. 
For $\cG=E_{n(n)}$\,, this decomposition is suitable for M-theory, but there exists another maximally isotropic subspace with $\mathfrak{n}=n-1$\,.
There, the generators are decomposed into representations of $\GL(n-1)\times \SL(2)$:\footnote{We use the indices $\sfa,\sfb$ in the type IIB picture, but also use $a,b$ when no confusion arises.}
\begin{align}
 E_{n(n)}: \ \{t_{\bm{\alpha}}\} &=\bigl\{ \tfrac{R_{\sfa_1\cdots \sfa_7,\sfa'}}{\sqrt{7!}},\,\tfrac{R^\alpha_{\sfa_1\cdots \sfa_6}}{\sqrt{6!}},\,\tfrac{R_{\sfa_1\cdots \sfa_4}}{\sqrt{4!}},\,\tfrac{R^\alpha_{\sfa_1\sfa_2}}{\sqrt{2!}},\,K^{\sfa}{}_{\sfb},\,R^\alpha{}_\beta,\,\tfrac{R_\alpha^{\sfa_1\sfa_2}}{\sqrt{2!}},\,\tfrac{R^{\sfa_1\cdots \sfa_4}}{\sqrt{4!}},\,\tfrac{R_\alpha^{\sfa_1\cdots \sfa_6}}{\sqrt{6!}}\,,\tfrac{R^{\sfa_1\cdots \sfa_7,\sfa'}}{\sqrt{7!}}\bigr\} 
\nn\\
 \text{\small(Type IIB)}~~~&\qquad (\sfa,\sfb=1,\dotsc,\mathfrak{n}=n-1\,,\quad \alpha=1,2)\,.
\end{align}
Here, $R^\alpha{}_\beta$ ($R^\alpha{}_\alpha=0$) are the generators of the $\SL(2)$\,, which is associated with the $S$-duality symmetry of type IIB theory. 

For generators in each decomposition, we introduce an ordering, called the level $\ell_{\bm{\alpha}}$\,, as
\begin{align}
 [K,\,t_{\bm{\alpha}}] = \sigma\, \ell_{\bm{\alpha}} \,t_{\bm{\alpha}}\qquad (K\equiv K^a{}_a)\,.
\end{align}
Here, the normalization constant $\sigma$ is given by $\sigma=2$, $3$, $2$ for $\OO(d,d)$, $E_{n(n)}$ (M-theory), and $E_{n(n)}$ (type IIB), respectively. 
This $\sigma$ is introduced such that the level coincides with the value known in the literature \cite{hep-th:0212291}, but it is not important in our discussion.
Explicitly, the level for each generator is given as follows:\footnote{By the definition, $\sigma\, \ell_{\bm{\alpha}}$ coincides with (number of upper $\GL$ indices)$-$(number of lower $\GL$ indices).}
\begin{align}
 \underline{\OO(d,d)}&\qquad 
\begin{array}{c|*3{|c}}
 \text{level} & -1 & 0 & 1 \\\hline
 t_{\bm{\alpha}} & R_{2} & K^1{}_1 & R^{2} 
\end{array}\,, 
\\
 \underline{\text{M-theory}}&\qquad
\begin{array}{c||c|c|c|c|c|c|c}
 \text{level} & -3 & -2 & -1 & 0 & 1 & 2 & 3 \\\hline
 t_{\bm{\alpha}} & R_{8,1} & R_{6} & R_{3} & K^1{}_1 & R^{3} & R^{6} & R^{8,1} 
\end{array}\,,
\\
 \underline{\text{Type IIB}}&\qquad
\begin{array}{c|*9{|c}}
 \text{level} & -4 & -3 & -2 & -1 & 0 & 1 & 2 & 3 & 4 \\\hline
 t_{\bm{\alpha}} & R_{7,1} & R^\alpha_{6} & R_{4} & R^\alpha_{2} & K^1{}_1\,,\, R^\alpha{}_\beta & R_\alpha^{2} & R^{4} & R_\alpha^{6} & R^{7,1} 
\end{array}\,.
\end{align}
Here we have denoted the $\GL(\mathfrak{n})$ indices schematically, for example, as $R_{8,1}\equiv R_{a_1\cdots a_8,a'}$\,. 
The Jacobi identity shows that the duality algebra
\begin{align}
 [t_{\bm{\alpha}},\,t_{\bm{\beta}}] = f_{\bm{\alpha}\bm{\beta}}{}^{\bm{\gamma}}\,t_{\bm{\gamma}}\,,
\end{align}
respects the level, in the sense that the commutator of a level-$p$ generator and a level-$q$ generator is a combination of level-($p+q$) generators. 

For later convenience, we denote the set of positive-level generators as $\{R^{\Pa}\}$ and the negative-level generators as $\{R_{\Pa}\}$\,. 
We also denote the set of level-0 generators other than $K^a{}_b$ as $\{R_{\cI}\}$\,. 
Then, for each duality group, we have
\begin{align}
 \text{$\OO(d,d)$: }\quad \{R^{\Pa}\} & = \{R^2\}\,,\qquad \{R_{\Pa}\} = \{R_2\}\,,\quad \{R_{\cI}\}=\emptyset\,, 
\\
 \text{M-theory: }\quad \{R^{\Pa}\} &= \{R^3,\,R^6,\,R^{8,1}\}\,,\qquad \{R_{\Pa}\} = \{R_3,\,R_6,\,R_{8,1}\}\,,\quad \{R_{\cI}\}=\emptyset\,, 
\\
\begin{split}
 \text{Type IIB: }\quad \{R^{\Pa}\} &= \{R_\alpha^2,\,R^4,\,R_\alpha^6,\,R^{7,1}\}\,,\qquad
 \{R_{\Pa}\} = \{R^\alpha_2,\,R_4,\,R^\alpha_6,\,R_{7,1}\}\,,
\\
 \{R_{\cI}\} &= \{R^\alpha{}_\beta\}\,. 
\end{split}
\end{align}

We note that the commutation relations among the $E_{n(n)}$ generators both in the M-theory and the type IIB picture are given in Appendix \ref{app:En-algebra}. 
The commutation relations for the $\OO(d,d)$ algebra are given in Eq.~\eqref{eq:Odd-algebra}. 
In each duality group, we see that the generators are representation of the subalgebra $\mathfrak{h}$ generated by the level-0 generators $\{K^a{}_b,R_{\cI}\}$.

The properties of the duality group $\cG$ that are important in the following discussion can be summarized as follows:
\begin{enumerate}
\item The generators $\{t_\alpha\}$ can be decomposed into the positive-level generators $\{R^{\Pa}\}$\,, negative-level generators $\{R_{\Pa}\}$\,, and the level-0 generators $\{K^a{}_b,\,R_{\cI}\}$.

\item $\{R^{\Pa}\}$ and $\{R_{\Pa}\}$ are representations of the subalgebra $\mathfrak{h}$ generated by $\{K^a{}_b,\,R_{\cI}\}$.
\end{enumerate}

\subsection{Vector representation $R_1$}
\label{sec:R1}

The usual momenta $P_a$ are related to various brane central charges via duality. 
For example, under $T$-duality, the momenta $P_a$ are mapped to the string winding charges $P^a$, and under $U$-duality, they are mapped to brane central charges.
All of the central charges are packaged to the generalized momenta $P_A$ and they transform in the vector representation of $\cG$ which we call the $R_1$ representation. 
In each duality group, we can summarize the contents of the generalized momenta as follows:\footnote{For $\cG=E_{8(8)}$, $P^{7,1}$ in M-theory and $P^{6,1}$ in type IIB theory are reducible. For example, $P^{7,1}$ can be decomposed into $P^{a_1\cdots a_7,a'}$ with $P^{[a_1\cdots a_7,a']}=0$ and the totally antisymmetric part $P^8$\,. However, this decomposition increase the matrix size of $(t_{\bm{\alpha}})_A{}^B$ and we do not consider such a decomposition.}
\begin{align}
 \text{$\OO(d,d)$: }\quad \{P_A\} &= \{P_1,\,P^1\}\,, 
\\
 \text{M-theory: }\quad \{P_A\} &=\{P_1,\,P^2,\,P^5,\,P^{7,1},\,P^{8,3},\,P^{8,6},\,P^{8,8,1}\}\,,
\\
 \text{Type IIB: }\quad \{P_A\} &=\{P_1,\,P^1_\alpha,\,P^3,\,P^5_\alpha,\,P^{6,1},\,P^{7,2}_\alpha,\,P^{7,4},\,P^{7,6}_\alpha,\,P^{7,7,1}\}\,. 
\end{align}
Since the generalized momenta are in a representation of the duality group, we have
\begin{align}
 [t_{\bm{\alpha}},\,P_A] = (t_{\bm{\alpha}})_A{}^B\,P_B\,.
\end{align}
By assuming the central charges are Abelian $[P_A,\,P_B]=0$ and using the Jacobi identities, the explicit forms of the matrices $(t_{\bm{\alpha}})_A{}^B$ were determined in \cite{hep-th:0307098,1111.0459,1303.2035,1405.7894} (see also \cite{1909.01335}).
They provide the matrix representation of the generators $\{t_{\bm{\alpha}}\}$ in the $R_1$ representation. 

We can also assign the level to the central charges $P_A$ as
\begin{align}
 [K,\,P_A] = \bigl(\sigma\,\ell_A - 1\bigr) \,P_A \,.
\end{align}
For example, $P_a$ has level 0 and other central charges have positive levels. 
Then, by considering the level, we find important properties
\begin{align}
 (R_{\Pa})_a{}^B = 0\,, \qquad
 (R^{\Pa})_A{}^b = 0\,, \qquad 
 (R_{\cI})_A{}^b = 0 \,. 
\label{eq:R-a-properties}
\end{align}
The first can be understood from
\begin{align}
 [R_{\Pa},\,P_a] = (R_{\Pa})_a{}^B\,P_B = \text{(sum of the central charges with negative level)}=0\,,
\end{align}
and the second can be also understood in a similar way.
The last property follows from the fact that $P_a$ is a singlet under the subalgebra generated by $\{R^{\cI}\}$: $[P_a,R^{\cI}]=0$\,.
These properties are very important in the construction of ExDA, and in the following discussion we frequently use them implicitly.
Moreover, since each component of the central charges are representations of $\mathfrak{h}$\,, the matrices $(K^c{}_d)_A{}^B$ and $(R_{\cI})_A{}^B$ have the block-diagonal form, which is also used frequently. 

\subsection{$Y$-tensor}
\label{sec:Y-tensor}

Now we explain the definition of the $Y$-tensor $Y^{IJ}_{KL}$\,. 
This is defined as \cite{1208.5884}
\begin{align}
 Y^{IJ}_{KL} \equiv \delta^I_K\,\delta^J_L + (t^{\bm{\alpha}})_K{}^J\, (t_{\bm{\alpha}})_L{}^I + \beta \,\delta^I_L\,\delta^J_K \,,
\label{eq:Y-tensor-def}
\end{align}
where $\beta$ is given by
\begin{align}
 \cG=\OO(d,d): \ \beta=0\,,\qquad \cG=E_{n(n)}:\ \beta= \frac{1}{9-n}\,,
\end{align}
and $\{t^{\bm{\alpha}}\}$ is defined as follows. 
For a while, we suppose that the duality group $\cG$ is simple, and define the Cartan--Killing form as
\begin{align}
 \kappa_{\bm{\alpha}\bm{\beta}} \equiv -\frac{1}{\alpha}\,\tr_{R_1}(t_{\bm{\alpha}}\,t_{\bm{\beta}})\,,\qquad 
 (\kappa^{\bm{\alpha}\bm{\beta}}) \equiv (\kappa_{\bm{\alpha}\bm{\beta}})^{-1}\,,
\end{align}
where the constant $\alpha$ for each duality group is
\begin{align}
\begin{array}{c||c|c|c|c|c|c}
 \cG & \OO(d,d) & E_{4(4)} & E_{5(5)} & E_{6(6)} & E_{7(7)} & E_{8(8)} \\\hline
 \alpha & 2 & 3 & 4 & 6 & 12 & 60
\end{array}\,. 
\end{align}
Then we define $t^{\bm{\alpha}}$ as 
\begin{align}
 t^{\bm{\alpha}} \equiv \kappa^{\bm{\alpha}\bm{\beta}}\, t_{\bm{\beta}} \,. 
\end{align}
In our convention, the are found as follows:
\begin{align}
\begin{split}
 \text{$\OO(n,n)$: }\quad \{t^{\bm{\alpha}}\} &= \{R^2,\, K_1{}^1,\, R_2 \}\,,
\\
 \text{M-theory: }\quad \{t^{\bm{\alpha}}\} &= \{R^{8,1},\,R^6,\, R^3,\, K_1{}^1,\, R_3,\, R_6,\,R_{8,1}\}\,,
\\
 \text{Type IIB: }\quad \{t^{\bm{\alpha}}\} &= \{R^{7,1},\,R_\alpha^6,\,R^4,\,R_\alpha^2,\,K_1{}^1,\,R_\alpha{}^\beta,\,R^\alpha_2,\,R_4,\,R^\alpha_6,\,R_{7,1}\}\,,
\end{split}
\label{eq:t^alpha-def}
\end{align}
where $K_1{}^1$ and $R_\alpha{}^\beta$ are defined as
\begin{align}
\begin{split}
 \text{$\OO(n,n)$: }\quad K_a{}^b &\equiv -K^b{}_a\,,
\\
 \text{M-theory: }\quad K_a{}^b &\equiv -\bigl(K^b{}_a-\tfrac{1}{9}\,\delta_a^b\,K\bigr) \,,
\\
 \text{Type IIB: }\quad K_{\sfa}{}^{\sfb} &\equiv -\bigl(K^{\sfb}{}_{\sfa}-\tfrac{1}{8}\,\delta_{\sfa}^{\sfb}\,K\bigr) \,,\qquad 
 R_\alpha{}^\beta \equiv -R^\beta{}_\alpha\,.
\end{split}
\end{align}
For non-simple $U$-duality groups $E_{n(n)}$ ($n\leq 3$), we adopt Eq.~\eqref{eq:t^alpha-def} as the definition of $\{t^{\bm{\alpha}}\}$. 
Then, we find the identities
\begin{align}
 \OO(d,d):&\quad (t^{\bm{\alpha}})_A{}^B\, (t_{\bm{\alpha}})_C{}^D = -2\,(P_{\tt adj})_A{}^B{}_C{}^D \,, 
\\
 n=2:&\quad (t^{\bm{\alpha}})_A{}^B\, (t_{\bm{\alpha}})_C{}^D = -(P_{\bm{3}})_A{}^B{}_C{}^D -\tfrac{17}{7}\,(P_{\bm{1}})_A{}^B{}_C{}^D \,, 
\\
 n=3:&\quad (t^{\bm{\alpha}})_A{}^B\, (t_{\bm{\alpha}})_C{}^D = -2\,(P_{\bm{(8,1)}})_A{}^B{}_C{}^D -3\,(P_{\bm{(1,3)}})_A{}^B{}_C{}^D \,, 
\\
 n\geq 4: &\quad (t^{\bm{\alpha}})_A{}^B\, (t_{\bm{\alpha}})_C{}^D = -\alpha\,(P_{\tt adj})_A{}^B{}_C{}^D \,, 
\end{align}
and $(t^{\bm{\alpha}})_A{}^B\, (t_{\bm{\alpha}})_C{}^D$ is always the projector on the adjoint representation (see Table \ref{tab:reps}).
Now, the right-hand side of \eqref{eq:Y-tensor-def} is defined and the $Y$-tensor is explicitly defined. 

For example, when $\cG=\OO(d,d)$, we can show
\begin{align}
 Y^{IJ}_{KL} =\delta^I_K\,\delta^J_L + (t^{\bm{\alpha}})_K{}^J\, (t_{\bm{\alpha}})_L{}^I = \eta^{IJ}\,\eta_{KL}\,,
\end{align}
where $\eta^{IJ}$ is the inverse matrix of $\eta_{IJ}$\,. 

For the $U$-duality group $\cG=E_{n(n)}$, the $Y$-tensor can be expressed as a sum of projectors on the section-condition multiplet $R_{{\tt SC}}$\footnote{The representation is called the section-condition multiplet because the section condition is expressed as \eqref{eq:SC} by using the $Y$-tensor.} (see Table \ref{tab:reps}) \cite{1208.5884}
\begin{align}
\begin{split}
 n=2 : &\quad Y^{IJ}_{KL}
 = 2\,(P_{\bm{2}})^{IJ}{}_{KL}\,,
\\
 n=3 : &\quad Y^{IJ}_{KL}
 = 4\,(P_{\bm{(3,1)}})^{IJ}{}_{KL}\,,
\\
 n=4 : &\quad Y^{IJ}_{KL}
 = 6\,(P_{\bm{\overline{5}}})^{IJ}{}_{KL}\,,
\\
 n=5 : &\quad Y^{IJ}_{KL}
 = 8\,(P_{\bm{10}})^{IJ}{}_{KL}\,,
\\
 n=6 : &\quad Y^{IJ}_{KL}
 = 10\,(P_{\bm{\overline{27}}})^{IJ}{}_{KL}\,,
\\
 n=7 : &\quad Y^{IJ}_{KL}
 = 12\,(P_{\bm{133}})^{IJ}{}_{KL} - 28\,(P_{\bm{1}})^{IJ}{}_{KL} \,,
\\
 n=8 : &\quad Y^{IJ}_{KL}
 = 14\,(P_{\bm{3875}})^{IJ}{}_{KL} -30\,(P_{\bm{248}})^{IJ}{}_{KL} + 62\,(P_{\bm{1}})^{IJ}{}_{KL} \,.
\end{split}
\end{align}
For each $n$, the projector on the largest irreducible representation in $R_{{\tt SC}}$ has the symmetries $(P_{\bm{\cdots}})^{IJ}{}_{KL}=(P_{\bm{\cdots}})^{(IJ)}{}_{KL}=(P_{\bm{\cdots}})^{IJ}{}_{(KL)}$\,. 
Moreover, similar to the DFT case, they are expressed as a quadratic form of the $\eta$-symbol, $\eta_{IJ}{}^{\cK}=\eta_{JI}{}^{\cK}$ \cite{1610.01620,1708.06342},
\begin{align}
 2\,(n-1)\,(P_{\bm{\cdots}})^{IJ}{}_{KL} = \eta^{IJ}{}_{\cI}\,\eta_{KL}{}^{\cI}\,.
\end{align}
For the second largest irreducible representation in $n=7,8$\,, the indices are antisymmetric
\begin{align}
 (P_{\bm{\cdots}})^{IJ}{}_{KL}=(P_{\bm{\cdots}})^{[IJ]}{}_{KL}=(P_{\bm{\cdots}})^{IJ}{}_{[KL]}\,,
\end{align}
and for the smallest irreducible representation in $n=8$, the indices are symmetric
\begin{align}
 (P_{\bm{1}})^{IJ}{}_{KL}=(P_{\bm{1}})^{(IJ)}{}_{KL}=(P_{\bm{1}})^{IJ}{}_{(KL)}\,.
\end{align}

\subsection{More on the generalized Lie derivative}

Here we explain the generalized Lie derivative in more detail. 
In ExFT, all fields and gauge parameters should satisfy the section condition \eqref{eq:SC}. 
As long as the section condition is satisfied, the algebra of the generalized Lie derivative is closed, in the sense that, for given generalized vector fields $V_1^I,\,V_2^I,\,V_3^I$\,, we can find a vector field $V_{12}^I$ satisfying
\begin{align}
 [\gLie_{V_1},\, \gLie_{V_2}] V_3^I = \gLie_{V_{12}} V_3^I \,.
\label{eq:gLie-closed}
\end{align}
However, the algebra is not closed in $E_{8(8)}$ EFT \cite{1208.5884}. 
In \cite{1406.3348}, by introducing additional gauge parameter $\Sigma_I$, the generalized Lie derivative was modified as
\begin{align}
 \gLie_{(V,\Sigma)} W^I \equiv \gLie_{V} W^I + \Sigma_K\,f^{KI}{}_J\,W^J\,,
\end{align}
where $f^{KI}{}_J$ corresponds to the structure constants of the $\mathfrak{e}_{8(8)}$ algebra. 
In this approach, by requiring that $\Sigma_I$ satisfies certain conditions similar to the section condition, it was shown that the modified algebra is closed in the following sense:
\begin{align}
 [\gLie_{(V_1,\Sigma_1)},\, \gLie_{(V_2,\Sigma_2)}] V_3^I = \gLie_{(V_{12},\Sigma_{12})} V_3^I \,.
\end{align}
This approach has the advantage that it can be applied to any curved space. 
However, the geometric meaning of $\Sigma_I$ is not clear. 
As explained in section \ref{sec:introduction}, we would like to define the structure constants $X_{AB}{}^C$ of the ExDA through the relation
\begin{align}
 \gLie_{E_A} E_B{}^I = - X_{AB}{}^C\,E_C{}^I \,,
\end{align}
but since the meaning of $\Sigma_I$ is not clear, it is not obvious how to extend this relation to the $E_{8(8)}$ case. 
We thus take another approach proposed in \cite{1410.8148}. 
There, by assuming the existence of certain generalized frame fields $E_A{}^I\in \cG\times\lR^+$, the gauge parameter $\Sigma_I$ was chosen as
\begin{align}
 \Sigma_I = -\frac{1}{\alpha}\,f_{IJ}{}^L\,\tilde{\Omega}_{KL}{}^J\, V^K \,,\qquad \tilde{\Omega}_{IJ}{}^K \equiv \Omega_I{}^{\bm{\dot\alpha}}\,(t_{\bm{\dot\alpha}})_J{}^K\,,
\end{align}
where $\Omega$ is the Weizenb\"ock connection
\begin{align}
 \Omega_{IJ}{}^K \equiv E_A{}^K\,\partial_I E_J{}^A \equiv \Omega_I{}^{\bm{\dot\alpha}}\,(t_{\bm{\dot\alpha}})_J{}^K + \Omega_I{}^0\,(t_0)_J{}^K \,,
\end{align}
and $(t_{\bm{\dot\alpha}})_I{}^J$ is the matrix representation of the $\mathfrak{e}_{8(8)}$ generator in the ``curved'' indices. 
This modified generalized Lie derivative
\begin{align}
 \gLie_{V} W^I \equiv V^J\,\partial_J W^I - W^J\,\partial_J V^I + Y^{IK}_{JL}\,\partial_K V^L\, W^J + \Sigma_K\,f^{KI}{}_J\,V^J\,,
\label{eq:gLie-E8}
\end{align}
was shown to be closed in the usual sense \eqref{eq:gLie-closed}. 
A disadvantage of this approach is that this requires the particular generalized frame fields $E_A{}^I$, but in the discussion of the ExDA, we have globally defined generalized frame fields $E_A{}^I$. 
Thus we use the corresponding Weizenb\"ock connection and adopt the generalized Lie derivative \eqref{eq:gLie-E8}. 

Since the expression \eqref{eq:gLie-E8} contains a quantity $f^{IJ}{}_K$ that is specific to the $E_{8(8)}$ EFT, we rewrite the definition as
\begin{align}
\begin{split}
 \gLie_{V} W^I &\equiv V^J\,\partial_J W^I - W^J\,\partial_J V^I + Y^{IK}_{JL}\,\partial_K V^L\, W^J - \Sigma^{\bm{\dot\alpha}} \, (t_{\bm{\dot\alpha}})_J{}^I\, W^J 
\\
 \bigl(\Sigma^{\bm{\dot\alpha}} &\equiv \chi^{I\bm{\dot\alpha}}\,\Omega_I{}^{\bm{\dot\beta}}\,\chi_{\bm{\dot\beta}K}\,V^K \bigr)\,,
\end{split}\end{align}
where the matrix $\chi_{\bm{\dot\alpha}I}$ and its inverse $\chi^{I\bm{\dot\alpha}}$ are given in Appendix \ref{app:chi}. 
In the $E_{8(8)}$ EFT, $\chi$ is an intertwiner connecting two representations $R_1$ and $R_{\tt adj}$, but it vanishes in the $E_{n(n)}$ EFT ($\leq 7$).
In DFT, we understand that $\chi$ is not present. 
This generalized Lie derivative reproduces \eqref{eq:gLie-E8} by considering $\Sigma_I \equiv \Sigma^{\bm{\dot\alpha}}\,\chi_{\bm{\dot\alpha}I}$ and $f^{IJ}{}_K\equiv -\chi^{I\bm{\dot\alpha}}\,(t_{\bm{\dot\alpha}})_K{}^J$ in the $E_{8(8)}$ EFT, and reduces to the ordinary generalized Lie derivative in the $E_{n(n)}$ EFT ($\leq 7$) and DFT. 

By using Eq.~\eqref{eq:Y-tensor-def}, it is convenient to express the generalized Lie derivative as
\begin{align}
\begin{split}
 &\gLie_V W^I \equiv V^J\,\partial_J W^I - \bigl[(\partial\times_{\text{ad}} V)^{\bm{\dot\alpha}} + \Sigma^{\bm{\dot\alpha}}\bigr] \, (t_{\bm{\dot\alpha}})_J{}^I\, W^J - \beta \,(\partial\cdot V)\,(t_0)_J{}^I \, W^J \,, 
\\
 &(V\times_{\text{ad}}W)^{\bm{\dot\alpha}} \equiv - (t^{\bm{\dot\alpha}})_L{}^K\,V_K\,W^L\,,\quad 
 (V\cdot W) \equiv V_I\,W^I\,,\quad
 (t_0)_I{}^J \equiv -\delta_I^J\,.
\end{split}
\label{eq:gen-Lie}
\end{align}
This expression clearly shows that the generalized Lie derivative generates an infinitesimal $\cG$ transformation and a scale transformation $\mathbb{R}^+$ generated by $t_0$. 
The constant $\beta$ can be understood as the density weight of the gauge parameter $V^I$\,. 

For a given set of generalized frame fields $E_A{}^I$ satisfying the section condition, we can expand the generalized Lie derivative as
\begin{align}
 \gLie_{E_A} E_B{}^I = - \bm{X}_{AB}{}^C\,E_C{}^I \,,
\end{align}
where the generalized fluxes $\bm{X}_{AB}{}^C$ are not constant in general. 
By using Eq.~\eqref{eq:gen-Lie}, the generalized fluxes can be expressed as
\begin{align}
 \bm{X}_{AB}{}^C &= \Omega_{AB}{}^C + (t^{\bm{\alpha}})_D{}^E\, (t_{\bm{\alpha}})_B{}^C\,\Omega_{EA}{}^D - \beta\, \Omega_{DA}{}^D\,(t_0)_B{}^C 
 +\chi^{D\bm{\alpha}}\,\Omega_D{}^{\bm{\beta}}\,\chi_{\bm{\beta}A}\, (t_{\bm{\alpha}})_B{}^C \,,
\end{align}
where we have converted the ``curved'' indices $I,J,\bm{\dot\alpha}$ to the ``flat'' indices $A,B,\bm{\alpha}$ by using the frame fields $E_A{}^I$\footnote{The flat and curved adjoint indices are connected by $\cE_{\bm{\dot\alpha}}{}^{\bm{\alpha}}$, which is defined as $E_A{}^I\,(t_{\bm{\dot\alpha}})_I{}^J\,E_J{}^B= \cE_{\bm{\dot\alpha}}{}^{\bm{\alpha}}\,(t_{\bm{\alpha}})_A{}^B$ such that the matrix elements of the $(t_{\bm{\alpha}})_B{}^C$ and $(t_{\bm{\dot\alpha}})_I{}^J$ are the same.} and, for example, the Weizenb\"ock connection with the ``flat'' indices is
\begin{align}
 \Omega_{AB}{}^C \equiv E_A{}^I\,E_B{}^J\,E_K{}^C\,\Omega_{IJ}{}^K = E_A{}^I\,E_B{}^K\,\partial_I E_K{}^C \,.
\end{align}
Using the expansion of the Weizenb\"ock connection,
\begin{align}
 \Omega_{AB}{}^C = \Omega_A{}^{\bm{\alpha}}\,t_{\bm{\alpha}} + \Omega_A{}^0\,t_0\,,
\label{eq:Weizenbock-expand}
\end{align}
we find
\begin{align}
\begin{split}
 \bm{X}_{AB}{}^C &= \bigl[ \delta^D_A \,\delta_{\bm{\alpha}}^{\bm{\beta}} + (t_{\bm{\beta}}\,t^{\bm{\alpha}})_A{}^D +\chi_{\bm{\beta}A}\,\chi^{D\bm{\alpha}} \bigr]\,\Omega_D{}^{\bm{\beta}}\,(t_{\bm{\alpha}})_B{}^C - \beta \, \Omega_{DA}{}^D\,(t_0)_B{}^C
\\
 &\quad - (t^{\bm{\alpha}})_A{}^D\,\Omega_D{}^0\,(t_{\bm{\alpha}})_B{}^C + \Omega_A{}^0\,(t_0)_B{}^C \,.
\end{split}
\end{align}
By introducing
\begin{align}
 \vartheta_A \equiv -\frac{1}{D}\,\bm{X}_{AB}{}^B = (1+\beta)\,\Omega_A{}^0 - \beta \, \Omega_D{}^{\bm{\alpha}}\,(t_{\bm{\alpha}})_A{}^D \,, 
\end{align}
we can further rewrite the generalized fluxes as
\begin{align}
 \bm{X}_{AB}{}^C = \Theta_A{}^{\bm{\alpha}}\,(t_{\bm{\alpha}})_B{}^C
 -\bigl[\tfrac{1}{1+\beta}\,(t^{\bm{\alpha}})_A{}^D\,(t_{\bm{\alpha}})_B{}^C + \delta_A^D\,\delta_B^C \bigr]\,\vartheta_D\,,
\label{eq:gen-flux}
\end{align}
where
\begin{align}
\begin{split}
 \Theta_A{}^{\bm{\alpha}} &\equiv \mathbb{P}_A{}^{\bm{\alpha}}{}^B{}_{\bm{\beta}}\,\Omega_B{}^{\bm{\beta}} \,,
\\
 \mathbb{P}_A{}^{\bm{\alpha}}{}^B{}_{\bm{\beta}}
 &\equiv \delta^B_A \,\delta_{\bm{\beta}}^{\bm{\alpha}} + (t_{\bm{\beta}}\,t^{\bm{\alpha}})_A{}^B - \tfrac{\beta}{1+\beta}\,(t^{\bm{\alpha}}\,t_{\bm{\beta}})_A{}^B + \chi_{\bm{\beta}A}\,\chi^{B\bm{\alpha}}\,.
\end{split}
\end{align}

In DFT (where $\beta=\vartheta=0$), $\mathbb{P}$ corresponds to the projector on the multiplet of the 3-form
\begin{align}
 \mathbb{P}_A{}^{\bm{\alpha}}{}^B{}_{\bm{\beta}}
 = \delta^B_A \,\delta_{\bm{\beta}}^{\bm{\alpha}} + (t_{\bm{\beta}}\,t^{\bm{\alpha}})_A{}^B = 3\,P_{{\binom{\bm{2d}}{\bm 3}}}\,,
\label{eq:DFT-flux}
\end{align}
and the embedding tensor is a 3-form. 
Accordingly, the generalized fluxes (with the last index lowered with $\eta_{AB}$) are totally antisymmetric $\bm{X}_{ABC} = \bm{X}_{[ABC]}$\,. 

In EFT with $n\leq 8$\,, $\mathbb{P}_A{}^{\bm{\alpha}}{}^B{}_{\bm{\beta}}$ is always a projector on the $R_{10-n}$ representation (see for example the Appendix of \cite{hep-th:0212239} for $n=5,6,7$)\footnote{In \cite{1410.8148}, the modification of the generalized Lie derivative in $n=8$ was motivated by this relation.}
\begin{align}
\begin{split}
 n=2 : &\quad \mathbb{P}_A{}^{\bm{\alpha}}{}^B{}_{\bm{\beta}}
 = (P_{\bm{3}})_A{}^{\bm{\alpha}}{}^B{}_{\bm{\beta}}
 + \tfrac{17}{8}\,(P_{\bm{2}})_A{}^{\bm{\alpha}}{}^B{}_{\bm{\beta}}\,,
\\
 n=3 : &\quad \mathbb{P}_A{}^{\bm{\alpha}}{}^B{}_{\bm{\beta}}
 = 2\,(P_{\bm{(6,2)}})_A{}^{\bm{\alpha}}{}^B{}_{\bm{\beta}}
   + \tfrac{24}{7}\,(P_{\bm{(\bar{3},2)}})_A{}^{\bm{\alpha}}{}^B{}_{\bm{\beta}}\,,
\\
 n=4 : &\quad \mathbb{P}_A{}^{\bm{\alpha}}{}^B{}_{\bm{\beta}}
 = 3\,(P_{\bm{\overline{40}}})_A{}^{\bm{\alpha}}{}^B{}_{\bm{\beta}}
   + 4\,(P_{\bm{\overline{15}}})_A{}^{\bm{\alpha}}{}^B{}_{\bm{\beta}}\,,
\\
 n=5 : &\quad \mathbb{P}_A{}^{\bm{\alpha}}{}^B{}_{\bm{\beta}}
 = 4\,(P_{\bm{144}})_A{}^{\bm{\alpha}}{}^B{}_{\bm{\beta}}\,,
\\
 n=6 : &\quad \mathbb{P}_A{}^{\bm{\alpha}}{}^B{}_{\bm{\beta}}
 = 5\,(P_{\bm{351}})_A{}^{\bm{\alpha}}{}^B{}_{\bm{\beta}}\,,
\\
 n=7 : &\quad \mathbb{P}_A{}^{\bm{\alpha}}{}^B{}_{\bm{\beta}}
 = 7\,(P_{\bm{912}})_A{}^{\bm{\alpha}}{}^B{}_{\bm{\beta}}\,,
\\
 n=8 : &\quad \mathbb{P}_A{}^{\bm{\alpha}}{}^B{}_{\bm{\beta}} 
 = 14\, (P_{\bm{3875}})_A{}^{\bm{\alpha}}{}^B{}_{\bm{\beta}} 
 + 62\, (P_{\bm{1}})_A{}^{\bm{\alpha}}{}^B{}_{\bm{\beta}} \,.
\end{split}
\end{align}
Thus, the embedding tensor $\Theta_A{}^{\bm{\alpha}}$ in each dimension transforms in the $R_{10-n}$ representation. 

Before closing this section, let us provide an additional useful information. 
In order to evaluate the generalized fluxes $\bm{X}_{AB}{}^C$, we need the explicit form of the matrix $(t^{\bm{\alpha}})_A{}^B$\,. 
In particular, the component $(K_c{}^d)_A{}^B$ frequently appears. 
In DFT and EFT both in the M-theory and the type IIB picture, $(K_c{}^d)_A{}^B$ always has the following form:
\begin{align}
 K_c{}^d &\equiv \begin{pmatrix}
 -\delta_a^d\,\delta_c^b & \bm{0} \\
 \bm{0} & * 
 \end{pmatrix},
\end{align}
where $*$ depends on the details of the ExFT. 

\section{General structure of ExDA}
\label{sec:ExDA}

\subsection{Definition of ExDA}

The ExDA is a Leibniz algebra with certain structure constants $X_{AB}{}^C$,
\begin{align}
 T_A\circ T_B = X_{AB}{}^C\,T_C\,.
\end{align}
As explained in section \ref{sec:E-model}, in order to realize the non-Abelian duality, we need a set of generalized frame fields $E_A{}^I$ that satisfies the algebra
\begin{align}
 \gLie_{E_A} E_B{}^I = - X_{AB}{}^C\,E_C{}^I \,.
\end{align}
This determine the possible forms of the structure constants $X_{AB}{}^C$\,.
Recalling the general form of the generalized Lie derivative \eqref{eq:gen-flux}, we find that $X_{AB}{}^C$ should be of the form
\begin{align}
 X_{AB}{}^C = \Theta_A{}^{\bm{\alpha}}\,(t_{\bm{\alpha}})_B{}^C
 -\bigl[\tfrac{1}{1+\beta}\,(t^{\bm{\alpha}})_A{}^D\,(t_{\bm{\alpha}})_B{}^C + \delta_A^D\,\delta_B^C\bigr]\,\vartheta_D\,,
\label{eq:X-Theta}
\end{align}
where the embedding tensors are given by the Weizenb\"ock connection,
\begin{align}
 \Theta_A{}^{\bm{\alpha}} \equiv \mathbb{P}_A{}^{\bm{\alpha}}{}^B{}_{\bm{\beta}}\,\Omega_B{}^{\bm{\beta}} \,,\qquad
 \vartheta_A = (1+\beta)\,\Omega_A{}^0 - \beta \, \Omega_D{}^{\bm{\alpha}}\,(t_{\bm{\alpha}})_A{}^D \,.
\label{eq:theta}
\end{align}
In general, the Weizenb\"ock connection $\Omega_{AB}{}^C$ on the right-hand side is not constant, but it turns out that the non-constant parts are canceled out on the right-hand side (see section \ref{sec:parallelizability}). 
Then, in order to evaluate the structure constants $X_{AB}{}^C$\,, it is enough to know the value of $\Omega_{AB}{}^C$ at a certain point $x_0$\,. 
Namely, in the embedding tensors \eqref{eq:theta}, we can replace $\Omega_{AB}{}^C$ with the constants
\begin{align}
 F_{AB}{}^C \equiv \Omega_{AB}{}^C\bigr\rvert_{x=x_0} \equiv \bigl(F_A{}^{\bm{\alpha}}\,t_{\bm{\alpha}} + F_A{}^0\,t_0\bigr)_B{}^C \,.
\label{eq:F-Omega}
\end{align}
In the following, we define the ExDA by requiring additional conditions on $F_{AB}{}^C$\,. 

\subsubsection{Requirement: Exitence of maximally isotropic subalgebra}

The important requirement is that the ExDA contains an $\mathfrak{n}$-dimensional subalgebra $\mathfrak{g}$
\begin{align}
 T_a \circ T_b = f_{ab}{}^c\,T_c\,,
\label{eq:subalgebra2}
\end{align}
that satisfies the maximal isotropicity
\begin{align}
 \langle T_a,\,T_b \rangle = 0\qquad \bigl[\ \langle T_A,\,T_B \rangle \equiv Y_{AB}^{CD}\,T_C\otimes T_D\ \bigr]\,.
\end{align}
Once the subalgebra $\mathfrak{g}$ is fixed, as explained in section \ref{sec:generators}, we can decompose the generators of the duality group $\cG$ such that the level-0 generators contain the $\GL(\mathfrak{n})$ generators $K^a{}_b$\,.
We can then decompose the $R_1$ representation into representations of the subalgebra $\mathfrak{h}$ (recall section \ref{sec:R1}): $\{T_A\}=\{T_a,\,T_{\tilde{C}}\}$ where $T_{\tilde{C}}$ are called the dual generators. 
In general, we have
\begin{align}
 T_a \circ T_b = X_{ab}{}^c\,T_c + X_{ab}{}^{\tilde{C}}\,T_{\tilde{C}} \,,
\end{align}
and the requirement \eqref{eq:subalgebra2} is equivalent to
\begin{align}
 X_{ab}{}^c=f_{ab}{}^c\,,\qquad X_{ab}{}^{\tilde{C}} = 0 \,. 
\label{eq:subalgebra3}
\end{align}
In order to consider the second condition, let us use an expression
\begin{align}
 X_{ab}{}^{\tilde{C}} &= \bigl[ \delta^D_a \,\delta_{\bm{\alpha}}^{\bm{\beta}} + (t_{\bm{\beta}}\,t^{\bm{\alpha}})_a{}^D +\chi_{\bm{\beta}a}\,\chi^{D\bm{\alpha}} \bigr]\,F_D{}^{\bm{\beta}}\,(t_{\bm{\alpha}})_b{}^{\tilde{C}} 
\nn\\
&= \bigl[ 2\,\delta^D_{[a|}\,(t_{\bm{\alpha}})_{|b]}{}^{\tilde{C}}
 + Y^{\tilde{C}D}_{Eb}\,(t_{\bm{\alpha}})_{a}{}^E
 + \chi_{\bm{\alpha}a}\,\chi^{D\bm{\beta}}\, (t_{\bm{\beta}})_b{}^{\tilde{C}} \bigr]\,F_{D}{}^{\bm{\alpha}}\,.
\end{align}
In general, the constants inside the square brackets do not vanish and in order to realize $X_{ab}{}^{\tilde{C}}=0$\,, we need to impose a condition on $F_A{}^{\bm{\alpha}}$\,. 
By considering $(R_{\Pa})_a{}^B=0$\,, $Y^{CD}_{eb}=0$ (i.e., the maximal isotropicity), and $\chi_{K^b{}_c a}=\chi_{R_{\cI} a}=\chi_{R_{\Pa} a}=0$\,, the (sufficient) condition is
\begin{align}
 F_A{}^{\bm{\alpha}}\,t_{\bm{\alpha}} = F_{Ab}{}^c\,K^b{}_c + F_A{}^{\cI}\,R_{\cI} + F_{A}{}^{\Pb}\, R_{\Pb} \qquad (\text{i.e., }F_{A\Pb} = 0)\,. 
\label{eq:subalgebra4}
\end{align}
Under this condition, we find
\begin{align}
 X_{ab}{}^c = 2\,F_{[ab]}{}^c\,, 
\end{align}
and the subalgebra $\mathfrak{g}$ is a Lie algebra
\begin{align}
 T_a \circ T_b = - T_b \circ T_a = f_{ab}{}^c\,T_c \qquad
 \bigl(f_{ab}{}^c = 2\,F_{[ab]}{}^c\bigr) \,.
\end{align}

Now let us explain the consequences of the condition \eqref{eq:subalgebra4} on the generalized frame fields $E_A{}^I$\,. 
Initially, the generalized frame fields $E_A{}^I$ were supposed to be a general element of $\cG\times\lR^+$ and then the Weizenb\"ock connection $\Omega_{AB}{}^C$ can be expanded as in Eq.~\eqref{eq:Weizenbock-expand}. 
However, $F_{AB}{}^C$ and $\Omega_{AB}{}^C$ are related as
\begin{align}
 F_A{}^{\bm{\alpha}}\,t_{\bm{\alpha}} + F_A{}^0\,t_0
 = \bigl[\Omega_A{}^{\bm{\alpha}}\,t_{\bm{\alpha}} + \Omega_A{}^0\,t_0 \bigr]_{x=x_0}
 = E_A{}^J\,E_B{}^I\,\partial_J E_I{}^C\bigr\rvert_{x=x_0}\,,
\label{eq:Weizenbock}
\end{align}
and Eq.~\eqref{eq:subalgebra4} means that $E_A{}^I$ should be generated by non-positive generators and $\lR^+$,\footnote{Naively, it appears to be more natural to introduce $\Delta$ as $E_A{}^I = \bm{\Pi}_A{}^B (\Exp{t_0 \Delta})_B{}^C\, \mathbb{E}_C{}^I= \Exp{-\Delta} \bm{\Pi}_A{}^B \mathbb{E}_B{}^I$ but here we introduced $\tilde{K}$ for later convenience (by following \cite{1911.07833}).}
\begin{align}
 E_A{}^I = \bm{\Pi}_A{}^B (\Exp{(\tilde{K}+t_0)\Delta})_B{}^C\, \mathbb{E}_C{}^I\,,\qquad 
 \mathbb{E} \equiv \Exp{-h^{\cI}\,R_{\cI}} \Exp{-(h_R)_a{}^b\,\tilde{K}^a{}_b}\,,\qquad
 \bm{\Pi} \equiv \prod_{\Pa}\Exp{-\pi^{\Pa}\, R_{\Pa}} \,,
\label{eq:E-twist}
\end{align}
where $\tilde{K}\equiv \tilde{K}^a{}_a$ and the function $\Delta$ is associated with the scale symmetry $\mathbb{R}^+$ and the functions $h$ and $\pi$ parameterize a twist by the non-positive generators. 
Here, the generator 
\begin{align}
 \tilde{K}^a{}_b\equiv K^a{}_b + \beta\,\delta^a_b\,t_0\,,
\end{align}
satisfies the same $\gl(\mathfrak{n})$ algebra as $K^a{}_b$ and the matrix representation has the form
\begin{align}
 \tilde{K}^c{}_d &\equiv \begin{pmatrix}
 \delta_a^c\,\delta_d^b & \bm{0} \\
 \bm{0} & * 
 \end{pmatrix}.
\end{align}
The details of $\Delta$, $\pi^{\Pa}$, and $\mathbb{E}_A{}^I$ are explained later in section \ref{sec:frame-def}.
Here, it is enough to know that $\Delta$ and $\pi^{\Pa}$ vanish at a certain point $x_0$:
\begin{align}
 \Delta\rvert_{x=x_0} = 0\,,\qquad \pi^{\Pa}\rvert_{x=x_0} = 0\,.
\label{eq:x=x0}
\end{align}
According to this property, we can easily compute the constants $F_{AB}{}^C$ by using Eq.~\eqref{eq:Weizenbock}.

\subsubsection{Section condition on $E_A{}^I$}

Another condition on $F_{AB}{}^C$ comes from the section condition of ExFT. 
The derivative $\partial_J$ on the right-hand side of Eq.~\eqref{eq:Weizenbock} should satisfy the section condition \eqref{eq:SC}. 
As we explain in section \ref{sec:frame-def}, the generalized frame fields $E_A{}^I$ are constructed from a group element $g\equiv \Exp{x^a\,T_a}$ where $T_a$ denotes the generators of the subalgebra $\mathfrak{g}$ and $x^a\,(=\delta^a_i\,x^i)$ are the corresponding $\mathfrak{n}$ coordinates (which is sometime called the physical coordinates). 
Accordingly, here we suppose that $E_A{}^I$ depends only on the physical coordinates $x^i$, and then the section condition \eqref{eq:SC} is trivially satisfied because $Y^{ij}_{KL}=0$\,. 
Then we find 
\begin{align}
 F_{AB}{}^C = E_A{}^J\,E_B{}^I\,\partial_j E_I{}^C \bigr\rvert_{x=x_0}
 = \mathbb{E}_A{}^j\,\mathbb{E}_B{}^I\,\partial_j E_I{}^C \bigr\rvert_{x=x_0}
 = \delta_A^d\, F_{dB}{}^C\,,
\end{align}
where we have used that $\mathbb{E}_A{}^j=\delta_A^d\,\mathbb{E}_d{}^j$\,. 
By using Eq.~\eqref{eq:E-twist}, $F_{aB}{}^C$ can be parameterized as
\begin{align}
\begin{split}
 F_{aB}{}^C &= \bigl[k_{ad}{}^e\, \tilde{K}^d{}_e + f_a{}^{\cI}\, R_{\cI} + f_a{}^{\Pa}\, R_{\Pa} - Z_a\,(\tilde{K} + t_0) \bigr]_B{}^C \,,
\\
 \bigl(k_{ad}{}^e\,\tilde{K}^d{}_e + f_a{}^{\cI}\, R_{\cI}\bigr)_B{}^C &\equiv \mathbb{E}_B{}^I\,D_a \mathbb{E}_I{}^C \bigr\rvert_{x=x_0}
\,,\quad 
 f_a{}^{\Pa} \equiv D_a\pi^{\Pa}\rvert_{x=x_0}\,,\quad 
 Z_a \equiv D_a \Delta\rvert_{x=x_0} \,,
\end{split}
\end{align}
where $D_A\equiv \mathbb{E}_A{}^I\,\partial_I$.
This can be equivalently expressed as
\begin{align}
 F_a{}^{\bm{\alpha}}\,t_{\bm{\alpha}} = (k_{ab}{}^c-Z_a\,\delta_b^c)\,K^b{}_c + f_a{}^{\cI}\,R_{\cI} + f_a{}^{\Pa}\, R_{\Pa} \,,\qquad 
 F_a{}^0 = \beta\,(k_{ab}{}^b-Z_a\,\delta_b^b) - Z_a \,.
\label{eq:F-alpha-0}
\end{align}
In the next subsection, we shall use a short-hand notation
\begin{align}
 \{R_{\hat{\Pa}}\} = \{R_{\cI},\,R_{\Pa}\}\,.
\end{align}

\subsubsection{Structure constants of ExDA}

Now, using the above setup, we can write down the structure constants $X_{AB}{}^C$ of the ExDA. 
By using the formula \eqref{eq:theta} and Eq.~\eqref{eq:F-alpha-0}, the embedding tensors are obtained as
\begin{align}
\begin{split}
 \Theta_A{}^{\bm{\alpha}}\,t_{\bm{\alpha}}
 &= (f_{Ac}{}^d -2\,Z_{[A}\,\delta_{c]}^d) \,K^c{}_d
 + (k_{bc}{}^d-Z_b\,\delta_c^d) \,\bigl[(\tilde{K}^c{}_d\,R_{\Pa})_A{}^b 
 + \chi_{K^c{}_d A}\,\chi^{b}{}_{\Pa}\bigr]\, R^{\Pa} 
\\
 &\quad + f_A{}^{\hat{\Pa}}\, R_{\hat{\Pa}}
 - f_c{}^{\Pb}\, (R_{\Pb})_A{}^d\, K^c{}_d 
 + f_b{}^{\hat{\Pb}} \,(R_{\hat{\Pb}}\,R_{\Pa})_A{}^b\, R^{\Pa} 
 + f_b{}^{\hat{\Pb}} \, \chi_{{\hat{\Pb}}A}\,\chi^{b}{}_{\Pa}\, R^{\Pa} 
\\
 &\quad + \tfrac{\beta}{1+\beta}\,\bigl[ (f_{bc}{}^c-2\,Z_{[b}\,\delta_{c]}^c) \,(t^{\bm{\alpha}})_A{}^b 
 - f_b{}^{\Pb} \,(t^{\bm{\alpha}}\,R_{\Pb})_A{}^b \bigr]\, t_{\bm{\alpha}}\,,
\end{split}
\\
 \vartheta_A &= \beta\, (f_{Ab}{}^b-2\,Z_{[A}\,\delta_{b]}^b) - \beta\, f_b{}^{\Pa}\, (R_{\Pa})_A{}^b - (1 + \beta) Z_A\,,
\end{align}
where $f_{Ab}{}^c\equiv \delta_A^d\,f_{db}{}^c$, $f_{ab}{}^c\equiv 2\,k_{[ab]}{}^c$, $f_A{}^{\hat{\Pb}}\equiv \{f_A{}^{\cI},\, f_A{}^{\Pb}\} \equiv \delta_A^c\,f_c{}^{\hat{\Pb}}$, and $Z_A\equiv \delta_A^b\,Z_b$. 
Substituting these embedding tensors into Eq.~\eqref{eq:X-Theta}, we obtain the structure constants of the ExDA as
\begin{align}
 X_{AB}{}^C 
 &= f_A{}^{\hat{\Pa}}\, (R_{\hat{\Pa}})_B{}^C 
 + \bigl[f_{Ad}{}^e - f_d{}^{\Pa}\,(R_{\Pa})_A{}^e - Z_A\,\delta_d^e \bigr]\,(K^d{}_e)_B{}^C
\nn\\
 &\quad 
 +\bigl\{ (k_{de}{}^f-Z_d\,\delta_e^f)\,\bigl[(\tilde{K}^e{}_f\,R_{\Pa})_A{}^d + \chi_{K^e{}_f A} \,\chi^{d}{}_{\Pa} \bigr]
\nn\\
 &\quad\quad\ + f_d{}^{\hat{\Pb}}\,\bigl[(R_{\hat{\Pb}}\,R_{\Pa})_A{}^d + \chi_{\hat{\Pb} A} \,\chi^{d}{}_{\Pa} \bigr]
 + (R_{\Pa})_A{}^d\, Z_d\bigr\}\,(R^{\Pa})_B{}^C
\nn\\
 &\quad 
 + \bigl[\beta\, (f_{Ad}{}^d -2\,Z_{[A}\,\delta_{d]}^d) - \beta\, f_d{}^{\Pa}\,(R_{\Pa})_A{}^d - (1+\beta)\,Z_A\bigr]\, (t_0)_B{}^C \,.
\end{align}
In the second line, the constants $k_{ab}{}^c$ appear without antisymmetrizing the lower indices. 
However, we find a general property
\begin{align}
 (\tilde{K}^e{}_f\,R_{\Pa})_A{}^d + \chi_{K^e{}_f A} \,\chi^{d}{}_{\Pa}=(\tilde{K}^{[e}{}_f\,R_{\Pa})_A{}^{d]} + \chi_{K^{[e}{}_f A} \,\chi^{d]}{}_{\Pa}\,,
\end{align}
and $k_{de}{}^f$ can be replaced by its antisymmetric part $k_{[de]}{}^f=\frac{1}{2}\,f_{de}{}^f$:\footnote{In $n=8$, $\chi_{\bm{\alpha} A}$ is indispensable to maintain this property.}
\begin{align}
\begin{split}
 X_{AB}{}^C 
 &= f_A{}^{\hat{\Pa}}\, (R_{\hat{\Pa}})_B{}^C 
 + \bigl[f_{Ad}{}^e - f_d{}^{\Pa}\,(R_{\Pa})_A{}^e \bigr]\,(\tilde{K}^d{}_e)_B{}^C
\\
 &\quad 
 +\bigl\{ (\tfrac{1}{2}\,f_{de}{}^f -Z_{[d}\,\delta_{e]}^f) \,\bigl[(\tilde{K}^e{}_f\,R_{\Pa})_A{}^d + \chi_{K^e{}_f A} \,\chi^{d}{}_{\Pa} \bigr]
\\
 &\quad\quad\ + f_d{}^{\hat{\Pb}}\,\bigl[(R_{\hat{\Pb}}\,R_{\Pa})_A{}^d + \chi_{\hat{\Pb}A} \,\chi^{d}{}_{\Pa} \bigr]
 + (R_{\Pa})_A{}^d\, Z_d\bigr\}\,(R^{\Pa})_B{}^C
 - Z_A \, (\tilde{K} + t_0)_B{}^C \,.
\end{split}
\label{eq:X-ABC}
\end{align}
This is the general formula for the structure constants.

For example, in DFT where $\beta=0$ and $Z_a=0$, Eq.~\eqref{eq:X-ABC} reduces to
\begin{align}
\begin{split}
 X_{AB}{}^C 
 &= \tfrac{1}{2}\,f_A{}^{d_1d_2}\, (R_{d_1d_2})_B{}^C + \bigl[f_{Ad}{}^e - \tfrac{1}{2}\,f_d{}^{f_1f_2}\,(R_{f_1f_2})_A{}^e \bigr]\,(K^d{}_e)_B{}^C
\\
 &\quad + \tfrac{1}{4}\,f_{de}{}^f\, (K^e{}_f\,R_{g_1g_2})_A{}^d \,(R^{g_1g_2})_B{}^C \,.
\end{split}
\end{align}
Using the matrix representations of the $\mathfrak{o}(d,d)$ generators,
\begin{align}
 K^c{}_d =\begin{pmatrix} \delta^c_a\,\delta_d^b & 0 \\ 0 & -\delta^a_d\,\delta_b^c
\end{pmatrix} ,\quad
 R^{c_1c_2} = \begin{pmatrix} 0 & 2\,\delta^{c_1c_2}_{ab} \\ 0 & 0 \end{pmatrix},\quad
 R_{c_1c_2} = \begin{pmatrix} 0 & 0 \\ 2\,\delta_{c_1c_2}^{ab} & 0 \end{pmatrix},
\end{align}
which satisfy
\begin{align}
\begin{split}
 &[K^a{}_b\,,R^{c_1c_2}] = 2\,\delta^{c_1c_2}_{bd}\,R^{ad}\,,\qquad
 [K^a{}_b\,,R_{c_1c_2}] = -2\,\delta_{c_1c_2}^{ad}\,R_{bd}\,,
\\
 &[R_{a_1a_2}\,,R^{b_1b_2}] = 4\,\delta_{[a_1}^{[b_1}\,K^{b_2]}{}_{a_2]}\,,
\end{split}
\label{eq:Odd-algebra}
\end{align}
we obtain the standard Drinfel'd algebra as $T_A\circ T_B=X_{AB}{}^C\,T_C$\,,
\begin{align}
 T_a\circ T_b = f_{ab}{}^c \, T_c \,,
\quad
 T_a\circ T^b = f_a{}^{bc}\, T_c - f_{ac}{}^b \, T^c = - T^b\circ T_a \,,
\quad
 T^a\circ T^b = f_c{}^{ab}\, T^c \,.
\end{align}

Similarly, by considering the $U$-duality group $\cG=E_{n(n)}$, we can reproduce the EDA from the general formula Eq.~\eqref{eq:X-ABC}. 
In sections \ref{sec:M} and \ref{sec:IIB}, this is worked out from the M-theory and the type IIB perspective, respectively. 
Before getting into the details about the EDA, we discuss a few more general aspects of ExDA.

\subsection{Leibniz identities}

In this section, we consider the Leibniz identity
\begin{align}
 T_A \circ (T_B\circ T_C) = (T_A\circ T_B)\circ T_C + T_B\circ (T_A\circ T_C) \,.
\end{align}
In terms of the structure constants $X_{AB}{}^C$\,, this is equivalently expressed as
\begin{align}
 X_{AC}{}^D{}\,X_{BD}{}^E-X_{BC}{}^D\,X_{AD}{}^E+X_{AB}{}^D\,X_{DC}{}^E = 0\,.
\label{eq:Leibniz-id-X}
\end{align}
If we define
\begin{align}
 X_A \equiv \widehat{\Theta}_A + \vartheta_A\, t_0 \,,\qquad
 \widehat{\Theta}_A\equiv \widehat{\Theta}_A{}^{\bm{\alpha}}\,t_{\bm{\alpha}} 
 \equiv \bigl[\Theta_A{}^{\bm{\alpha}} - \tfrac{1}{1+\beta}\,(t^{\bm{\alpha}})_A{}^B \,\vartheta_B\bigr]\,t_{\bm{\alpha}} \,,
\end{align}
the Leibniz identity can also be expressed as
\begin{align}
 [X_A,\,X_B] = - X_{AB}{}^C\,X_C \qquad \bigl(\,\Rightarrow\ Z_{AB}{}^C\,X_C = 0\bigr)\,,
\label{eq:XXXX}
\end{align}
where $Z_{AB}{}^C\equiv 2\,X_{(AB)}{}^C$\,. 
Since $t_0$ commutes with all generators, the left-hand side is an element of the duality algebra. 
Then we can decompose \eqref{eq:XXXX} into the following two identities:
\begin{align}
 &[\widehat{\Theta}_A,\,\widehat{\Theta}_B] = - X_{AB}{}^C\,\widehat{\Theta}_C \,,
\label{eq:Theta-identity}
\\
 &X_{AB}{}^C\, \vartheta_C = 0\,.
\end{align}

As was studied in \cite{1911.06320,1911.07833,2007.08510}, the Leibniz identity consists of several types of conditions on the structure constants $f_{ab}{}^c$, $f_a{}^{\cI}$, $f_a{}^{\Pb}$, and $Z_a$. 
Here, we show that the identity Eq.~\eqref{eq:Theta-identity} contains the so-called the cocycle conditions and the fundamental identities. 
They are not all of the constraints on the structure constants and, in sections \ref{sec:M} and \ref{sec:IIB}, some of additional constraints for the EDA are reproduced from $Z_{AB}{}^C\,X_C = 0$ in Eq.~\eqref{eq:XXXX}. 

\subsubsection{Cocycle conditions}
\label{sec:cocycle}

Let us consider a particular component of the Leibniz identity,
\begin{align}
 [\widehat{\Theta}_a,\,\widehat{\Theta}_b] = - f_{ab}{}^c\,\widehat{\Theta}_c \,. 
\label{eq:Theta-abc}
\end{align}
For convenience, we decompose the embedding tensor into the level-0 part $\widehat{\Theta}^{(0)}_a$ (that is a linear combination of $K^a{}_b$ and $R_{\cI}$) and the negative-level part $\widetilde{\Theta}_a$: 
\begin{align}
 \widehat{\Theta}_a = \widehat{\Theta}^{(0)}_a + \widetilde{\Theta}_a\,.
\end{align}
Then the level-$0$ part of Eq.~\eqref{eq:Theta-abc} gives
\begin{align}
 [\widehat{\Theta}^{(0)}_a,\,\widehat{\Theta}^{(0)}_b] = - f_{ab}{}^c\,\widehat{\Theta}^{(0)}_c \,.
\label{eq:cocycle-0}
\end{align}
In particular, if we consider the coefficients of $K^a{}_b$\,, we find
\begin{align}
 f_{ac}{}^d\,f_{bd}{}^e - f_{bc}{}^d\,f_{ad}{}^e = - f_{ab}{}^d\,f_{dc}{}^e \,,\qquad f_{ab}{}^c\,Z_c =0\,.
\end{align}
The former is the standard Bianchi identity $f_{[ab}{}^d\,f_{c]d}{}^{e} = 0$ and the latter was found in \cite{1911.07833}. 
When there are additional level-0 generators $\{R_{\cI}\}$ (e.g., the type IIB case), we find additional identities that come from Eq.~\eqref{eq:cocycle-0}.

If we consider the negative-level part of Eq.~\eqref{eq:Theta-abc}, we obtain
\begin{align}
 [\widehat{\Theta}^{(0)}_a,\, \widetilde{\Theta}_b] - [\widehat{\Theta}^{(0)}_b,\, \widetilde{\Theta}_a] + f_{ab}{}^c\,\widetilde{\Theta}_c + [\widetilde{\Theta}_a,\,\widetilde{\Theta}_b] = 0 \,. 
\label{eq:cocycle}
\end{align}
To clarify the meaning of Eq.~\eqref{eq:cocycle}, we introduce a notation that is similar to the one used in the Lie algebra cohomology. 
An $n$-cochain of $\mathfrak{g}$ is a skew-symmetric linear map $f: \mathfrak{g}^n \to N_-$ where $N_-$ denotes the subalgebra generated by the negative-level generators $R_{\Pa}$,
\begin{align}
 f(x_1,\dotsc,x_n) = x_1^{a_1}\cdots x_n^{a_n}\,f_{a_1\cdots a_n}{}^{\Pa}\,R_{\Pa} \,,
\end{align}
where $x_i=x_i^a\,T_a$ and $f_{a_1\cdots a_n}{}^{\Pa}$ are certain constants satisfying $f_{a_1\cdots a_n}{}^{\Pa}=f_{[a_1\cdots a_n]}{}^{\Pa}$\,.
We denote the vector space of $n$-cochains by $C^n$ and define the coboundary operator $\delta_n : C^n\to C^{n+1}$ as
\begin{align}
 \delta_0 f(x) &\equiv x^a\,(\Exp{\text{ad}_f}-1)\,\widehat{\Theta}^{(0)}_a = x\cdot f + \tfrac{1}{2!}\,\bigl[f,\,\bigl[f,\, x^a\,\widehat{\Theta}^{(0)}_a \bigr] \bigr] + \cdots \quad \bigl(f \equiv f^{\Pa}\,R_{\Pa}\bigr)\,,
\\
 \delta_1 f(x,y) &\equiv x\cdot f(y) - y\cdot f(x) - f([x,y]) - [f(x),\,f(y)] \,,
\\
 \delta_2 f(x,y,z) &\equiv x \cdot f(y,z) + y \cdot f(z,x) + z \cdot f(x,y)
 - f([x,y],z) - f([y,z],x) - f([z,x],y)
\nn\\
 &\quad - [f(x),\,f(y,z)] - [f(y),\,f(z,x)] - [f(z),\,f(x,y)] \,,
\end{align}
where
\begin{align}
 x\cdot f(y_1,\cdots,y_n) \equiv \bigl[f(y_1,\cdots,y_n),\,x^a\,\widehat{\Theta}^{(0)}_a\bigr] \,.
\end{align}
We can check $\delta_{n+1}\delta_n =0$ by using Eq.~\eqref{eq:cocycle-0}, namely,
\begin{align}
 x\cdot \bigl(y\cdot f(z_1,\dotsc,z_n)\bigr) - y\cdot \bigl(x\cdot f(z_1,\dotsc,z_n)\bigr) = [x,y]\cdot f(z_1,\dotsc,z_n)\,.
\label{eq:cocycle-0-dot}
\end{align}
The most non-trivial coboundary operator is $\delta_0$\,, which is defined so as to satisfy $\delta_1\delta_0 =0$\,. 

By using the above definitions, the Leibniz identity \eqref{eq:cocycle} is interpreted as the cocycle condition \cite{1911.07833,2007.08510},
\begin{align}
 \delta_1f(x,y) = x\cdot f(y) - y\cdot f(x) - f([x,y]) - [f(x),\,f(y)] = 0 \,,
\label{eq:1-cocycle}
\end{align}
where we have identified the constants $f_a{}^{\Pa}$ as the dual structure constants, and thus
\begin{align}
 f(x)=x^a\,f_a{}^{\Pa}\,R_{\Pa} = x^a\,\widetilde{\Theta}_a\,.
\end{align}
In summary, the Leibniz identity \eqref{eq:Theta-abc} can be rewritten as Eqs.~\eqref{eq:cocycle-0-dot} and \eqref{eq:1-cocycle}. 

In DFT, $N_-$ is Abelian and we have $[f(\cdots),\, f(\cdots)]=0$ and $\bigl[f,\,\bigl[f,\, x^a\,\widehat{\Theta}^{(0)}_a \bigr] \bigr]=0$\,. 
Then the coboundary operators reduce to the usual ones. 
In particular, we have
\begin{align}
 x\cdot f(y) = -x^a\,y^{b} \,f_{ac}{}^{[d_1} \,f_b{}^{d_2]c}\,R_{d_1d_2} \,,
\end{align}
and the cocycle condition \eqref{eq:1-cocycle} becomes
\begin{align}
 2\,f_{ac}{}^{[d_1} \,f_b{}^{d_2]c} 
 - 2\,f_{bc}{}^{[d_1}\,f_a{}^{d_2]c} 
 + f_{ab}{}^c\, f_c{}^{d_1d_2} = 0 \,.
\end{align}

For completeness, it may be interesting to find the definition of the coboundary operators $\delta_n$ for $n\geq 3$ that satisfy $\delta_{n+1}\delta_n =0$. 
We find that the coboundary operator can be generally defined as
\begin{align}
\begin{split}
 \delta_n f(x_1,\dotsc,x_{n+1}) &\equiv \sum_i (-1)^{i-1}x_i\cdot f(x_1,\dotsc,\hat{x}_i,\dotsc,x_{n+1})
\\
 &\quad +\sum_{i<j}(-1)^{i+j} f([x_i,x_j],x_1,\dotsc,\hat{x}_i,\dotsc,\hat{x}_j,\dotsc,x_{n+1})
\\
 &\quad + \sum_i(-1)^{i}[f(x_i),\,f(x_1,\dotsc,\hat{x}_i,\dotsc,x_{n+1})] \,,
\end{split}
\end{align}
where the hat $\hat{x}_i$ denotes the omission of $x_i$. 
By using the properties Eqs.~\eqref{eq:cocycle-0-dot} and \eqref{eq:1-cocycle}, we find that this satisfies the property $\delta_{n+1}\delta_n =0$. 
In the Abelian case, the 1-cocycle condition \eqref{eq:1-cocycle} is not necessary for the derivative of $\delta_{n+1}\delta_n =0$ and $f(x)$ can be arbitrary. 
However, as we can see from the definition of $\delta_n$\,, $f(x)$ plays a distinguished role in the non-Abelian case, and in order to show $\delta_{n+1}\delta_n =0$ for $n\geq 2$, $f(x)$ needs to satisfy the cocycle condition \eqref{eq:1-cocycle}. 

\subsubsection{Fundamental identities}

Here we consider the Leibniz identity
\begin{align}
 [\widehat{\Theta}_{\mathring{A}},\,\widehat{\Theta}_{\mathring{B}}] + X_{\mathring{A}\mathring{B}}{}^C \,\widehat{\Theta}_C = 0\,,
\end{align}
where $\mathring{A}$ and $\mathring{B}$ denotes the dual components with the same level. 
The fundamental identities appear as the coefficients of $K^c{}_d$ in this Leibniz identity. 
By using the property
\begin{align}
 \widehat{\Theta}_{\tilde{A}c}{}^d = - f_c{}^{\Pb}\,(R_{\Pb})_{\tilde{A}}{}^{d}\,, 
\label{eq:k-symmetric}
\end{align}
that follow from Eq.~\eqref{eq:X-ABC}, we obtain
\begin{align}
 f_c{}^{\Pc}\,f_e{}^{\Pd}\,\bigl[(R_{\Pc})_{\mathring{A}}{}^e\,(R_{\Pd})_{\mathring{B}}{}^d
 - (R_{\Pc})_{\mathring{B}}{}^{e}\,(R_{\Pd})_{\mathring{A}}{}^{d}\bigr] 
= X_{\mathring{A}\mathring{B}}{}^{C}\, (R_{\Pd})_{C}{}^d\, f_c{}^{\Pd} \,.
\label{eq:FI}
\end{align}
This reproduces the so-called fundamental identities, although we need to specify the duality group $\cG$ in order to obtain the more explicit form.

For example, in DFT, the fundamental identity becomes
\begin{align}
\begin{split}
 f_c{}^{e_1e_2}\,f_e{}^{f_1f_2}\,\bigl[\delta_{e_1e_2}^{ae}\,\delta_{f_1f_2}^{bd}-\delta_{e_1e_2}^{be}\,\delta_{f_1f_2}^{ad} \bigr] 
= X^{ab}{}_e \, f_c{}^{ed} 
\quad \Leftrightarrow\quad 
 f_c{}^{e[a}\,f_e{}^{bd]} = 0\,.
\end{split}
\end{align}
This is the familiar Bianchi identity for the dual structure constants. 

\subsection{Generalized frame fields}
\label{sec:frame-def}

In this section, we explain the construction of the generalized frame fields $E_A{}^I$ when an ExDA and its maximally isotropic subalgebra $\mathfrak{g}$ are given. 
Using the generators of the Lie algebra $\mathfrak{g}$, we define a group element $g\equiv \Exp{x^a\,T_a}$\,. 
Then we define the left-/right-invariant 1-form as
\begin{align}
 \ell \equiv \ell^a_i\,\rmd x^i\,T_a\equiv g^{-1}\,\rmd g\,,\qquad
 r \equiv r^a_i\,\rmd x^i\,T_a\equiv \rmd g\,g^{-1}\,.
\end{align}
We also define their dual vector fields as
\begin{align}
 v_a\equiv v_a^i\,\partial_i\,,\qquad
 e_a\equiv e_a^i\,\partial_i \qquad
 \bigl(v_a^i\,\ell^b_i=\delta_a^b= e_a^i\,r^b_i\bigr)\,,
\end{align}
which satisfy the algebra
\begin{align}
 [v_a,\,v_b]=f_{ab}{}^c\,v_c\,,\qquad 
 [e_a,\,e_b]=-f_{ab}{}^c\,e_c\,,\qquad
 [v_a,\,e_b]=0\,.
\end{align}
These can be extended to the generalized vector fields (with the natural weight) as
\begin{align}
 \hat{V}_A{}^I \equiv (\Exp{-(h_L)_c{}^d\,\tilde{K}^c{}_d})_A{}^B\,\delta_B^I\,,\qquad
 \hat{E}_A{}^I \equiv (\Exp{-(h_R)_c{}^d\,\tilde{K}^c{}_d})_A{}^B\,\delta_B^I\,,
\label{eq:V-E-def}
\end{align}
where $v_a^i\equiv (\Exp{-h_L})_a{}^b\,\delta_b^i$ and $e_a^i\equiv (\Exp{-h_R})_a{}^b\,\delta_b^i$.
They satisfy
\begin{align}
 \gLie_{\hat{V}_A}\hat{V}_B = \hat{X}_{AB}{}^C\,\hat{V}_C\,,\qquad
 \gLie_{\hat{E}_A}\hat{E}_B = -\hat{X}_{AB}{}^C\,\hat{E}_C\,,
\end{align}
where $\hat{X}_{AB}{}^C$ is the structure constant $X_{AB}{}^C$ with vanishing dual structure constants:
\begin{align}
 \hat{X}_{AB}{}^C \equiv f_{Ad}{}^{e}\,(\tilde{K}^d{}_e)_B{}^C
 + \tfrac{1}{2}\,f_{de}{}^f\,\bigl[(\tilde{K}^e{}_f\,R_{\Pa})_A{}^d + \chi_{K^e{}_f A} \,\chi^{d}{}_{\Pa} \bigr] \,(R^{\Pa})_B{}^C \,.
\end{align}

Now, we consider a natural action of the group element $g$ on the generators $T_A$
\begin{align}
 g\,\triangleright T_A \equiv \Exp{x^b\,T_b\circ} T_A 
 = T_A + x^b\,T_b\circ T_A + \tfrac{1}{2!}\, x^b\,T_b\circ \bigl(x^c\,T_c\circ T_A\bigr) + \cdots\,,
\end{align}
and define a matrix $M_A{}^B$ as
\begin{align}
 g^{-1}\,\triangleright T_A \equiv M_A{}^B(g)\,T_B\,.
\label{eq:M-def}
\end{align}
Since the left action of $T_a$ on $T_A$ is generated by $\tilde{K}^a{}_b$, $R_{\cI}$, $R_{\Pa}$, and $t_0$\,, the exponential action can be parameterized as
\begin{align}
 M_A{}^B(g) &= \bigl[\bm{\Pi}(x) \Exp{(\tilde{K}+t_0)\Delta(x)} \,\bm{A}(x) \bigr]_A{}^B\,,
\label{eq:M-param}
\\
 \bm{\Pi}_A{}^B(x) &= \bigl({\textstyle\prod_{\Pa}}\Exp{-\pi^{\Pa}\,R_{\Pa}}\bigr)_A{}^B\,,\qquad 
 \bm{A}_A{}^B(x) = \bigl(\Exp{-h^{\cI}\,R_{\cI}}\Exp{-\alpha_c{}^d\,\tilde{K}^c{}_d}\bigr)_A{}^B \,.
\end{align}
In the following, we define $a_a{}^b\equiv \bm{A}_a{}^b=(\Exp{-\alpha})_a{}^b$ by following the standard notation.
By multiplying the matrix $M_A{}^B$ by $\hat{V}_A{}^I$\,, we define the generalized frame fields as
\begin{align}
 E_A{}^I(x) \equiv M_A{}^B(g)\,\hat{V}_B{}^I = \bm{\Pi}_A{}^B(x)\, (\Exp{(\tilde{K}+t_0)\Delta(x)})_B{}^C\, \mathbb{E}_C{}^I \,,
\label{eq:frame-def}
\end{align}
where $\mathbb{E}_A{}^I=(\Exp{-h^{\cI}\,R_{\cI}})_A{}^B\,\hat{E}_B{}^I$ and we have used an identity $e_a^i=a_a{}^b\,v_b^i$\,.

\subsubsection{Important identities}

Here, we explain several important properties of $\bm{A}_A{}^B$, $\pi^{\Pa}$, and $\Delta$\,. 
All of these properties are straightforward extensions of the ones known in the Poisson--Lie $T$-duality. 

\paragraph{Values at the unit element\\}

The first property follows from the definition \eqref{eq:M-def}.
By choosing $g=e$, where $e$ is the unit element of the physical subgroup $G$, we find
\begin{align}
 M_A{}^B(e)=\delta_A^B\qquad 
 \Leftrightarrow \qquad 
 \pi^{\Pa}\rvert_{g=e}=0\,,\quad 
 \Delta\rvert_{g=e}=0\,,\quad 
 \bm{A}_A{}^B\rvert_{g=e}=\delta_A^B\,.
\end{align}
They reproduce Eq.~\eqref{eq:x=x0} if we denote the coordinate value of the unit element $e$ as $x_0$\,. 

\paragraph{Algebraic identity\\}

The second property comes from the Leibniz identity, which shows
\begin{align}
 g \,\triangleright (T_A\circ T_B) = (g \,\triangleright T_A)\circ (g \,\triangleright T_B)\,.
\end{align}
In terms of $M_A{}^B$, this reads
\begin{align}
 (M^{-1})_D{}^C\, X_{AB}{}^D = (M^{-1})_A{}^D\,(M^{-1})_B{}^E\,X_{DE}{}^C\,.
\label{eq:alg-id}
\end{align}
From this identity, we find non-trivial algebraic identities for $\bm{A}_A{}^B$, $\pi^{\Pa}$, and $\Delta$\,. 

For example, in DFT, the algebraic identity \eqref{eq:alg-id} gives
\begin{align}
\begin{split}
 &f_{ab}{}^c = a_a{}^{a'}\,a_b{}^{b'}\,(a^{-1})_{c'}{}^{c}\,f_{a'b'}{}^{c'}\,,\qquad
 f_{d}{}^{[ab}\,\pi^{c]d} + f_{de}{}^{[a}\,\pi^{b|d|}\,\pi^{c]e} = 0\,,
\\
 &f_a{}^{bc} = a_a{}^d\,(a^{-1})_e{}^b\,(a^{-1})_f{}^c\,f_d{}^{ef} + 2\,f_{ad}{}^{[b}\,\pi^{c]d}\,,
\end{split}
\label{eq:DFT-algebraic-id}
\end{align}
which are presented in Appendix A of \cite{hep-th:9710163}. 

\paragraph{Differential identity\\}

The third property follows from the identity
\begin{align}
 (hg)^{-1}\,\triangleright T_A = g^{-1}\,\triangleright (h^{-1}\,\triangleright T_A)\,,
\label{eq:BCH}
\end{align}
which can be shown as follows. 
Since $\mathfrak{g}$ is a Lie algebra, for $g=\Exp{g^a\,T_a}$ and $h=\Exp{h^a\,T_a}$, we have
\begin{align}
 hg = \Exp{h^a\,T_a} \Exp{g^b\,T_b} = \Exp{(hg)^{a}\,T_a} \,,
\end{align}
where $(hg)^{a}$ can be determined by using the Baker--Campbell--Hausdorff formula. 
Since $X_a$ satisfies the same algebra as $-T_a$, we find
\begin{align}
 \bigl(\Exp{-(hg)^a\,X_a} \bigr)_A{}^B\,T_B
 = \bigl(\Exp{-h^a\,X_a} \Exp{-g^b\,X_b}\bigr)_A{}^B\,T_B\,,
\end{align}
and this shows the identity \eqref{eq:BCH}. 

In terms of $M_A{}^B$\,, the identity \eqref{eq:BCH} is equivalent to
\begin{align}
 M_A{}^B(hg) = M_A{}^C(h)\,M_C{}^B(g) \,.
\label{eq:M-product}
\end{align}

For example, in DFT, this relation gives
\begin{align}
 a_a{}^b(hg) = a_a{}^c(h)\,a_c{}^b(g)\,,\qquad
 \pi^{ab}(hg) = \pi^{ab}(h) + \pi^{cd}(g)\,(a^{-1})_c{}^a(h)\,(a^{-1})_d{}^b(h) \,.
\end{align}
The second equation shows that the bi-vector $\pi^{ij}\equiv \pi^{ab}\,e_a^i\,e_b^j$ is multiplicative (see e.g., \cite{1511.02491}). 
We thus consider that the relation \eqref{eq:M-product} defines that the multi-vectors $\pi^{\Pa}$ are multiplicative. 

Now, we can derive a differential identity from the identity \eqref{eq:M-product}. 
Supposing $h=e+\epsilon^a\,T_a$\,, where $e$ is the unit element of $G$ and $\epsilon^a$ are infinitesimal parameters, we have
\begin{align}
 M_A{}^B(h) = \delta_A^B - \epsilon^c\,X_{cA}{}^B \,.
\end{align}
Then, since an infinitesimal left multiplication $g(x)\to (1+\epsilon_a\,T^a)\,g(x)$ corresponds to the infinitesimal coordinate transformation $\delta x^i= \epsilon^a\,e_a^i$, we obtain the differential identity
\begin{align}
 D_c M_A{}^B(g) &= \lim_{\epsilon^a\to 0} \frac{M_A{}^B(hg) - M_A{}^B(g)}{\epsilon^a} = - X_{cA}{}^D\,M_D{}^B(g)\,,
\\
 &\Leftrightarrow \quad M_A{}^D\,D_c (M^{-1})_D{}^B = X_{cA}{}^B \,.
\label{eq:diff-id-M}
\end{align}
Recalling the decomposition $M= \bm{\Pi}\Exp{(\tilde{K} + t_0)\Delta}\bm{A}$, we find that the block-diagonal components of this relation (i.e., matrix elements $(\cdots)_A{}^B$ with $A$ and $B$ having the same level) give
\begin{align}
 \bm{A}\,D_c\bm{A}^{-1}= f_{cd}{}^e\,\tilde{K}^d{}_e + f_c{}^{\cI}\,R_{\cI}\,,\qquad 
 D_a \Delta = Z_a \,.
\label{eq:AdA}
\end{align}
The other components give the differential identity for $\bm{\Pi}$ as
\begin{align}
 \bigl(\bm{\Pi}\,D_c \bm{\Pi}^{-1}\bigr)_A{}^B = (\widehat{\Theta}_c)_{A}{}^B - \bigl(\bm{\Pi}\,\widehat{\Theta}^{(0)}_c\,\bm{\Pi}^{-1}\bigr)_A{}^B \,.
\label{eq:diff-id}
\end{align}
They determine the derivatives of $\bm{A}_A{}^B$, $\Delta$, and $\pi^{\Pa}$. 
Another useful expression that follows from \eqref{eq:diff-id} and the algebraic identity is
\begin{align}
 v_a^k\,\mathbb{E}_I{}^A\, (\Exp{-K\,\Delta}\partial_k\bm{\Pi}^{-1}\,\bm{\Pi}\Exp{K\,\Delta})_A{}^B\,\mathbb{E}_B{}^J
 = \hat{V}_I{}^A\,\bigl(f_a{}^{\Pa}\,R_{\Pa}\bigr)_A{}^B\,\hat{V}_B{}^J \,.
\label{eq:partial-pi}
\end{align}

In DFT, the differential identity \eqref{eq:diff-id} gives
\begin{align}
 D_c\pi^{ab} = f_c{}^{ab} - 2\,f_{cd}{}^{[a}\,\pi^{b]d}\,.
\end{align}
This leads to the condition that $\pi$ is multiplicative,
\begin{align}
 \Lie_{v_a}\pi^{mn} = f_a{}^{bc}\,v_b^m\,v_c^n\,. 
\label{eq:multiplicative-DFT}
\end{align}
Here, we used the identities \eqref{eq:DFT-algebraic-id}, but Eq.~\eqref{eq:multiplicative-DFT} can be directly obtained from Eq.~\eqref{eq:partial-pi}. 
In the EDA, we use Eq.~\eqref{eq:partial-pi} when we derive the relations similar to Eq.~\eqref{eq:multiplicative-DFT}. 

\subsection{Generalized parallelizability}
\label{sec:parallelizability}

Now we are ready to show that the generalized fluxes $E_A{}^I$ defined in Eq.~\eqref{eq:frame-def} indeed satisfies the algebra
\begin{align}
 \gLie_{E_A} E_B{}^I = - X_{AB}{}^C\,E_B{}^I \,.
\label{eq:Frame-algebra}
\end{align}
This property for the Drinfel'd algebra was shown in \cite{1810.11446} by using DFT, and the same property for the $E_{n(n)}$ EDA $(n\leq 6)$ was shown in \cite{1911.06320,1911.07833,2007.08510}. 
There, the proof depends on the details of the duality group $\cG$ and the choice of the subalgebra $\mathfrak{g}$. 
Consequently, the proof becomes more complicated as the dimension of the ExDA becomes larger. 
However, the general proof to be presented here is simple and does not depend on $\cG$ and the choice of $\mathfrak{g}$.
It can be applied even to the $E_{8(8)}$ EDA, both in the M-theory and the type IIB picture. 

Before going into the details of the technical computation, let us explain the outline of our proof. 
The point is that, as we show below, the Weizenb\"ock connection
\begin{align}
 \Omega_{AB}{}^C = E_A{}^I\,E_B{}^K\,\partial_I E_K{}^C\,,
\label{eq:Weizenbock-again}
\end{align}
can be expressed as a duality twist of some functions $\mathbb{F}_{AB}{}^C$:
\begin{align}
 \Omega_{AB}{}^C = M_A{}^D \,M_B{}^E \,(M^{-1})_F{}^C \, \mathbb{F}_{DE}{}^F\,,
\label{eq:W-M-F}
\end{align}
where $M_A{}^B$ is the matrix defined in Eq.~\eqref{eq:M-def}. 
Since $M_A{}^B\in \cG \times \mathbb{R}^+$ and the generalized fluxes $\bm{X}_{AB}{}^C$ are constructed from $\Omega_{AB}{}^C$ and $\cG$-invariant tensors (recall Eqs.~\eqref{eq:X-Theta} and \eqref{eq:theta}), this means that the generalized fluxes have the form
\begin{align}
 \bm{X}_{AB}{}^C = M_A{}^D \,M_B{}^E \,(M^{-1})_F{}^C \,\mathbb{X}_{DE}{}^F \,,
\end{align}
where $\mathbb{X}_{AB}{}^C$ are the generalized fluxes \eqref{eq:gen-flux} with $\Omega_{AB}{}^C$ replaced with $\mathbb{F}_{AB}{}^C$\,. 

As we show below, the non-constant part of $\mathbb{F}_{AB}{}^C$ does not contribute to $\mathbb{X}_{AB}{}^C$ and the generalized fluxes $\mathbb{X}_{AB}{}^C$ are constant. 
Evaluating the constants $\mathbb{X}_{AB}{}^C$ at a particular point $x=x_0$ (where $M_A{}^B=\delta_A^B$ and $\bm{X}_{AB}{}^C=X_{AB}{}^C$), we can easily see $\mathbb{X}_{AB}{}^C=X_{AB}{}^C$\,. 
Then, the algebraic identity \eqref{eq:alg-id} shows
\begin{align}
 \bm{X}_{AB}{}^C = M_A{}^D \,M_B{}^E \,(M^{-1})_F{}^C \,X_{DE}{}^F =X_{AB}{}^C \,.
\end{align}
By recalling the definition of the generalized fluxes, this is equivalent to the desired relation
\begin{align}
 \gLie_{E_A} E_B{}^I = - \bm{X}_{AB}{}^C\,E_B{}^I = - X_{AB}{}^C\,E_B{}^I \,.
\end{align}
Thus, our remaining tasks are to derive Eq.~\eqref{eq:W-M-F} with some functions $\mathbb{F}_{AB}{}^C$ and to show that $\mathbb{X}_{AB}{}^C$ are constant. 

Using the definition of the generalized frame fields given in Eq.~\eqref{eq:frame-def}, and the definition of the Weizenb\"ock connection \eqref{eq:Weizenbock-again}, we obtain
\begin{align}
 \Omega_{AB}{}^C &= M_A{}^d\,\bigl\{M_B{}^E\,(M^{-1})_F{}^C\, \hat{\Omega}_{dE}{}^F + (a^{-1})_d{}^e\,\,\bigl(M\,D_e M^{-1}\bigr)_B{}^C \bigr\}\,,
\label{eq:Omega-ABC}
\\
 \hat{\Omega}_{AB}{}^C &\equiv \hat{V}_A{}^I\,\hat{V}_B{}^K\,\partial_I \hat{V}_K{}^C \,. 
\label{eq:Omega-hat}
\end{align}
Here, recalling Eq.~\eqref{eq:V-E-def} and using $\hat{V}_A{}^i=\delta_A^b\,v_b^i$, we have
\begin{align}
 \hat{\Omega}_{AB}{}^C = \delta_A^e\, \bm{k}_{ef}{}^g\, \tilde{K}^f{}_g\,,\qquad 
 \bm{k}_{ab}{}^c\equiv v_b^k\,v_a^i\,\partial_i \ell_k^c \,.
\end{align}
In general, $\bm{k}_{ab}{}^c$ is non-constant, but its antisymmetric part is given by the structure constants $\bm{k}_{[ab]}{}^c = - \frac{1}{2}\,f_{ab}{}^c$\,. 
Then, using the differential identity \eqref{eq:diff-id-M}, we can rewrite Eq.~\eqref{eq:Omega-ABC} as
\begin{align}
 \Omega_{AB}{}^C = M_A{}^d\, \bigl[ \bm{k}_{de}{}^f\,(M\,\tilde{K}^e{}_f\,M^{-1})_B{}^C + (a^{-1})_d{}^e\,X_{eB}{}^C\bigr]\,.
\end{align}
Then, by using $\bm{k}_{[ab]}{}^c = -\frac{1}{2}\,f_{ab}{}^c$\,, $M_A{}^b = \bm{\Pi}_A{}^c\,a_c{}^b$\,, and the identity
\begin{align}
 (a^{-1})_a{}^d\,X_{dB}{}^{C} &= M_B{}^{D}\,(M^{-1})_{E}{}^C\, X_{aD}{}^{E} \,,
\label{eq:X-a-id}
\end{align}
which can be obtained from the algebraic identity \eqref{eq:alg-id}, we obtain
\begin{align}
 \Omega_{AB}{}^C &= M_A{}^d\, M_{B}{}^{B'}\, (M^{-1})_{C'}{}^{C}\,\mathbb{F}_{aB}{}^C \,,
\\
 \mathbb{F}_{aB}{}^C &\equiv \bigl(\bm{k}_{(ad)}{}^e-\tfrac{1}{2}\,f_{ad}{}^e\bigr)\,(\tilde{K}^d{}_e)_B{}^C + X_{aB}{}^{C} \,.
\end{align}
Finally, by introducing
\begin{align}
 \mathbb{F}_{AB}{}^C \equiv \delta_A^d\,\mathbb{F}_{dB}{}^C\,,
\end{align}
the Weizenb\"ock connection can be expressed in the desired form
\begin{align}
 \Omega_{AB}{}^C = M_A{}^D\, M_B{}^E\,(M^{-1})_F{}^C\, \mathbb{F}_{DE}{}^F\,.
\end{align}

Now, let us explain the constancy of $\mathbb{X}_{AB}{}^C$. 
The generalized fluxes $\mathbb{X}_{AB}{}^C$ are constructed from $\mathbb{F}_{AB}{}^C$ but its non-constant part is coming only from the symmetric part $\bm{k}_{(ab)}{}^c$ of $\bm{k}_{ab}{}^c$\,. 
Namely, the non-constant part of $\mathbb{X}_{AB}{}^C$ becomes
\begin{align}
 \mathbb{X}_{AB}{}^C\rvert_{\text{non-constant}}
 &= \bm{k}_{(de)}{}^f \,\bigl[(\tilde{K}^e{}_f\,R_{\Pa})_A{}^d + \chi_{K^e{}_f A} \,\chi^{d}{}_{\Pa} \bigr] \,(R^{\Pa})_B{}^C \,.
\end{align}
Using the identity \eqref{eq:k-symmetric}, we find that this vanishes.
Namely, we have shown that $\mathbb{X}_{AB}{}^C$ is constant, and thus $\mathbb{X}_{AB}{}^C = X_{AB}{}^C$\,. 

This completes the proof that for any choice of the maximally isotropic subalgebra $\mathfrak{g}$\,, the ExDA always gives a generalized parallelizable space. 

\subsubsection{Nambu--Lie structures}

In ExFT, the generalized fluxes are decomposed into representations of the subgroup $\mathfrak{h}$. 
The most familiar example is the case of DFT, where $\mathbb{X}_{AB}{}^C$ are decomposed as
\begin{align}
 H_{abc} \equiv \mathbb{X}_{abc}\,,\quad
 F_{ab}{}^c \equiv \mathbb{X}_{ab}{}^c\,,\quad
 Q_a{}^{bc} \equiv \mathbb{X}_a{}^{bc}\,,\quad
 R^{abc} \equiv \mathbb{X}^{abc}\,,
\end{align}
because $\mathbb{X}_{ABC}$ are totally antisymmetric. 
Using the generalized frame fields
\begin{align}
 E_a = e_a \,,\qquad E^a = -\pi^{ab}\,e_b + r^a \qquad \bigl(r^a\equiv r^a_m\,\rmd x^m\bigr)\,,
\end{align}
we can compute the fluxes as
\begin{align}
\begin{split}
 H_{abc} &= 0\,,\quad
 F_{ab}{}^c = f_{ab}{}^c\,,\quad
 Q_a{}^{bc} = D_a\pi^{bc} + 2\,f_{ad}{}^{[b}\,\pi^{c]d}\,,
\\
 R^{abc} &= 3\,\pi^{d[a}\,D_d\pi^{bc]} + 3\,f_{d_1d_2}{}^{[a}\,\pi^{b|d_1|}\,\pi^{c]d_2}\,.
\end{split}
\end{align}
Then, the generalized parallelizability means that
\begin{align}
 Q_a{}^{bc} = D_a\pi^{bc} + 2\,f_{ad}{}^{[b}\,\pi^{c]d}= f_a{}^{bc}\,,\quad 
 R^{abc} = 3\,\pi^{d[a}\,D_d\pi^{bc]} + 3\,f_{d_1d_2}{}^{[a}\,\pi^{b|d_1|}\,\pi^{c]d_2} = 0\,.
\end{align}
Namely, the dual structure constants are identified as the $Q$-flux (also known as the globally non-geometric flux) and the absence of the $R$-flux (called the locally non-geometric flux) gives a differential identity for $\pi^{ab}$\,. 
For the bi-vector field $\pi^{mn}=\pi^{ab}\,e_a^m\,e_b^n$, the latter reads
\begin{align}
 \pi^{qm}\,\partial_q\pi^{np} + \pi^{qn}\,\partial_q\pi^{pm} + \pi^{qp}\,\partial_q\pi^{mn} = 0\,,
\end{align}
which shows that $\pi^{mn}$ is a Poisson tensor. 

In a general ExFT, the correspondent of the $R$-fluxes are $\mathbb{X}_{\mathring{A}\mathring{B}}{}^{c}$ and we observe that they always vanish,
\begin{align}
 \mathbb{X}_{\mathring{A}\mathring{B}}{}^{c}= X_{\mathring{A}\mathring{B}}{}^{c}=0 \,.
\label{eq:Nambu-Poisson}
\end{align}
The computation of the flux $\mathbb{X}_{\mathring{A}\mathring{B}}{}^{c}$ is very hard in the $E_{8(8)}$ EFT, and we do not compute the explicit components in this paper.
Here we just show an example, a locally non-geometric flux $\mathbb{X}^{a_1a_2b_1b_2c}$ in M-theory picture.
Under some algebraic identities, this is expressed as
\begin{align}
 \mathbb{X}^{a_1a_2b_1b_2c} &= -\pi^{a_1a_2 d}\, \nabla_d \pi^{b_1b_2c} + 3\, \pi^{d[b_1b_2}\,\nabla_d \pi^{c]a_1a_2}
  + f_{d_1d_2}{}^{[a_1}\, \pi^{a_2] b_1b_2 c d_1d_2} = 0\,,
\end{align}
where $\nabla_b\pi^{a_1a_2a_3} \equiv D_b \pi^{a_1a_2a_3} - \tfrac{3}{2}\, f_{bc}{}^{[a_1|}\, \pi^{c|a_2a_3]}$\,. 
This may be further simplified by using certain relations among $\pi$ and the structure constants, but it is obvious that this is a natural extension of the $R$-flux. 
As noted in \cite{1911.06320}, at least if we assume $f_{ab}{}^c=0$, this is precisely the condition that the $\pi^{i_1i_2i_3}$ is the Nambu--Poisson tensor. 
Suggested by this, we regard Eq.~\eqref{eq:Nambu-Poisson} as a definition that the multi-vectors $\pi$ are the generalized Nambu--Poisson tensors. 
Moreover, combining this with the fact that $\pi$ are multiplicative, we call $\pi$ the Nambu--Lie structure \cite{math/9812064}. 

\subsection{Generalized classical Yang--Baxter equation}

As discussed in section \ref{sec:cocycle}, the Leibniz identity requires that the dual structure constants $f_a{}^{\Pa}$ satisfy the cocycle condition \eqref{eq:1-cocycle}, $\delta_1f(x,y) = 0$.
This condition is trivially satisfied by assuming that the cocycle $f(x)$ is a coboundary \cite{1911.07833,2007.08510},
\begin{align}
 f(x) = \delta_0 \rr(x) \equiv x^a\,(\Exp{\text{ad}_{\bm{\rr}}}-1)\,\widehat{\Theta}^{(0)}_a \qquad \bigl(\bm{\rr} \equiv \tilde{\rr}^{\Pa}\, R_{\Pa}\bigr)\,.
\end{align}
By recalling $f(x) = x^a\,\widetilde{\Theta}_a$ and $\widehat{\Theta}_a=\widehat{\Theta}^{(0)}_a+\widetilde{\Theta}_a$, this coboundary ansatz corresponds to
\begin{align}
 \widehat{\Theta}_a = \Exp{\text{ad}_{\bm{\rr}}}\, \widehat{\Theta}^{(0)}_a \,.
\end{align}
The right-hand side can be rewritten as
\begin{align}
 \widehat{\Theta}_a = \Exp{\text{ad}_{\bm{\rr}}}\, \widehat{\Theta}^{(0)}_a 
 = \cR\,\widehat{\Theta}^{(0)}_a \,\cR^{-1} \,,\qquad 
 \cR\equiv \prod_{\Pa}\Exp{\rr^{\Pa}\, R_{\Pa}}\,, 
\label{eq:cR-def}
\end{align}
where $\rr^{\Pa}$ is a redefinition of $\tilde{\rr}^{\Pa}$\,. 
By comparing this with the general expression $\widehat{\Theta}_a = \widehat{\Theta}^{(0)}_a + f_a{}^{\Pa}\,R_{\Pa}$, the dual structure constants $f_a{}^{\Pa}$ can be expressed symbolically as
\begin{align}
 f_a{}^{\Pa}\,R_{\Pa} = \cR\,\widehat{\Theta}^{(0)}_a \,\cR^{-1} - \widehat{\Theta}^{(0)}_a \equiv \mathfrak{f}_a{}^{\Pa}\,R_{\Pa}\,.
\end{align}

For example, in DFT, we have $\widehat{\Theta}^{(0)}_a = f_{ab}{}^c \,K^b{}_c$ and $\bm{\rr}=\frac{1}{2!}\,\rr^{a_1a_2}\,R_{a_1a_2}$\,, and the coboundary ansatz becomes
\begin{align}
 \widehat{\Theta}_a = f_{ab}{}^c \,K^b{}_c - f_{ac}{}^{[b_1}\,\rr^{b_2]c}\, R_{b_1b_2} \,.
\end{align}
Comparing this with the Drinfel'd algebra
\begin{align}
 \widehat{\Theta}_a = f_{ab}{}^c \,K^b{}_c + \tfrac{1}{2!}\,f_a{}^{b_1b_2}\,R_{b_1b_2} \,,
\end{align}
we can identify the dual structure constants as
\begin{align}
 f_a{}^{b_1b_2} = 2\, f_{ac}{}^{[b_1|}\,\rr^{c|b_2]} \equiv \mathfrak{f}_a{}^{b_1b_2}\,.
\end{align}

Since all of the dual structure $f_a{}^{\Pb}$ are replaced by $\mathfrak{f}_a{}^{\Pb}$\,, we obtain an ExDA whose structure constants $\cX_{AB}{}^C$ are characterized by $f_{ab}{}^c$, $f_{a}{}^{\cI}$, $Z_a$, and $\rr^{\Pa}$\,.
We call this the coboundary ExDA. 
The cocycle conditions are trivially satisfied but other Leibniz identities are not satisfied in general, and we need to impose further conditions on $\rr^{\Pa}$\,. 
Below, following the approach discussed in \cite{1911.06320}, we explain the procedure to find such additional conditions on $\rr^{\Pa}$\,. 

Let us note that Eq.~\eqref{eq:cR-def} can be expressed as
\begin{align}
 \cX_{aB}{}^C = \cR_a{}^{D}\,\cR_B{}^{E}\,(\cR^{-1})_{F}{}^C\,X^{(0)}_{DE}{}^{F}\,,\qquad 
 X^{(0)}_{AB}{}^{C} \equiv (\cX^{\rr=0})_{AB}{}^{C} \,,
\end{align}
where $\cR_A{}^B$ is the matrix representation of $\cR$ in the $R_1$ representation.
This suggests us to extend this relation to define a new Leibniz algebra twisted by $\cR$,
\begin{align}
 \mathfrak{X}_{AB}{}^C \equiv \cR_A{}^{D}\,\cR_B{}^{E}\,(\cR^{-1})_{F}{}^C\,X^{(0)}_{DE}{}^{F} \,.
\end{align}
The untwisted structure constants $X^{(0)}_{AB}{}^{C}$ satisfy the Leibniz identity (when the Bianchi identities \eqref{eq:cocycle-0} are satisfied) and it is clear that $\mathfrak{X}_{AB}{}^C$ also satisfy the Leibniz identity \eqref{eq:Leibniz-id-X}. 
However, the algebra defined by the structure constants $\mathfrak{X}_{AB}{}^C$ is not an ExDA in general. 
Only the particular components, namely, $\mathfrak{X}_{aB}{}^C$ coincide with $\cX_{aB}{}^C$ of an ExDA.
Thus, to ensure that $\mathfrak{X}_{AB}{}^C$ are the structure constants of a certain ExDA, we require
\begin{align}
 \mathfrak{X}_{\tilde{A}B}{}^C = \cX_{\tilde{A}B}{}^C \,.
\label{eq:gen-YB}
\end{align}
This requires certain conditions on $\rr^{\Pa}$ and if these are satisfied, we obtain a coboundary ExDA that automatically satisfies the Leibniz identity. 

For example, in the case of DFT, using $\cR=\Exp{\frac{1}{2!}\,\rr^{a_1a_2}\,R_{a_1a_2}}$ we find
\begin{align}
 \cX^a{}_{B}{}^C &= 2\, f_{df}{}^{[a}\,\rr^{e]f} \,(K^d{}_e)_B{}^C
 + \tfrac{1}{2}\,f_{d_1d_2}{}^a\, (R^{d_1d_2})_B{}^C \,.
\\
 \mathfrak{X}^a{}_{B}{}^C &= \rr^{af}\,f_{fd}{}^e \,(\cR\,K^d{}_e\,\cR^{-1})_B{}^C
  + \tfrac{1}{2}\,f_{d_1d_2}{}^a\, (\cR\,R^{d_1d_2}\,\cR^{-1})_B{}^C 
\nn\\
 &= 2\, f_{df}{}^{[a}\,\rr^{e]f} \,(K^d{}_e)_B{}^C
 + \tfrac{1}{2}\,f_{d_1d_2}{}^a\, (R^{d_1d_2})_B{}^C 
\nn\\
 &\quad + \tfrac{3}{2}\, \rr^{d_1[a}\,\rr^{|d_2|e}\,f_{d_1d_2}{}^{f]}\, (R_{ef})_B{}^C \,.
\end{align}
Then the requirement \eqref{eq:gen-YB} is equivalent to the classical Yang--Baxter equation (CYBE),
\begin{align}
 \rr^{d_1a}\,\rr^{d_2e}\,f_{d_1d_2}{}^f + \rr^{d_1e}\,\rr^{d_2f}\,f_{d_1d_2}{}^a + \rr^{d_1f}\,\rr^{d_2a}\,f_{d_1d_2}{}^e = 0\,.
\label{eq:CYBE}
\end{align}

The definition of the generalized CYBE \eqref{eq:gen-YB} is very simple, but in practice, we find that it is difficult to identify the full set of the independent conditions on $\rr^{\Pa}$\,. 
In general, we can express $X^{\rr}$ and $X^{\cR}$ as
\begin{align}
 \cX_{\tilde{A}} &= \underbrace{\cdots}_{\text{non-negative-level generators}} {} - {} \beta\, \mathfrak{f}_d{}^{\Pa}\,(R_{\Pa})_{\tilde{A}}{}^d \, t_0 \,,
\\
 \mathfrak{X}_{\tilde{A}} &= \underbrace{\cdots}_{\text{non-negative-level generators}} {} - {} \beta\, \mathfrak{f}_d{}^{\Pa}\,(R_{\Pa})_{\tilde{A}}{}^d \, t_0 + \text{CYBE}_{\tilde{A}}{}^{\Pa}\, R_{\Pa} + \text{CYBE}_{\tilde{A}}{}^0\,t_0 \,, 
\end{align}
where
\begin{align}
\begin{split}
 \text{CYBE}_{\tilde{A}}{}^{\Pb}
 &\equiv \cR_{\tilde{A}}{}^{b}\,\mathfrak{f}_b{}^{\Pb} 
 + \cR_{\tilde{A}}{}^B\,\bigl\{ \bigl(\tfrac{1}{2}\,f_{de}{}^f -Z_{[d}\,\delta_{e]}^f\bigr)\,\bigl[(\tilde{K}^e{}_f\,R_{\Pa})_{B}{}^d + \chi_{K^e{}_f B} \,\chi^{d}{}_{\Pa} \bigr] 
\\
 &\qquad\qquad\qquad +f_d{}^{\cI}\,(R_{\cI}\,R_{\Pa})_{B}{}^d + \chi_{\cI B} \,\chi^{d}{}_{\Pa}\bigr) +(R_{\Pa})_B{}^d\,Z_d \bigr\}\,[\cR\,R^{\Pa}\,\cR^{-1}]^{\Pb} \,,
\end{split}
\\
 \text{CYBE}_{\tilde{A}}{}^0 &\equiv \beta\, \mathfrak{f}_c{}^{\Pb}\,(R_{\Pb})_{\tilde{A}}{}^c + \cR_{\tilde{A}}{}^b\, \bigl[\beta\,(f_{bc}{}^c-Z_b\,\delta_c^c)-Z_b \bigr]\,,
\end{align}
and $[\cdots]^{\Pb}$ denotes the coefficient of $R_{\Pb}$: namely, $\cR\,R^{\Pa}\,\cR^{-1}=\cdots +[\cR\,R^{\Pa}\,\cR^{-1}]^{\Pb}\,R_{\Pb}$\,. 
Then the condition \eqref{eq:gen-YB} at least requires
\begin{align}
 \text{CYBE}_{\tilde{A}}{}^{\Pb} = 0\,,\qquad 
 \text{CYBE}_{\tilde{A}}{}^0 = 0\,. 
\label{eq:gen-CYBE}
\end{align}
In general, terms including the non-negative-level generators in $\cX_{AB}{}^C$ and $\mathfrak{X}_{AB}{}^C$ do not coincide and we find additional conditions, but extensions of the quadratic equation \eqref{eq:CYBE} are contained in $\text{CYBE}_{\tilde{A}}{}^{\Pb} = 0$ as we see in the case of the EDA.

\section{Non-Abelian duality}
\label{sec:NAD}

In this section, we explain the procedure of the non-Abelian duality. 
For a given ExDA, we choose a maximally isotropic subalgebra $\mathfrak{g}$ and construct the generalized frame fields $E_A{}^I$ as explained in the previous section. 
For simplicity, if we suppose $\theta_A=0$, the generalized frame field has unit determinant and then we define the generalized metric as
\begin{align}
 \cM_{IJ} = E_I{}^A\,E_J{}^B\,\hat{\cM}_{AB}\qquad (\det \hat{\cM}=1)\,.
\label{eq;cM-EEM}
\end{align}
In general, $\hat{\cM}_{AB}$ can depend on the external coordinates (see \cite{1903.12175} for details), but here we suppose that it is constant for simplicity and consider only the internal part of the spacetime. 
According to the flux formulation \cite{1304.1472}, the equations of motion of the ExFT are covariant equations for the fluxes $\bm{X}_{AB}{}^C$ and $\hat{\cM}_{AB}$.\footnote{In DFT, we also have the dilaton and we need to consider the corresponding flux $\cF_A$ as well (see \cite{1903.12175}).}
In our situation, the generalized fluxes are constants $X_{AB}{}^C$ and the equations of motion are algebraic equations for $X_{AB}{}^C$ and $\hat{\cM}_{AB}$. 
If we find a solution for $\hat{\cM}_{AB}$, the background \eqref{eq;cM-EEM} is a solution of ExFT. 
Since $\cM_{IJ}$ depends only on the physical coordinates $x^i$, it is also a solution of the standard supergravity.\footnote{When $\theta_A$ does not vanish, it will be a solution of a certain gauged supergravity.}

In general, the ExDA has another maximally isotropic subalgebra $\mathfrak{g}'$ whose generators are denoted as $T'_a$\,. 
We also identify the full set of generators $\{T'_A\}=\{T_a,\cdots\}$ such that $T'_A$ satisfy an ExDA. 
This is just a redefinition of generators
\begin{align}
 T'_A = C_A{}^B\,T_B\qquad (C_A{}^B:\text{constant})\,,
\label{eq:NAD}
\end{align}
and the structure constants of the new algebra are
\begin{align}
 X'_{AB}{}^C=C_A{}^{A'}\,C_B{}^{B'}\,(C^{-1})_{C'}{}^{C}\,X_{A'B'}{}^{C'}\,.
\label{eq:X'-X}
\end{align}
Using the new set of generators $T'_A$, we can construct new generalized frame fields $E'_A{}^I$ which depend on the physical coordinates in the primed system $x'^i$\,. 
Here, the dimension of the physical space can be different, in which case $C_A{}^B$ is not an element of the duality group $\cG$\,.\footnote{The linear map of \cite{1701.07819}, which relates the M-theory and the type IIB picture, is such an example.} 
In general, the frame fields $E'_A{}^I$ are completely different from the original ones $E_A{}^I$ and they are not related by $C_A{}^B$: $E'_A{}^I\neq C_{A}{}^B\,E_B{}^I$\,. 
However, the generalized fluxes in the original and the dualized frame are related as in Eq.~\eqref{eq:X'-X} and since the equations of motion of ExFT are covariant equations of the fluxes and the constant matrix $\hat{\cM}_{AB}$, by introducing
\begin{align}
 \hat{\cM}'_{AB} \equiv C_A{}^{A'}\,C_B{}^{B'}\,\hat{\cM}_{A'B'}\,,
\end{align}
the new background
\begin{align}
 \cM'_{IJ} = E'_I{}^A\,E'_J{}^B\,\hat{\cM}'_{AB} \,,
\end{align}
is a solution of ExFT. 
This is a basic procedure to perform the non-Abelian duality. 

\subsection{Yang--Baxter deformation}

Let us consider the Yang--Baxter deformation \cite{hep-th:0210095} as a particular class of non-Abelian duality. 
In this case, we assume that the original ExDA is a semi-Abelian algebra, meaning that the dual structure constants $f_a{}^{\Pa}$ are absent.\footnote{Here, we suppose that the structure constants $f_a{}^{\cI}$ and $Z_a$ can be present.} 
Then the generalized frame fields are simply given by $E_A{}^I=\Exp{(\tilde{K}+t_0)\Delta}\mathbb{E}_A{}^I$\,. 
We now consider the redefinition \eqref{eq:NAD} with $C_A{}^B=\cR_A{}^B$ where $\cR$ is the twist matrix defined in Eq.~\eqref{eq:cR-def}.
As long as $\rr^{\Pa}$ satisfies the generalized CYBE, the redefined generators $T'_A$ satisfy a coboundary ExDA.
Using the ExDA, we can straightforwardly compute the generalized frame fields $E'_A{}^I$\,. 
Since $f_{ab}{}^c$, $f_a{}^{\cI}$, and $Z_a$ are not deformed under the redefinition $T'_A=\cR_A{}^B\,T_B$\,, we find
\begin{align}
 E'_A{}^I = \bm{\Pi}'_A{}^B E_B{}^I \,,
\end{align}
where $E_A{}^I$ are the original generalized frame fields.
Then we obtain a new solution of ExFT,
\begin{align}
 \cM'_{IJ} &\equiv E'_I{}^A\,E'_J{}^B\,\hat{\cM}'_{AB} = U_I{}^K\,U_J{}^L\,\cM_{KL}\,,
\\
 U_I{}^J &\equiv E_I{}^A\,(\bm{\Pi}'^{-1}\,\cR)_A{}^B\,E_B{}^J \,.
\label{eq:twist-U}
\end{align}

For example, in DFT, we obtain
\begin{align}
 (U_I{}^J) \equiv \begin{pmatrix} \delta_m^n & 0 \\
 (\pi'^{ab} + \rr^{ab})\,e_a^m\,e_b^n & \delta^m_n \end{pmatrix} .
\end{align}

For a general ExDA, it is a hard task to obtain $\pi'^{\Pa}$ because we need to compute the exponential action \eqref{eq:M-def}.
However, the usefulness of the Yang--Baxter deformation is that we can generally express $\pi^{\Pa}$ in terms of the classical $r$-matrix. 
For example, in DFT, we find
\begin{align}
 \pi'^{mn} = \rr^{ab}\,(v_a^m\,v_b^n - e_a^m\,e_b^n)\,.
\end{align}
Indeed, this vanishes at the unit element $x=x_0$ and also satisfies the differential equation \eqref{eq:multiplicative-DFT}. 
Then, the matrix $U_I{}^J$ is simplified as
\begin{align}
 (U_I{}^J) \equiv \begin{pmatrix} \delta_m^n & 0 \\
 \rho^{mn} & \delta^m_n \end{pmatrix} ,\qquad 
 \rho^{mn} \equiv \rr^{ab}\, v_a^m\,v_b^n\,.
\end{align}
Thus, for a given solution $\rr^{ab}$ of CYBE \eqref{eq:CYBE}, we can easily obtain the deformed background $\cM'_{IJ}$ simply by acting the coordinate-dependent matrix $U_I{}^J$. 

The same story can be extended to a general coboundary ExDA. 
In sections \ref{sec:M} and \ref{sec:IIB}, we find the explicit form of the Nambu--Lie structures $\pi'$ for coboundary EDAs. 
We then find that the matrix $U_I{}^J$ has a very simple form.
Again, $U_I{}^J$ is parameterized by the multi-vectors $\rho$ which is given by the classical $r$-matrix $\rr^{\Pa}$ and the left-invariant vector fields $v_a^i$. 

\section{Exceptional Drinfel'd algebra: M-theory section}
\label{sec:M}

As explained in section \ref{sec:R1}, the $R_1$ representation in the M-theory picture is decomposed as
\begin{align}
 (T_A) =\bigl( T_a,\,\tfrac{T^{a_1a_2}}{\sqrt{2!}},\,\tfrac{T^{a_1\cdots a_5}}{\sqrt{5!}},\,\tfrac{T^{a_1\cdots a_7,a'}}{\sqrt{7!}},\,\tfrac{T^{8,a_1a_2a_3}}{\sqrt{3!}},\,\tfrac{T^{8,a_1\cdots a_6}}{\sqrt{6!}},\,T^{8,8,a} \bigr)\,,
\end{align}
where $8$ denotes eight totally antisymmetric indices $1\cdots 8$, e.g., $T^{8,a_1a_2a_3}=T^{1\cdots 8,a_1a_2a_3}$.
When we consider the case $n=7$, generators including eight indices automatically disappear, and when we consider $n=6$, $T^{a_1\cdots a_7,a'}$ additionally disappears because it requires seven antisymmetric indices. 
Then we find that the dimension of the EDA coincides with the dimension of the $R_1$ representation given in Table \ref{tab:reps}. 

Using the general results given in the previous section, the embedding tensors for $E_{n(n)}$ EDA ($n\leq 8$) in the M-theory picture are obtained as follows:
\begin{align}
 X_a
 &= \tfrac{1}{3!}\,f_a{}^{b_1b_1b_3}\, R_{b_1b_1b_3} 
 + \tfrac{1}{6!}\,f_a{}^{b_1\cdots b_6}\, R_{b_1\cdots b_6} 
 + \tfrac{1}{8!}\,f_a{}^{b_1\cdots b_8,c}\, R_{b_1\cdots b_8,c} 
\nn\\
 &\quad + f_{ab}{}^c \, \tilde{K}^b{}_c - Z_a\, (\tilde{K} + t_0) \,,
\\
 X^{a_1a_2} 
 &= - f_c{}^{d a_1a_2}\, \tilde{K}^c{}_d 
 -f_{c_1c_2}{}^{[a_1}\,R^{a_2]c_1c_2} + 3\,Z_d\,R^{da_1a_2} \,,
\\
 X^{a_1\cdots a_5}
 &= - f_c{}^{d a_1\cdots a_5} \, \tilde{K}^c{}_d 
 -\tfrac{5!}{3!\,2!}\, f_b{}^{[a_1a_2a_3}\, R^{a_4a_5]b} 
\nn\\
 &\quad 
 - \tfrac{5!}{2!\,4!}\,f_{c_1c_2}{}^{[a_1}\,R^{a_2\cdots a_5]c_1c_2} + 6\,Z_b\,R^{ba_1\cdots a_5} \,,
\\
 X^{a_1\cdots a_7,a'}
 &= - \bigl(f_c{}^{d a_1\cdots a_7,a'} - \tfrac{1}{4}\,f_c{}^{a_1\cdots a_7a',d}\bigr) \, \tilde{K}^c{}_d
\nn\\
 &\quad 
 + 7\,f_b{}^{[a_1\cdots a_7}\,R^{a_7]a'b} -14\,f_b{}^{[a_1\cdots a_6}\,R^{a_7a']b}
\nn\\
 &\quad 
 + 21\,f_b{}^{a'[a_1a_2}\,R^{a_3\cdots a_7]b} + 14\,f_b{}^{[a_1a_2a_3}\,R^{a_4\cdots a_7a']b}
\nn\\
 &\quad 
 - \tfrac{7}{2}\,f_{b_1b_2}{}^{[a_1}\,R^{a_2\cdots a_7]b_1b_2,a'} + f_{bc}{}^{a'}\,R^{a_1\cdots a_7b,c}
\nn\\
 &\quad + Z_b\,\bigl(9\,R^{b a_1\cdots a_7,a'} + \tfrac{3}{4}\, R^{a_1\cdots a_7a',b} \bigr)\,,
\\
 X^{a_1\cdots a_8, a'_1a'_2a'_3} 
 &= 3\,f_b{}^{a_1\cdots a_8,[a'_1}\,R^{a'_2a'_3]b}
 - 56\,f_b{}^{a'_1a'_2a'_3 [a_1a_2a_3}\,R^{a_4\cdots a_8] b} 
\nn\\
 &\quad + 3\,f_b{}^{[a'_1a'_2|b}\,R^{a_1\cdots a_8,|a'_3]} + f_b{}^{a'_1a'_2a'_3} \, R^{a_1\cdots a_8,b} \,,
\\
 X^{8, a_1\cdots a_6} 
 &= -6\,f_b{}^{8,[a_1}\,R^{a_2\cdots a_6]b}
  + 6\,f_b{}^{[a_1\cdots a_5|b}\,R^{8,|a_6]} + f_b{}^{a_1\cdots a_6}\,R^{8,b} \,,
\\
 X^{8, 8,a} 
 &= 2\,f_d{}^{8,a}\, R^{8,d} + f_d{}^{8,d}\, R^{8,a} \,.
\end{align}
In particular we have
\begin{align}
\begin{split}
 &\vartheta_a = \beta\,f_{ab}{}^b -(1+\beta\,n)\,Z_a\,,\quad 
 \vartheta^{a_1a_2} = -\beta\,f_b{}^{ba_1a_2} \,,\quad 
 \vartheta^{a_1\cdots a_5} = -\beta\,f_b{}^{ba_1\cdots a_5} \,,
\\
 &\vartheta^{a_1\cdots a_7,a'} = -\beta\,\bigl(f_b{}^{ba_1\cdots a_7,a'} - \tfrac{1}{4}\,f_b{}^{a_1\cdots a_7a',b}\bigr) \,,\quad 
 \vartheta^{8,3} =
 \vartheta^{8,6} =
 \vartheta^{8,8,1} = 0\,.
\end{split}
\end{align}
By using these, we can easily write down the explicit form of the EDA,
\begin{align}
 T_A\circ T_B = (X_A)_B{}^C\, T_C \,.
\end{align}
We note that the vector $Z_{a}$ corresponds to $-L_a/9$ used in \cite{1911.07833} (or $-Z_{a}/3$ of \cite{1911.06320}). 

For simplicity, we show the explicit form of the EDA for $n\leq 7$:
\begin{align}
 T_a \circ T_b &= f_{ab}{}^c\,T_c \,,
\nn\\
 T_a \circ T^{b_1b_2} &= f_a{}^{b_1b_2c}\,T_c + 2\,f_{ac}{}^{[b_1}\,T^{b_2]c}
 +3\,Z_a\,T^{b_1b_2}\,,
\nn\\
 T_a \circ T^{b_1\cdots b_5} &= -f_a{}^{b_1\cdots b_5c}\,T_c - 10\,f_{a}{}^{[b_1b_2b_3}\,T^{b_4b_5]} - 5\,f_{ac}{}^{[b_1}\,T^{b_2\cdots b_5]c} 
 +6\,Z_a\,T^{b_1\cdots b_5}\,,
\nn\\
 T_a \circ T^{b_1\cdots b_7,b'} &= 7\,f_a{}^{[b_1\cdots b_6}\,T^{b_7] b'} 
 - 21\,f_a{}^{b'[b_1b_2}\,T^{b_3\cdots b_7]}
\nn\\
 &\quad - 7\,f_{ac}{}^{[b_1}\,T^{b_2\cdots b_7]c,b'}
 - f_{ac}{}^{b'}\,T^{b_1\cdots b_7,c} 
 +9\,Z_a\,T^{b_1\cdots b_7,b'}\,,
\nn\\
 T^{a_1a_2} \circ T_b &= -f_b{}^{a_1a_2c}\,T_c + 3\,f_{[c_1c_2}{}^{[a_1}\,\delta^{a_2]}_{b]}\,T^{c_1c_2}
 -9\,Z_c\,\delta_b^{[c}\,T^{a_1a_2]}\,,
\nn\\
 T^{a_1a_2} \circ T^{b_1b_2} &= -2\, f_c{}^{a_1a_2[b_1}\, T^{b_2]c} - f_{c_1c_2}{}^{[a_1}\,T^{a_2]b_1b_2c_1c_2}
 +3\,Z_c\,T^{a_1a_2b_1b_2c}\,,
\nn\\
 T^{a_1a_2} \circ T^{b_1\cdots b_5} &= 5\,f_c{}^{a_1a_2[b_1}\, T^{b_2\cdots b_5]c} -3\,\delta^{a_1a_2}_{de}\,f_{c_1c_2}{}^d\,T^{b_1\cdots b_5[c_1c_2,e]} 
 + 9\,Z_c\,T^{b_1\cdots b_5[a_1a_2,c]}\,,
\nn\\
 T^{a_1a_2} \circ T^{b_1\cdots b_7,b'} &= 7\,f_c{}^{a_1a_2[b_1}\, T^{b_2\cdots b_7]c,b'} + f_c{}^{a_1a_2b'}\, T^{b_1\cdots b_7,c} \,,
\nn\\
 T^{a_1\cdots a_5} \circ T_b &= f_b{}^{a_1\cdots a_5c}\,T_c + 10\,f_b{}^{[a_1a_2a_3}\,T^{a_4a_5]} + 20\,f_c{}^{[a_1a_2a_3}\,\delta_b^{a_4}\,T^{a_5]c} 
\\
 &\quad + 5\,f_{bc}{}^{[a_1}\,T^{a_2\cdots a_5]c} + 10\,f_{c_1c_2}{}^{[a_1}\,\delta^{a_2}_b\,T^{a_3a_4a_5]c_1c_2} 
 -36\,Z_c\,\delta_b^{[c}\,T^{a_1\cdots a_5]}\,,
\nn\\
 T^{a_1\cdots a_5} \circ T^{b_1b_2} &= 2\,f_c{}^{a_1\cdots a_5[b_1}\,T^{b_2]c} - 10\,f_c{}^{[a_1a_2a_3}\, T^{a_4a_5]b_1b_2c}
\nn\\
 &\quad + 5\,f_{c_1c_2}{}^{[a_1}\,T^{a_2\cdots a_5]c_1c_2[b_1,b_2]} 
 -12\,Z_c\,T^{ca_1\cdots a_5[b_1,b_2]}\,,
\nn\\
 T^{a_1\cdots a_5} \circ T^{b_1\cdots b_5} &= -5\,f_c{}^{a_1\cdots a_5[b_1}\, T^{b_2\cdots b_5]c} - 30\,f_c{}^{[a_1a_2a_3}\,\delta^{a_4a_5]c}_{d_1d_2e}\,T^{b_1\cdots b_5 d_1d_2,e} \,,
\nn\\
 T^{a_1\cdots a_5} \circ T^{b_1\cdots b_7,b'} &= -7\,f_c{}^{a_1\cdots a_5 [b_1}\,T^{b_2\cdots b_7]c,b'} - f_c{}^{a_1\cdots a_5 b'}\,T^{b_1\cdots b_7,c} \,,
\nn\\
 T^{a_1\cdots a_7,a'} \circ T_b &= 
 -21\,f_c{}^{[a_1\cdots a_6}\, \delta^{a_7]a'c}_{b d_1d_2} \,T^{d_1d_2}
 -126\, f_c{}^{a'[a_1a_2}\, \delta_{b d_1\cdots d_5}^{a_3\cdots a_7] c}\,T^{d_1\cdots d_5} \,,
\nn\\
 T^{a_1\cdots a_7,a'} \circ T^{b_1b_2} &= 7\, f_c{}^{[a_1\cdots a_6}\,T^{a_7]a' c b_1b_2} 
 -42\,f_c{}^{a'[a_1a_2}\,T^{a_3\cdots a_7] c [b_1,b_2]}\,,
\nn\\
 T^{a_1\cdots a_7,a'} \circ T^{b_1\cdots b_5} &= 21\,f_c{}^{[a_1\cdots a_6}\,\delta^{a_7]a'c}_{d_1d_2e}\,T^{b_1\cdots b_5 d_1d_2,e} \,,
\nn\\
 T^{a_1\cdots a_7,a'} \circ T^{b_1\cdots b_7,b'} &= 0 \,.
\nn
\end{align}
We can easily reproduce the $E_{6(6)}$ EDA of \cite{2007.08510} (up to sign convention) by a truncation.

Now, let us consider the Leibniz identities.
If we define
\begin{align}
\begin{split}
 f^3(x) &\equiv \tfrac{1}{3!}\,x^a\,f_a{}^{b_1b_2b_3}\,R_{b_1b_2b_3}\,,\qquad
 f^6(x) \equiv \tfrac{1}{6!}\,x^a\,f_a{}^{b_1\cdots b_6}\,R_{b_1\cdots b_6}\,,
\\
 f^{8,1}(x)& \equiv \tfrac{1}{8!}\,x^a\,f_a{}^{b_1\cdots b_8,b'}\,R_{b_1\cdots b_8,b'}\,,
\end{split}
\end{align}
the cocycle conditions \eqref{eq:1-cocycle} are decomposed as follows:
\begin{align}
 &\rmd f^3 (x,y) = 0\,,
\\
 &\rmd f^{6}(x,y) - [f^3(x),\,f^3(y)] =0 \,,
\\
 &\rmd f^{8,1}(x,y) - [f^3(x),\,f^6(y)] - [f^6(x),\,f^3(y)] =0 \,,
\end{align}
where
\begin{align}
\begin{split}
 \rmd f^{*}(x,y) &\equiv x\cdot f^{*}(y) - y\cdot f^{*}(x) - f^{*}([x,y])\,,
\\
 x\cdot f^{*}(y) &\equiv x^{\sfa}\, [f^{*}(y),\,f_{ab}{}^{c}\,K^{b}{}_{c}-Z_a\,K]\,.
\end{split}
\end{align}
If we introduce
\begin{align}
 \bm{\rr}^3 \equiv \tfrac{1}{3!}\,\rr^{a_1a_2a_3}\,R_{a_1a_2a_3}\,,\qquad
 \bm{\rr}^6 \equiv \tfrac{1}{6!}\,\rr^{a_1\cdots a_6}\,R_{a_1\cdots a_6}\,,\qquad
 \bm{\rr}^{8,1} \equiv \tfrac{1}{8!}\,\rr^{a_1\cdots a_8,a'}\,R_{a_1\cdots a_8,a'}\,,
\end{align}
and define $\cR\equiv \Exp{\bm{\rr}^3}\Exp{\bm{\rr}^6}\Exp{\bm{\rr}^{8,1}}$, the coboundary ansatz \eqref{eq:cR-def} are decomposed as
\begin{align}
 \mathfrak{f}^3(x) &= \rmd\bm{\rr}^3(x)\,,
\\
 \mathfrak{f}^6(x) &= \rmd\bm{\rr}^6(x) + \tfrac{1}{2!}\,\bigl[\bm{\rr}^3,\,f^3(x)\bigr]\,,
\\
 \mathfrak{f}^{8,1}(x) &= \rmd\bm{\rr}^{8,1}(x) + \bigl[\bm{\rr}^3,\, f^6(x)\bigr] - \tfrac{1}{3}\, \bigl[\bm{\rr}^3,\,\bigl[\bm{\rr}^3,\,f^3(x)\bigr]\bigr] \,,
\end{align}
where
\begin{align}
 \rmd\bm{\rr}^*(x) \equiv x^a\,\bigl[\bm{\rr}^*,\,f_{ab}{}^c\,K^b{}_c -Z_a\,K\bigr]\,.
\end{align}
More explicitly, the cocycle conditions are expressed as
\begin{align}
 0&= 6\,f_{[a|d}{}^{[c_1|}\,f_{|b]}{}^{d|c_2c_3]} - f_{ab}{}^{d}\,f_{d}{}^{c_1c_2c_3} -6\,Z_{[a}\,f_{b]}{}^{c_1c_2c_3}\,,
\\
 0&= 12\,f_{[a|d}{}^{[c_1|}\,f_{|b]}{}^{d|c_2\cdots c_6]} - f_{ab}{}^{d}\,f_{d}{}^{c_1\cdots c_6} -20\,f_{[a}{}^{[c_1c_2c_3}\,f_{b]}{}^{c_4c_5c_6]}
 -12\,Z_{[a}\,f_{b]}{}^{c_1\cdots c_6}\,,
\\
\begin{split}
 0&= 16\,f_{[a|d}{}^{[c_1|}\,f_{|b]}{}^{d|c_2\cdots c_8],c'} + 2\,f_{[a|d}{}^{c'}\,f_{|b]}{}^{c_1\cdots c_8,d} - f_{ab}{}^{d}\,f_{d}{}^{c_1\cdots c_8,c'}
\\
 &\quad -112\,f_{[a}{}^{[c_1c_2c_3}\,f_{b]}{}^{c_4\cdots c_8]c'}
 -18\,Z_{[a}\,f_{b]}{}^{c_1\cdots c_8,c'}\,,
\end{split}
\end{align}
and the coboundary ansatzes are
\begin{align}
 \mathfrak{f}_a{}^{b_1b_2b_3} &= 3\,f_{ac}{}^{[b_1|}\,\rr^{c|b_2b_3]} -3\,Z_a\,\rr^{b_1b_2b_3}\,,
\\
 \mathfrak{f}_a{}^{b_1\cdots b_6} &= 6\,f_{ac}{}^{[b_1|}\,\rr^{c|b_2\cdots b_6]} -10\,f_{a}{}^{[b_1b_2b_3}\,\rr^{b_4b_5b_6]} -6\,Z_a\,\rr^{b_1\cdots b_6}\,,
\\
\begin{split}
 \mathfrak{f}_a{}^{b_1\cdots b_8,b'} &= 8\,f_{ac}{}^{[b_1|}\,\rr^{c|b_2\cdots b_8],b'} + f_{ac}{}^{b'}\,\rr^{b_1\cdots b_8,c}
 + 56\,\rr^{[b_1b_2b_3}\,f_a{}^{b_4\cdots b_8]b'}
\\
 &\quad 
  -\tfrac{560}{3}\,f_{a}{}^{[b_1b_2b_3}\,\rr^{b_4b_5b_6}\,\rr^{b_7b_8]b'} 
  -9\,Z_a\,\rr^{b_1\cdots b_8,b'} \,.
\end{split}
\end{align}

To compute the fundamental identities \eqref{eq:FI}, we need the explicit components of the structure constants $X_{\mathring{A}\mathring{B}}{}^{\tilde{C}}$. 
For simplicity, if we consider the case $n\leq 7$, we find
\begin{align}
 f_c{}^{da_1a_2}\, f_d{}^{b_1b_2b_3}
 - 3\,f_d{}^{a_1a_2 [b_1}\,f_c{}^{b_2b_3]d}
 &= f_{d_1d_2}{}^{[a_1} \, f_c{}^{a_2]b_1b_2b_3d_1d_2} 
 + 3\,f_c{}^{a_1a_2b_1b_2b_3d}\,Z_d\,,
\\
 f_c{}^{da_1\cdots a_5}\, f_d{}^{b_1\cdots b_6}
 - 6\,f_d{}^{a_1\cdots a_5[b_1}\,f_c{}^{b_2\cdots b_6]d}
 &= - 30\,f_d{}^{[a_1a_2a_3}\,\delta^{a_4a_5]d}_{e_1e_2e_3} \, f_c{}^{b_1\cdots b_6 e_1e_2,e_3} \,,
\end{align}
which reduce to the known results \cite{2007.08510} if we consider the reduction to $n\leq 6$\,. 

As was found in \cite{1911.06320,1911.07833,2007.08510}, there is an additional constraint in the $E_{n(n)}$ EDA for $n\leq 6$\,. 
This comes from $Z_{AB}{}^C\,X_C = 0$ of Eq.~\eqref{eq:XXXX}. 
For example, the non-vanishing components of $Z^{a_1a_2}{}_b{}^C$ are determined as
\begin{align}
 Z^{a_1a_2}{}_{bc_1c_2} = 2\,f_{c_1c_2}{}^{[a_1}\,\delta^{a_2]}_b -12\,\delta^{a_1a_2}_{b[c_1}\,Z_{c_2]} \,,
\end{align}
and then we find
\begin{align}
\begin{split}
 Z^{a_1a_2}{}_{b}{}^C\,X_C
 &= \tfrac{1}{2!}\, Z^{a_1a_2}{}_{bc_1c_2} \,X^{c_1c_2}
\\
 &= \bigl(- f_{e_1e_2}{}^f +6\,Z_{e_1}\,\delta_{e_2}^f\bigr)\,\delta^{a_1a_2}_{fb} \,f_c{}^{d e_1e_2} \, \tilde{K}^c{}_d =0\,,
\end{split}
\end{align}
where we have used the Bianchi identities $f_{[ab}{}^d\,f_{c]d}{}^{e} = 0$ and $f_{ab}{}^c\,Z_c=0$. 
This is equivalent to
\begin{align}
 f_{c_1c_2}{}^a\,f_b{}^{d c_1c_2} - 6\,Z_c\, f_b{}^{adc} = 0\,.
\end{align}
As studied in \cite{2007.08510}, when $n\leq 6$ and $Z_a=0$, this condition, the cocycle condition, the fundamental identities, and the usual Bianchi identity $f_{[ab}{}^d\,f_{c]d}{}^{e} = 0$ are all of the Leibniz identities. 
In general, we will find additional identities, and in order to find the full set of identities, it may be useful to employ the results of the quadratic constraints studied in the gauged supergravity \cite{0809.5180}.
Here we do not study further on the Leibniz identities. 

For a given EDA, we can compute the adjoint action \eqref{eq:M-def} to obtain the matrix $M_A{}^B$ and using that we can get the generalized frame fields $E_A{}^I$\,. 
We parameterize the matrix $\bm{\Pi}$ as
\begin{align}
 \bm{\Pi} = \Exp{-\frac{1}{3!}\,\pi^{a_1a_2a_3}\,R_{a_1a_2a_3}} \Exp{-\frac{1}{6!}\,\pi^{a_1\cdots a_6}\,R_{a_1\cdots a_6}} \Exp{-\frac{1}{8!}\,\pi^{a_1\cdots a_8,a'}\,R_{a_1\cdots a_8,a'}}\,.
\end{align}
In the $E_{8(8)}$ case, the matrix size of $E_A{}^I$ is large, so here we show the explicit form of the generalized frame fields for $n\leq 7$,
\begin{align}
\begin{split}
 E_a &= e_a\,,
\\
 E^{a_1a_2} &= - \pi^{ba_1a_2}\,e_b + \Exp{-3\Delta} r^{a_1a_2}\,,
\\
 E^{a_1\cdots a_5} &= -\bigl(\pi^{ba_1\cdots a_5} + 5\,\pi^{[a_1a_2a_3}\,\pi^{a_4a_5]b}\bigr)\,e_b + 10\Exp{-3\Delta} \pi^{[a_1a_2a_3}\,r^{a_4a_5]} + \Exp{-6\Delta} r^{a_1\cdots a_5}\,,
\\
 E^{a_1\cdots a_7,a'} &=
 - \bigl(21\,\pi^{b[a_1\cdots a_5}\,\pi^{a_6a_7]a'} + 35\,\pi^{a'[a_1a_2}\,\pi^{a_3a_4a_5}\,\pi^{a_6a_7]b}\bigr)\,e_b
\\
 &\quad - 7\Exp{-3\Delta} \pi^{[a_1\cdots a_6}\,r^{a_7]a'}
 + 105\Exp{-3\Delta} \pi^{a'[a_1a_2}\,\pi^{a_3a_4a_5}\,r^{a_6a_7]}
\\
 &\quad + 21\Exp{-6\Delta}r^{[a_1\cdots a_5} \,\pi^{a_6a_7]a'} + \Exp{-9\Delta}r^{a_1\cdots a_7} \otimes r^{a'}\,,
\end{split}
\end{align}
where $r^{a_1\cdots a_p}\equiv r^{a_1}\wedge\cdots\wedge r^{a_p}$ ($r^a\equiv r^a_i\,\rmd x^i$) and $\pi^{\Pa}$ correspond to the Nambu--Lie structures. 

To find the explicit form of the differential identities \eqref{eq:diff-id}, we use the identity,
\begin{align}
 \Exp{-X}D_c \Exp{X} = D_c X - \tfrac{1}{2!}\,[X,D_cX] + \tfrac{1}{3!}\,[X,[X,D_cX]] - \tfrac{1}{4!}\,[X,[X,[X,D_cX]]] + \cdots\,,
\end{align}
and then we obtain
\begin{align}
\begin{split}
 D_a\bm{\pi}^3 &= f_a{}^3 + f_{ab}{}^c\,[\bm{\pi}^3,\,K^b{}_c] - 3\,Z_a\,\bm{\pi}^3\,,\qquad D_a \Delta = Z_a\,,
\\
 D_a\bm{\pi}^6 &= f_a{}^6 + f_{ab}{}^c\,[\bm{\pi}^6,\,K^b{}_c] + \tfrac{1}{2}\,[\bm{\pi}^3,\,f^3] - 6\,Z_a\,\bm{\pi}^6\,,
\\
 D_a\bm{\pi}^{8,1} &= f_a{}^{8,1} + f_{ab}{}^c\,[\bm{\pi}^{8,1},\,K^b{}_c] + [\bm{\pi}^3,\,f^6] + \tfrac{1}{3}\,[\bm{\pi}^3,\,[\bm{\pi}^3,\,f^3]] - 9\,Z_a\,\bm{\pi}^{8,1}\,,
\end{split}
\end{align}
where $f_a{}^3 \equiv f^3(T_a)$, $f_a{}^6 \equiv f^6(T_a)$, $f_a{}^{8,1} \equiv f^{8,1}(T_a)$, and
\begin{align}
 \bm{\pi}^3 &\equiv \tfrac{1}{3!}\,\pi^{b_1b_2b_3}\,R_{b_1b_2b_3}\,,\quad
 \bm{\pi}^6 \equiv \tfrac{1}{6!}\,\pi^{b_1\cdots b_6}\,R_{b_1\cdots b_6}\,,
\quad
 \bm{\pi}^{8,1} \equiv \tfrac{1}{8!}\,\pi^{b_1\cdots b_8,b'}\,R_{b_1\cdots b_8,b'}\,.
\end{align}
In components, they are expressed as
\begin{align}
\begin{split}
 D_a\pi^{b_1b_2b_3} &= f_a{}^{b_1b_2b_3} + 3\,f_{ac}{}^{[b_1|}\, \pi^{c|b_2b_3]} - 3\,Z_a\,\pi^{b_1b_2b_3}\,,\qquad D_a \Delta = Z_a\,,
\\
 D_a\pi^{b_1\cdots b_6} &= f_a{}^{b_1\cdots b_6} + 6\,f_{ac}{}^{[b_1|}\, \pi^{c|b_2\cdots b_6]} - 6\,Z_a\,\pi^{b_1\cdots b_6} - 10\,f_a{}^{[b_1b_2b_3}\,\pi^{b_4b_5b_6]} \,,
\\
 D_a\pi^{b_1\cdots b_8,b'} &= f_a{}^{b_1\cdots b_8,b'} + 8\,f_{ac}{}^{[b_1|}\,\pi^{c|b_2\cdots b_8,b'} + f_{ac}{}^{b'}\,\pi^{b_1\cdots b_8,c} - 9\,Z_a\,\pi^{b_1\cdots b_8,b'}
\\
 &\quad + 56\,f_a{}^{b'[b_1\cdots b_5}\,\pi^{b_6b_7b_8]} + \tfrac{560}{3}\,f_a{}^{[b_1b_2b_3}\,\pi^{b_4b_5b_6}\,\pi^{b_7b_8]b'}\,.
\end{split}
\end{align}
For the Nambu--Lie structures (e.g., $\pi^{i_1i_2i_3}=\pi^{a_1a_2a_3}\,e_{a_1}^{i_1}\,e_{a_2}^{i_2}\,e_{a_3}^{i_3}$), we find
\begin{align}
\begin{split}
 &\Lie_{v_a}\pi^{i_1i_2i_3} = \Exp{-3 \Delta} f_{a}{}^{b_1b_2b_3}\,v_{b_1}^{i_1}\,v_{b_2}^{i_2}\,v_{b_3}^{i_3}\,, \qquad
 \Lie_{v_a} \Delta = Z_a\,,
\\
 &\Lie_{v_a}\pi^{i_1\cdots i_6}+10\,\pi^{[i_1i_2i_3}\,\Lie_{v_a}\pi^{i_4i_5i_6]} = \Exp{-6 \Delta} f_{a}{}^{b_1\cdots b_6}\,v_{b_1}^{i_1}\cdots v_{b_6}^{i_6}\,, 
\\
 &\Lie_{v_a}\pi^{i_1\cdots i_8,i'}-56\,\bigl(\pi^{i'[i_1\cdots i_5}+\tfrac{10}{3!}\,\pi^{i'[i_1i_2}\,\pi^{i_3i_4i_5} \bigr)\,\Lie_{v_a}\pi^{i_6i_7i_8]} = \Exp{-9 \Delta} f_{a}{}^{b_1\cdots b_8,b'} v_{b_1}^{i_1}\cdots v_{b_8}^{i_8}\,v_{b'}^{i'}\,,
\end{split}
\label{eq:d_pi-M}
\end{align}
by using Eq.~\eqref{eq:partial-pi}.
They are the properties for a general EDA. 

In particular, for a coboundary EDA, we can find a solution of the differential equations \eqref{eq:d_pi-M} that satisfies $\pi=0$ at $x=x_0$,
\begin{align}
 \pi^{i_1i_2i_3} &= \rr^{a_1a_2a_3} \,\bigl(\Exp{-3\Delta}v_{a_1}^{i_1}\,v_{a_2}^{i_2}\,v_{a_3}^{i_3} - e_{a_1}^{i_1}\,e_{a_2}^{i_2}\,e_{a_3}^{i_3}\bigr) \,,
\label{eq:M-pi-sol1}
\\
 \pi^{i_1\cdots i_6} &= \rr^{a_1\cdots a_6} \,\bigl(\Exp{-6\Delta}v_{a_1}^{i_1}\cdots v_{a_6}^{i_6} - e_{a_1}^{i_1}\cdots e_{a_6}^{i_6}\bigr)
 - 10\,\pi^{[i_1i_2i_3}\,\rr^{i_4i_5i_6]} \,,
\\
 \pi^{i_1\cdots i_8,i'} &= \rr^{a_1\cdots a_8,b'} \,\bigl(\Exp{-9\Delta}v_{a_1}^{i_1}\cdots v_{a_8}^{i_8}\,v_{a'}^{i'} - e_{a_1}^{i_1}\cdots e_{a_8}^{i_8}\,e_{a'}^{i'}\bigr)
 +56\Exp{-3\Delta} \rr^{a_1a_2a_3}\,\pi^{i'[i_1\cdots i_5}\,v^{i_6}_{a_1}\,v^{i_7}_{a_2}\,v^{i_8]}_{a_3}
\nn\\
 &\quad 
 -\tfrac{280}{3}\Exp{-3\Delta}\rr^{a_1a_2a_3}\,\pi^{i'[i_1i_2}\,v^{i_3}_{a_1}\,v^{i_4}_{a_2}\,v^{i_5}_{a_3}\,\rr^{i_6i_7i_8]} 
 +\tfrac{560}{3}\, \rr^{i'[i_1i_2}\,\rr^{i_3i_4i_5}\,\pi^{i_6i_7i_8]}\,,
\label{eq:M-pi-sol3}
\end{align}
where $\rr^{i_1i_2i_3}\equiv \rr^{a_1a_2a_3}\,e_{a_1}^{i_1}\,e_{a_2}^{i_2}\,e_{a_3}^{i_3}$ and $n\leq 8$ is assumed. 
Namely, in this case, we can easily get the Nambu--Lie structures without computing the adjoint action \eqref{eq:M-def}. 

In order to perform the Yang--Baxter deformation, we need the matrix $U_I{}^J$ defined in \eqref{eq:twist-U}. 
Since the above expression for $\pi$ is intricate, one may consider that $U_I{}^J$ is also complicated, but in fact $U_I{}^J$ is surprisingly simple,
\begin{align}
 U_I{}^J = \bigl(\Exp{\Exp{-3\Delta}\rho^3} \Exp{\Exp{-6\Delta}\rho^6} \Exp{\Exp{-9\Delta}\rho^{8,1}}\bigr)_I{}^J\,,
\label{eq:U-M}
\end{align}
where we have defined $\rho^3 \equiv \tfrac{1}{3!}\,\rho^{i_1i_2i_3}\,R_{i_1i_2i_3}$, $\rho^6\equiv \tfrac{1}{6!}\,\rho^{i_1\cdots i_6}\,R_{i_1\cdots i_6}$, and $\rho^{8,1}\equiv \tfrac{1}{8!}\,\rho^{i_1\cdots i_8,i'}\,R_{i_1\cdots i_8,i'}$, and the multi-vectors $\rho$ are defined, for example, as $\rho^{i_1i_2i_3}\equiv r^{a_1a_2a_3}\,v_{a_1}^{i_1}\,v_{a_2}^{i_2}\,v_{a_3}^{i_3}$\,. 

This simple expression is not a coincidence and can be derived as follows.
For convenience, let us define a matrix $U_A{}^B\equiv E^{(0)}_A{}^I\,U_I{}^J\,E^{(0)}_J{}^B = (\bm{\Pi}^{-1}\,\cR)_A{}^B$\,, where $E^{(0)}_A{}^I\equiv E_A{}^I\rvert_{\pi=0}$\,. 
The differential identity \eqref{eq:diff-id-M} shows
\begin{align}
 (\cR\,U^{-1}\Exp{(\tilde{K}+t_0)\Delta}\bm{A})_A{}^D\,D_c (\bm{A}^{-1} \Exp{-(\tilde{K}+t_0)\Delta}U\,\cR^{-1})_D{}^B = X_{cA}{}^B\,,
\end{align}
and by using the coboundary ansatz, $X_{cA}{}^B = (\cR\,X^{(0)}_c\,\cR^{-1})$, this becomes
\begin{align}
 (U^{-1}\Exp{(\tilde{K}+t_0)\Delta}\bm{A})_A{}^D\,D_c (\bm{A}^{-1}\Exp{-(\tilde{K}+t_0)\Delta}U)_D{}^B = X^{(0)}_c \,.
\end{align}
Using the differential identities Eq.~\eqref{eq:AdA}, we can easily find a solution of this differential equation. 
In fact, $\bm{A}^{-1} \Exp{-(\tilde{K}+t_0)\Delta}U$ should coincide with $\bm{A}^{-1}\Exp{-(\tilde{K}+t_0)\Delta}$ up to the left multiplication by a constant matrix. 
The constant matrix can be found by considering $U_A{}^B=\cR_A{}^B$, $\bm{A}_A{}^B=\delta_A^B$, and $\Delta=0$ at $x=x_0$, and then we get
\begin{align}
 \bm{A}^{-1}\Exp{-(\tilde{K}+t_0)\Delta}U = \cR\,\bm{A}^{-1}\Exp{-(\tilde{K}+t_0)\Delta}.
\end{align}
This shows that $U_A{}^B = \bigl(\bm{A}\Exp{(\tilde{K}+t_0)\Delta}\cR\Exp{-(\tilde{K}+t_0)\Delta}\bm{A}^{-1}\bigr)_A{}^B$ and then we find
\begin{align}
 U_I{}^J = (\Omega_\Delta)_I{}^J \,,\qquad 
 (\Omega_\Delta)_I{}^J \equiv \hat{V}_I{}^A\,\bigl(\Exp{(\tilde{K}+t_0)\Delta}\cR\Exp{-(\tilde{K}+t_0)\Delta}\bigr)_A{}^B\,\hat{V}_B{}^I\,.
\label{eq:U-formula}
\end{align}
This generally explains the simple result \eqref{eq:U-M}. 

Similar to the case of DFT, for a given solution of CYBE for $\rr^{a_1a_2a_3}$, $\rho^{a_1\cdots a_6}$, and $\rho^{a_1\cdots a_8,a'}$, we can easily generate a new solution of EFT simply by multiplying the coordinate-dependent twist matrix $U_I{}^J$ to the original generalized metric. 

In the following, we consider the generalized CYBE by assuming $Z_a=0$ for simplicity.
Using $\cR=\Exp{\bm{\rr}^3}\Exp{\bm{\rr}^6}\Exp{\bm{\rr}^{8,1}}$, we can compute some components of Eq.~\eqref{eq:gen-CYBE} as
\begin{align}
 \text{CYBE}^{a_1a_2 b_1b_2b_3} &= -\bigl[3\,f_{cd}{}^{[b_1}\,\rr^{|c|b_2b_3]}\,\rr^{a_1a_2d} + f_{cd}{}^{[a_1}\, \bigl(\rr^{a_2]cd b_1b_2b_3} - 5\,\rr^{a_2][cd}\,\rr^{b_1b_2b_3]} \bigr)\bigr]\,,
\label{eq:CYBE-3}
\\
 \text{CYBE}^{a_1a_2}{}^0 &= 2\,\beta\,f_{cd}{}^{[a_1}\,\rr^{a_2]cd} = 0\,. 
\end{align}
The first equation is precisely the CYBE found in \cite{2007.08510} but the second one is weaker than the known one
\begin{align}
 f_{cd}{}^{a}\,\rr^{bcd} = 0\,.
\label{eq:rho-3-trace}
\end{align}
Actually, this condition comes from the non-negative-level part (coefficients of $K^a{}_b$) of the relation $\mathfrak{X}^{a_1a_2} = \cX^{a_1a_2}$ and thus the conditions \eqref{eq:gen-CYBE} are not sufficient for Eq.~\eqref{eq:gen-YB}. 

When $n\leq 6$, we find $\mathfrak{f}_a{}^{b_1\cdots b_6}=0$ and Eqs.~\eqref{eq:CYBE-3} and \eqref{eq:rho-3-trace} will be the only constraints on $\rr^{a_1a_2a_3}$ and $\rr^{a_1\cdots a_6}$\,. 
In $n\leq 7$, we obtain further conditions. 
Here, we do not try to find the full set of constraints, but consider the CYBE for $\rr^{a_1\cdots a_6}$\,. 
For simplicity, we truncate other multi-vectors and consider $\cR=\Exp{\bm{\rr}^6}$.
Then, we have $\text{CYBE}^{a_1a_2}{}^0=0$\,, and $\text{CYBE}^{a_1a_2}{}^{\Pb}=0$ and $\text{CYBE}^{a_1\cdots a_5}{}^0=0$ follow from $\text{CYBE}^{a_1a_2 b_1b_2b_3} = - f_{cd}{}^{[a_1}\, \rr^{a_2]cd b_1b_2b_3}=0$\,. 
The next non-trivial component is
\begin{align}
 \text{CYBE}^{a_1\cdots a_5}{}^{b_1\cdots b_6} &=
 -\bigl[6\,f_{cd}{}^{[b_1}\,\rr^{|c|b_2\cdots b_6]}\,\rr^{da_1\cdots a_5}
\nn\\
 &\quad +5\,f_{cd}{}^{[a_1} \bigl(9\,\delta^{a_2\cdots a_5]cd}_{e_1\cdots e_6}\,\delta_f^{[e_1}\rr^{e_2\cdots e_6][b_1}\,\rr^{b_2\cdots b_6]f} - \rr^{a_2\cdots a_5]cd}\,\rr^{b_1\cdots b_6} \bigr)\bigr]=0 .
\end{align}
By using $f_{cd}{}^{[a_1}\, \rr^{a_2]cd b_1b_2b_3}=0$ this is simplified as
\begin{align}
 \text{CYBE}^{a_1\cdots a_5}{}^{b_1\cdots b_6} =
 -\bigl(6\,f_{cd}{}^{[b_1}\,\rr^{|c|b_2\cdots b_6]}\,\rr^{da_1\cdots a_5}
 - 15\,f_{cd}{}^{[a_1}\,\rr^{a_2\cdots a_5]c[b_1}\,\rr^{b_2\cdots b_6]d} \bigr)=0\,,
\end{align}
and this can be regarded as the generalized CYBE for $\rr^{a_1\cdots a_6}$\,. 
We can also do the same computation for $\rr^{8,1}$ but we leave further details for future work. 

\section{Exceptional Drinfel'd algebra: type IIB section}
\label{sec:IIB}

In this section, we consider the case where the dimension of the subalgebra $\mathfrak{g}$ is $n-1$\,. 
In this case, we can decompose the generators as
\begin{align}
 (T_A) =\bigl(T_{\sfa},\, T^{\sfa}_\alpha,\, \tfrac{T^{\sfa_{123}}}{\sqrt{3!}},\, \tfrac{T^{\sfa_{1\cdots 5}}_\alpha}{\sqrt{5!}},\, \tfrac{T^{\sfa_{1\cdots 6},\sfa}}{\sqrt{6!}},\, T^7_{(\alpha_1\alpha_2)} ,\,
 \tfrac{T^{7,\sfa_{12}}_\alpha}{\sqrt{2!}},\, \tfrac{T^{7,\sfa_{1\cdots 4}}}{\sqrt{4!}},\, \tfrac{T^{7,\sfa_{1\cdots 6}}_\alpha}{\sqrt{6!}},\, T^{7,7,\sfa}\bigr)\,,
\end{align}
where $7$ denotes eight totally antisymmetric indices and we also used a shorthand notation $\sfa_{1\cdots p}\equiv \sfa_1\cdots \sfa_p$\,. 
Similar to the case of the M-theory picture, we can check that the dimension of the EDA coincides with the dimension of the $R_1$ representation given in Table \ref{tab:reps}. 

Similar to the M-theory case, the embedding tensors for $E_{n(n)}$ EDA ($n\leq 8$) in the type IIB picture are obtained as follows:
\begin{align}
 X_{\sfa} &= f_{\sfa\sfb}{}^{\sfc}\,\tilde{K}^{\sfb}{}_{\sfc}
 + f_{\sfa\beta}{}^\gamma \,R^\beta{}_\gamma
 + \tfrac{1}{2!}\,f_{\sfa}{}_{\beta}^{\sfb_1\sfb_2}\,R^\beta_{\sfb_1\sfb_2}\,,
 + \tfrac{1}{4!}\,f_{\sfa}{}^{\sfb_1\cdots \sfb_4}\,R_{\sfb_1\cdots \sfb_4} 
\nn\\
 &\quad 
 + \tfrac{1}{6!}\,f_{a}{}_{\beta}^{\sfb_1\cdots \sfb_6}\,R^\beta_{\sfb_1\cdots \sfb_6}
 + \tfrac{1}{7!}\,f_{a}{}^{\sfb_1\cdots \sfb_7,\sfb'}\,R_{\sfb_1\cdots \sfb_7,\sfb'}
 -Z_{\sfa}\,(\tilde{K}+t_0)\,, 
\\
 X^{\sfa}_{\alpha} &= -f_{\sfb}{}_{\alpha}^{\sfc\sfa}\,\tilde{K}^{\sfb}{}_{\sfc}
 -\tfrac{1}{2!}\,\bigl(\delta_\alpha^\beta\,f_{\sfb_1\sfb_2}{}^{\sfa} + 2\,\delta^{\sfa}_{[\sfb_1}\,f_{\sfb_2]\alpha}{}^\beta\bigr)\,R_\beta^{\sfb_1\sfb_2}
 -2\,Z_{\sfb}\,R_\alpha^{\sfa\sfb}\,,
\\
 X^{\sfa_{123}} &= -f_{\sfb}{}^{\sfc \sfa_1\sfa_2\sfa_3}\,\tilde{K}^{\sfb}{}_{\sfc}
 -3\,\epsilon^{\beta\gamma}\,f_{\sfb}{}_{\beta}^{[\sfa_1\sfa_2}\, R_\gamma^{\sfa_3]\sfb}
 -\tfrac{3}{2}\,f_{\sfb_1\sfb_2}{}^{[\sfa_1}\, R^{\sfa_2\sfa_3]\sfb_1\sfb_2}
 -4\,Z_{\sfb}\,R^{\sfa_{123}\sfb}\,,
\\
 X_\alpha^{\sfa_1\cdots \sfa_5} &= -f_{\sfb}{}_{\alpha}^{\sfc\sfa_1\cdots \sfa_5}\,\tilde{K}^{\sfb}{}_{\sfc}
 - 5\, f_{\sfb}{}^{[\sfa_1\cdots \sfa_4}\, R_\alpha^{\sfa_5]\sfb}
 + 10\,f_{\sfb}{}_{\alpha}^{[\sfa_1\sfa_2}\, R^{\sfa_3\sfa_4\sfa_5]\sfb} 
\nn\\
 &\quad -\tfrac{5}{2}\, f_{\sfb_1\sfb_2}{}^{[\sfa_1}\, R_\alpha^{\sfa_2\cdots \sfa_5]\sfb_1\sfb_2}
 - f_{\sfb\alpha}{}^\beta\, R_\beta^{\sfa_1\cdots\sfa_5\sfb} 
 - 6\,Z_{\sfb}\, R_\alpha^{\sfa_1\cdots\sfa_5\sfb}\,, 
\\
 X^{\sfa_1\cdots \sfa_6,\sfa'} &=
 -\bigl(f_{\sfb}{}^{\sfc\sfa_1\cdots\sfa_6,\sfa'} 
 - c_{7,1}\,f_{\sfb}{}^{\sfa_1\cdots\sfa_6\sfa',\sfc}\bigr)\,\tilde{K}^{\sfb}{}_{\sfc}
\nn\\
 &\quad + \epsilon^{\beta\gamma}\, \bigl(f_{\sfb}{}_{\beta}^{\sfa_1\cdots\sfa_6}\,R_\gamma^{\sfa' \sfb} -7\,c_{6}\,f_{\sfb}{}^{[\sfa_1\cdots\sfa_6}_\beta\,R^{\sfa']\sfb}_\gamma \bigr)
\nn\\
 &\quad -20\,f_{\sfb}{}^{\sfa'[\sfa_1\sfa_2\sfa_3}\,R^{\sfa_4\sfa_5\sfa_6]\sfb}
 +35\,c_4\,f_{\sfb}{}^{[\sfa_1\cdots\sfa_4}\,R^{\sfa_5\sfa_6\sfa']\sfb}
\nn\\
 &\quad +\epsilon^{\beta\gamma}\,\bigl(6\,f_{\sfb}{}_{\beta}^{\sfa'[\sfa_1}\,R^{\sfa_2\cdots\sfa_6]\sfb}_\gamma 
 -21\,c_2\,f_{\sfb}{}^{[\sfa_1\sfa_2}_\beta\,R_\gamma^{\sfa_3\cdots\sfa_6\sfa']\sfb} \bigr) 
\nn\\
 &\quad -3\,f_{\sfb_1\sfb_2}{}^{[\sfa_1}\,R^{\sfa_2\cdots\sfa_6]\sfb_1\sfb_2,\sfa'}
 - f_{\sfb\sfc}{}^{\sfa'}\,R^{\sfa_1\cdots\sfa_6\sfb,\sfc}
\nn\\
 &\quad 
 +8\,Z_{\sfb}\,R^{\sfa_1\cdots\sfa_6\sfb,\sfa'}-(1+c_{7,1})\,Z_{\sfb}\,R^{\sfa_1\cdots\sfa_6\sfa',\sfb} \,,
\\
 X^{\sfa_{1\cdots 7}}_{(\alpha_1\alpha_2)}
 &= -7\,f_{\sfb}{}^{[\sfa_1\cdots\sfa_6}_{(\alpha_1}\,R^{\sfa_7]\sfb}_{\alpha_2)}
 + 21\,f_{\sfb}{}^{\sfa_1\sfa_2}_{(\alpha_1}\,R^{\sfa_3\cdots\sfa_7]\sfb}_{\alpha_2)} 
 + f_{\sfb(\alpha_1}{}^\beta\,\epsilon_{\alpha_2)\beta}\,R^{\sfa_1\cdots\sfa_7,\sfb}\,,
\\
 X^{\sfa_{1\cdots 7},\sfa'_{1 2}}_\alpha
 &= -2\,f_{\sfb}{}^{\sfa_1\cdots\sfa_7,[\sfa'_1}\,R^{\sfa'_2]\sfb}_\alpha
 - 7\, f_{\sfb}{}^{[\sfa_1\cdots\sfa_6}_{\alpha}\,R^{\sfa_7]\sfa'_1\sfa'_2\sfb}
\nn\\
 &\quad + 21\,f_{\sfb}{}^{\sfa'_1\sfa'_2[\sfa_1\sfa_2}\,R^{\sfa_3\cdots\sfa_7]\sfb}_\alpha
 + 2\,f_{\sfb}{}^{[\sfa'_1|\sfb}_\alpha\,R^{\sfa_1\cdots\sfa_7,|\sfa'_2]} 
 + f_{\sfb}{}^{\sfa'_1\sfa'_2}_\alpha\,R^{\sfa_1\cdots\sfa_7,\sfb} \,,
\\
 X^{\sfa_{1\cdots 7},\sfa'_{1\cdots 4}} 
 &= -4\,f_{\sfb}{}^{\sfa_1\cdots\sfa_7,[\sfa'_1}\,R^{\sfa'_2\sfa'_3\sfa'_4]\sfb}
 - 7\,\epsilon^{\beta\gamma}\,f_{\sfb}{}^{[\sfa_1\cdots\sfa_6}_{\beta}\,R^{\sfa_7]\sfa'_1\cdots\sfa'_4\sfb}_\gamma
\nn\\
 &\quad + 4\,f_{\sfb}{}^{[\sfa'_1\sfa'_2\sfa'_3|\sfb}\,R^{\sfa_1\cdots\sfa_7,|\sfa'_4]}
 + f_{\sfb}{}^{\sfa'_1\cdots\sfa'_4}\,R^{\sfa_1\cdots\sfa_7,\sfb} \,,
\\
 X^{\sfa_{1\cdots 7},\sfa'_{1\cdots 6}}_\alpha &= -6\,f_{\sfb}{}^{\sfa_{1\cdots 7},[\sfa'_1}\,R_\alpha^{\sfa'_2\cdots\sfa'_6]\sfb}
 + 6\,f_{\sfb}{}^{[\sfa'_1\cdots\sfa'_5|\sfb}_{\alpha}\,R^{\sfa_{1\cdots 7},|\sfa'_6]} 
 + f_{\sfb}{}^{\sfa'_1\cdots\sfa'_6}_\alpha\,R^{\sfa_{1\cdots 7},\sfb} \,,
\\
 X^{7,7,\sfa} &= f_{\sfb}{}^{7,\sfb}\,R^{7,\sfa} +2\, f_{\sfb}{}^{7,\sfa}\,R^{7,\sfb}\,,
\end{align}
where $\epsilon^{12}=\epsilon_{12}=1$ and constants $c_2$, $c_4$, $c_6$, and $c_{7,1}$ are defined in Appendix \ref{app:IIB-algebra}.
From this expression, we find the vector $\vartheta_A$ as follows:
\begin{align}
 &\vartheta_{\sfa} = \beta\,f_{\sfa\sfb}{}^{\sfb} -Z_{\sfa}\,(1+\beta\,\delta_{\sfb}^{\sfb})\,, \quad
 \vartheta^{\sfa}_{\alpha} = -\beta\,f_{\sfb}{}_{\alpha}^{\sfb\sfa} \,,\quad
 \vartheta^{\sfa_{123}} = -\beta\,f_{\sfb}{}^{\sfb \sfa_{123}} \,,\quad
 \vartheta_\alpha^{\sfa_{1\cdots 5}} = -\beta\,f_{\sfb}{}_{\alpha}^{\sfb\sfa_{1\cdots 5}} \,,
\nn\\
 &\vartheta^{\sfa_{1\cdots 6},\sfa'} =
 -\beta\,\bigl(f_{\sfb}{}^{\sfb\sfa_{1\cdots 6},\sfa'} 
 - c_{7,1}\,f_{\sfb}{}^{\sfa_{1\cdots 6}\sfa',\sfb}\bigr) \,,\quad
 \vartheta^{7}_{\alpha\beta}=\vartheta^{7,2}_\alpha=\vartheta^{7,4}=\vartheta^{7,6}_\alpha =\vartheta^{7,7,1} = 0\,.
\end{align}

We show a more concise form of the EDA for a particular case, $n\leq 7$:
{\small
\begin{align}
\begin{split}
 T_{\sfa}\circ T_{\sfb} &=f_{\sfa\sfb}{}^{\sfc}\,T_{\sfc}\,,
\\
 T_{\sfa}\circ T^{\sfb}_\beta &= f_{\sfa}{}_{\beta}^{\sfc\sfb}\,T_{\sfc}
 + f_{\sfa\beta}{}^{\gamma}\,T_\gamma^{\sfb} - f_{\sfa\sfc}{}^{\sfb}\,T_\beta^{\sfc} 
 +2\,Z_{\sfa}\,T^{\sfb}_\beta \,,
\\
 T_{\sfa}\circ T^{\sfb_1\sfb_2\sfb_3} &=
 f_{\sfa}{}^{\sfc\sfb_1\sfb_2\sfb_3}\, T_{\sfc} + 3\,\epsilon^{\gamma\delta}\,f_{\sfa}{}_{\gamma}^{[\sfb_1\sfb_2}\, T_{\delta}^{\sfb_3]} - 3\,f_{\sfa\sfc}{}^{[\sfb_1}\, T^{\sfb_2\sfb_3]\sfc} 
 +4\,Z_{\sfa}\,T^{\sfb_1\sfb_2\sfb_3}\,,
\\
 T_{\sfa}\circ T_\beta^{\sfb_1\cdots \sfb_5} &=
 f_{\sfa}{}_{\beta}^{\sfc\sfb_1\cdots \sfb_5}\,T_{\sfc}
 + 5\, f_{\sfa}{}^{[\sfb_1\cdots \sfb_4}\,T_\beta^{\sfb_5]}
 - 10\, f_{\sfa}{}_{\beta}^{[\sfb_1\sfb_2}\,T^{\sfb_3\sfb_4\sfb_5]} 
\\
 &\quad 
 + f_{\sfa\beta}{}^{\gamma}\, T_\gamma^{\sfb_1\cdots \sfb_5} 
 - 5\,f_{\sfa\sfc}{}^{[\sfb_1}\,T_\beta^{\sfb_2\cdots \sfb_5]\sfc} 
 + 6\,Z_{\sfa}\,T^{\sfb_1\cdots\sfb_5}_\beta\,,
\\
 T_{\sfa}\circ T^{\sfb_1\cdots \sfb_6,\sfb'} &=
 -\epsilon^{\gamma\delta}\,f_{\sfa}{}_{\gamma}^{\sfb_1\cdots \sfb_6}\, T_\delta^{\sfb'}
 + 20\, f_{\sfa}{}^{\sfb' [\sfb_1\sfb_2\sfb_3}\,T^{\sfb_4\sfb_5\sfb_6]}
 - 6\,\epsilon^{\gamma\delta}\,f_{\sfa}{}_{\gamma}^{\sfb'[\sfb_1}\,T_\delta^{\sfb_2\cdots \sfb_6]}
\\
 &\quad 
 - 6\,f_{\sfa\sfc}{}^{[\sfb_1|}\,T^{\sfc|\sfb_2\cdots \sfb_6],\sfb'}
 - f_{\sfa\sfc}{}^{\sfb'}\,T^{\sfb_1\cdots \sfb_6,\sfc}
 + 8\,Z_{\sfa}\,T^{\sfb_1\cdots\sfb_6,\sfb'}\,,
\\
 T^{\sfa}_\alpha\circ T_{\sfb} &= 
 f_{\sfb}{}_{\alpha}^{\sfa\sfc} \, T_{\sfc} 
 + 2\,\delta^{\sfa}_{[\sfb}\,f_{\sfc]\alpha}{}^{\gamma}\, T_\gamma^{\sfc}
 + f_{\sfb\sfc}{}^{\sfa}\,T_\alpha^{\sfc} 
 +4\,Z_{\sfc}\,\delta^{[\sfa}_{\sfb}\,T^{\sfc]}_\alpha\,,
\\
 T^{\sfa}_\alpha\circ T^{\sfb}_\beta &= - f_{\sfc}{}_{\alpha}^{\sfa\sfb}\,T_\beta^{\sfc} 
 - f_{\sfc\alpha}{}^\gamma\,\epsilon_{\gamma\beta}\,T^{\sfc \sfa\sfb} 
 + \tfrac{1}{2}\,\epsilon_{\alpha\beta}\,f_{\sfc_1\sfc_2}{}^{\sfa}\,T^{\sfc_1\sfc_2\sfb}
 -2\,\epsilon_{\alpha\beta}\,Z_{\sfc}\,T^{\sfa\sfb\sfc} \,,
\\
 T^{\sfa}_\alpha\circ T^{\sfb_1\sfb_2\sfb_3}
 &= -3\,f_{\sfc}{}_{\alpha}^{\sfa[\sfb_1}\,T^{\sfb_2\sfb_3]\sfc} 
  - f_{\sfc\alpha}{}^\gamma\,T_\gamma^{\sfa\sfc \sfb_1\sfb_2\sfb_3} 
  - \tfrac{1}{2}\,f_{\sfc_1\sfc_2}{}^{\sfa}\,T_\alpha^{\sfc_1\sfc_2 \sfb_1\sfb_2\sfb_3}
 +2\,Z_{\sfc}\,T^{\sfa\sfb_1\sfb_2\sfb_3\sfc}_\alpha \,,
\\
 T^{\sfa}_\alpha\circ T_\beta^{\sfb_1\cdots \sfb_5}
 &= -5\, f_{\sfc}{}_{\alpha}^{\sfa[\sfb_1}\,T_\beta^{\sfb_2\cdots \sfb_5]\sfc}
 - \epsilon_{\alpha\beta}\,f_{\sfc\sfd}{}^{\sfa}\,T^{\sfb_1\cdots \sfb_5 \sfc,\sfd}
 - 2\,f_{\sfc\alpha}{}^\gamma\,\epsilon_{\gamma\beta}\,T^{\sfb_1\cdots \sfb_5[\sfa,\sfc]}
\\
 &\quad +4\,\epsilon_{\alpha\beta}\,Z_{\sfc}\,T^{\sfa\sfb_1\cdots\sfb_5,\sfc}\,,
\\
 T^{\sfa}_\alpha\circ T^{\sfb_1\cdots \sfb_6,\sfb'} &=
 -6\, f_{\sfc}{}_{\alpha}^{\sfa[\sfb_1|}\,T_\beta^{\sfc|\sfb_2\cdots \sfb_6],\sfb'}
 - f_{\sfc}{}_{\alpha}^{\sfa\sfb'}\,T_\beta^{\sfb_1\cdots \sfb_6,\sfc}\,,
\\
 T^{\sfa_1\sfa_2\sfa_3} \circ T_{\sfb} 
 &= -f_{\sfb}{}^{\sfc \sfa_1\sfa_2\sfa_3}\,T_{\sfc} 
 - 6\,\epsilon^{\gamma\delta}\,f_{[\sfb|}{}_{\gamma}^{[\sfa_1\sfa_2}\,\delta_{|\sfc]}^{\sfa_3]}\,T_\delta^{\sfc}
\\
 &\quad + 3\,f_{\sfb\sfc}{}^{[\sfa_1}\, T^{\sfa_2\sfa_3]\sfc}
 + 3\, f_{\sfc_1\sfc_2}{}^{[\sfa_1}\,\delta_{\sfb}^{\sfa_2}\,T^{\sfa_3]\sfc_1\sfc_2}
 +16\,Z_{\sfc}\,\delta_{\sfb}^{[\sfa_1}\,T^{\sfa_2\sfa_3\sfc]} \,,
\\
 T^{\sfa_1\sfa_2\sfa_3} \circ T_\beta^{\sfb} 
 &= -f_{\sfc}{}^{\sfa_1\sfa_2\sfa_3 \sfb}\,T_\beta^{\sfc}
  + 3\, f_{\sfc}{}_{\beta}^{[\sfa_1\sfa_2}\,T^{\sfa_3] \sfb\sfc}
  + \tfrac{3}{2}\, f_{\sfc_1\sfc_2}{}^{[\sfa_1}\,T_\beta^{\sfa_2\sfa_3] \sfb \sfc_1\sfc_2}
 -4\,Z_{\sfc}\,T_\beta^{\sfa_1\sfa_2\sfa_3\sfb\sfc}\,,
\\
 T^{\sfa_1\sfa_2\sfa_3} \circ T^{\sfb_1\sfb_2\sfb_3} 
 &= -3\, f_{\sfc}{}^{\sfa_{123} [\sfb_1}\, T^{\sfb_{23}] \sfc}
   + 3\, \epsilon^{\gamma\delta}\,f_{\sfc}{}_{\gamma}^{[\sfa_{12}}\,T_\delta^{\sfa_3]\sfb_{123} \sfc}
\\
 &\quad + 3\,f_{\sfd_{12}}{}^{[\sfa_1}\, \delta_{\sfc}^{\sfa_2}\, T^{\sfa_3] \sfb_{123} \sfd_{12}, \sfc}
   + 3\,f_{\sfc\sfd}{}^{[\sfa_1}\,T^{\sfa_{23}] \sfb_{123} \sfc, \sfd} 
 +16\,Z_{\sfc}\,T^{\sfb_{123}[\sfa_{123},\sfc]} \,,
\\
 T^{\sfa_1\sfa_2\sfa_3} \circ T_\beta^{\sfb_1\cdots \sfb_5} 
 &= -5\,f_{\sfc}{}^{\sfa_1\sfa_2\sfa_3 [\sfb_1}\,T_\beta^{\sfb_2\cdots \sfb_5] \sfc} 
 + 6\,f_{\sfc}{}_{\beta}^{[\sfa_1\sfa_2}\,\delta^{\sfa_3]\sfc}_{\sfd\sfe}\,T^{\sfb_1\cdots \sfb_5 \sfd,\sfe} \,,
\\
 T^{\sfa_1\sfa_2\sfa_3} \circ T^{\sfb_1\cdots \sfb_6,\sfb'} &=
  -6\,f_{\sfc}{}^{\sfa_1\sfa_2\sfa_3 [\sfb_1|}\,T^{\sfc|\sfb_2\cdots \sfb_6], \sfb'}
  - f_{\sfc}{}^{\sfa_1\sfa_2\sfa_3 \sfb'}\,T^{\sfb_1\cdots \sfb_6, \sfc} \,,
\\
 T_\alpha^{\sfa_1\cdots \sfa_5} \circ T_{\sfb} 
 &= f_{\sfb}{}_{\alpha}^{\sfa_1\cdots \sfa_5 \sfc}\,T_{\sfc}
 - 10\,f_{[\sfb}{}^{[\sfa_1\cdots \sfa_4}\, \delta_{\sfc]}^{\sfa_5]}\,T_\alpha^{\sfc}
 - 30\,f_{\sfc}{}_{\alpha}^{[\sfa_1\sfa_2}\, \delta_{\sfb}^{\sfa_3}\, T^{\sfa_4\sfa_5]\sfc}
\\
 &\quad
 + 10\,f_{\sfb}{}_{\alpha}^{[\sfa_1\sfa_2}\,T^{\sfa_3\sfa_4\sfa_5]}
 + 5\,f_{\sfc\alpha}{}^{\gamma}\, \delta^{[\sfa_1}_{\sfb}\, T_\gamma^{\sfa_2\cdots \sfa_5]\sfc}
 - f_{\sfb\alpha}{}^{\gamma}\,T_\gamma^{\sfa_1\cdots \sfa_5}
\\
 &\quad
 + 5\,f_{\sfb\sfc}{}^{[\sfa_1}\, T_\alpha^{\sfa_2\cdots \sfa_5]\sfc}
 + 10\,f_{\sfc_1\sfc_2}{}^{[\sfa_1}\, \delta^{\sfa_2}_{\sfb}\,T_\alpha^{\sfa_3\sfa_4\sfa_5]\sfc_1\sfc_2}
 +36\,Z_{\sfc}\,\delta_{\sfb}^{[\sfa_1}\,T_\alpha^{\sfa_2\cdots\sfa_5\sfc]}\,,
\\
 T_\alpha^{\sfa_1\cdots \sfa_5} \circ T_\beta^{\sfb} 
 &= -f_{\sfc}{}_{\alpha}^{\sfa_1\cdots \sfa_5\sfb}\,T_\beta^{\sfc}
 -5\,\epsilon_{\alpha\beta}\,f_{\sfc}{}^{[\sfa_1\cdots \sfa_4}\,T^{\sfa_5] \sfb\sfc}
 + 10\,f_{\sfc}{}_{\alpha}^{[\sfa_1\sfa_2}\,T_\beta^{\sfa_3\sfa_4\sfa_5] \sfb\sfc} 
\\
 &\quad + f_{\sfc\alpha}{}^\gamma\,\epsilon_{\gamma\beta}\,T^{\sfc\sfa_1\cdots\sfa_6,\sfb}
 - \tfrac{5}{2}\,\epsilon_{\alpha\beta}\,f_{\sfc_1\sfc_2}{}^{[\sfa_1}\,T^{\sfa_2\cdots \sfa_5]\sfc_1\sfc_2,\sfb}
 -6\,\epsilon_{\alpha\beta}\,Z_{\sfc}\,T^{\sfa_1\cdots\sfa_5\sfc,\sfb}\,,
\\
 T_\alpha^{\sfa_1\cdots \sfa_5} \circ T^{\sfb_1\sfb_2\sfb_3} 
 &=
 -3\,f_{\sfc}{}_{\alpha}^{\sfa_1\cdots \sfa_5 [\sfb_1}\, T^{\sfb_2\sfb_3]\sfc}
 + 5\,f_{\sfc}{}^{[\sfa_1\cdots \sfa_4}\, T_\alpha^{\sfa_5] \sfb_1\sfb_2\sfb_3\sfc} 
\\
 &\quad - 40\,f_{\sfc}{}_{\alpha}^{[\sfa_1\sfa_2}\,\delta^{\sfa_3\sfa_4\sfa_5]\sfc}_{\sfd_1\sfd_2\sfd_3\sfe}\,T^{\sfb_1\sfb_2\sfb_3\sfd_1\sfd_2\sfd_3,\sfe}\,,
\\
 T_\alpha^{\sfa_1\cdots \sfa_5} \circ T_\beta^{\sfb_1\cdots \sfb_5} 
 &= -5\,f_{\sfc}{}_{\alpha}^{\sfa_1\cdots \sfa_5 [\sfb_1}\, T_\beta^{\sfb_2\cdots \sfb_5]\sfc}
 -10\,\epsilon_{\alpha\beta}\,f_{\sfc}{}^{[\sfa_1\cdots \sfa_4}\,\delta^{\sfa_5]\sfc}_{\sfd\sfe}\,T^{\sfb_1\cdots \sfb_5\sfd,\sfe} \,,
\\
 T^{\sfa_1\cdots \sfa_5} \circ T^{\sfb_1\cdots \sfb_6,\sfb'} &= -6\,f_{\sfc}{}_{\alpha}^{\sfa_1\cdots \sfa_5 [\sfb_1|}\, T^{\sfc|\sfb_2\cdots \sfb_6],\sfb'}
 - f_{\sfc}{}_{\alpha}^{\sfa_1\cdots \sfa_5 \sfb'}\, T^{\sfb_1\cdots \sfb_6,\sfc} \,,
\\
 T^{\sfa_1\cdots \sfa_6,\sfa'} \circ T_{\sfb} &= 
 -2\,\epsilon^{\gamma\delta}\,f_{\sfc}{}_{\gamma}^{\sfa_1\cdots\sfa_6}\,\delta^{[\sfa'}_{\sfb}\,T^{\sfc]}_\delta
 - 20\,f_{\sfb}{}^{\sfa'[\sfa_1\sfa_2\sfa_3}\,T^{\sfa_4\sfa_5\sfa_6]}
 \rlap{${}\displaystyle + 60\,f_{\sfc}{}^{\sfa'[\sfa_1\sfa_2\sfa_3}\,\delta_{\sfb}^{\sfa_4}\,T^{\sfa_5\sfa_6]\sfc}$}
\\
 &\quad +6\,\epsilon^{\gamma\delta}\,f_{\sfb}{}_{\gamma}^{\sfa'[\sfa_1}\,T^{\sfa_2\cdots\sfa_6]}_\delta
 -30\,\epsilon^{\gamma\delta}\,f_{\sfc}{}_{\gamma}^{\sfa'[\sfa_1}\,\delta_{\sfb}^{\sfa_2}\,T^{\sfa_3\cdots\sfa_6]\sfc}_\delta \,,
\\
 T^{\sfa_1\cdots \sfa_6,\sfa'} \circ T_\beta^{\sfb} &= -f_{\sfc}{}_{\beta}^{\sfa_1\cdots \sfa_6}\,T^{\sfa'\sfb\sfc} 
 - 20\,f_{\sfc}{}^{\sfa'[\sfa_1\sfa_2\sfa_3}\,T_\beta^{\sfa_4\sfa_5\sfa_6]\sfb\sfc}
 -6\,f_{\sfc}{}_{\beta}^{\sfa'[\sfa_1}\,T^{\sfa_2\cdots\sfa_6]\sfc,\sfb} \,,
\\
 T^{\sfa_1\cdots \sfa_6,\sfa'} \circ T^{\sfb_1\sfb_2\sfb_3} &= 
 -\epsilon^{\gamma\delta}\,f_{\sfc}{}_{\gamma}^{\sfa_1\cdots\sfa_6}\,T_\delta^{\sfa'\sfb_1\sfb_2\sfb_3\sfc} 
 + 80\,f_{\sfc}{}^{\sfa'[\sfa_1\sfa_2\sfa_3}\,\delta^{\sfa_4\sfa_5\sfa_6]\sfc}_{\sfd_1\sfd_2\sfd_3\sfe}\,T^{\sfb_1\sfb_2\sfb_3\sfd_1\sfd_2\sfd_3,\sfe} \,,
\\
 T^{\sfa_1\cdots \sfa_6,\sfa'} \circ T_\beta^{\sfb_1\cdots \sfb_5} &= -2\,f_{\sfc}{}_{\beta}^{\sfa_1\cdots\sfa_6}\,T^{\sfb_1\cdots\sfb_5[\sfa',\sfc]} \,,
\\
 T^{\sfa_1\cdots \sfa_6,\sfa'} \circ T^{\sfb_1\cdots \sfb_6,\sfb'} &= 0 \,.
\end{split}
\end{align}}
This is complicated and it may be also useful to consider the reduction to $n\leq 6$:
\begin{align}
\begin{split}
 T_{\sfa}\circ T_{\sfb} &=f_{\sfa\sfb}{}^{\sfc}\,T_{\sfc}\,,
\\
 T_{\sfa}\circ T^{\sfb}_\beta &= f_{\sfa}{}_{\beta}^{\sfc\sfb}\,T_{\sfc}
 + f_{\sfa\beta}{}^{\gamma}\,T_\gamma^{\sfb} - f_{\sfa\sfc}{}^{\sfb}\,T_\beta^{\sfc} 
 +2\,Z_{\sfa}\,T^{\sfb}_\beta \,,
\\
 T_{\sfa}\circ T^{\sfb_1\sfb_2\sfb_3} &=
 f_{\sfa}{}^{\sfc\sfb_1\sfb_2\sfb_3}\, T_{\sfc} + 3\,\epsilon^{\gamma\delta}\,f_{\sfa}{}_{\gamma}^{[\sfb_1\sfb_2}\, T_{\delta}^{\sfb_3]} - 3\,f_{\sfa\sfc}{}^{[\sfb_1}\, T^{\sfb_2\sfb_3]\sfc} 
 +4\,Z_{\sfa}\,T^{\sfb_1\sfb_2\sfb_3}\,,
\\
 T_{\sfa}\circ T_\beta^{\sfb_1\cdots \sfb_5} &=
 5\, f_{\sfa}{}^{[\sfb_1\cdots \sfb_4}\,T_\beta^{\sfb_5]}
 - 10\, f_{\sfa}{}_{\beta}^{[\sfb_1\sfb_2}\,T^{\sfb_3\sfb_4\sfb_5]} 
\\
 &\quad 
 + f_{\sfa\beta}{}^{\gamma}\, T_\gamma^{\sfb_1\cdots \sfb_5} 
 - 5\,f_{\sfa\sfc}{}^{[\sfb_1}\,T_\beta^{\sfb_2\cdots \sfb_5]\sfc} 
 + 6\,Z_{\sfa}\,T^{\sfb_1\cdots\sfb_5}_\beta\,,
\\
 T^{\sfa}_\alpha\circ T_{\sfb} &= 
 f_{\sfb}{}_{\alpha}^{\sfa\sfc} \, T_{\sfc} 
 + 2\,\delta^{\sfa}_{[\sfb}\,f_{\sfc]\alpha}{}^{\gamma}\, T_\gamma^{\sfc}
 + f_{\sfb\sfc}{}^{\sfa}\,T_\alpha^{\sfc} 
 +4\,Z_{\sfc}\,\delta^{[\sfa}_{\sfb}\,T^{\sfc]}_\alpha\,,
\\
 T^{\sfa}_\alpha\circ T^{\sfb}_\beta &= - f_{\sfc}{}_{\alpha}^{\sfa\sfb}\,T_\beta^{\sfc} 
 - f_{\sfc\alpha}{}^\gamma\,\epsilon_{\gamma\beta}\,T^{\sfc \sfa\sfb} 
 + \tfrac{1}{2}\,\epsilon_{\alpha\beta}\,f_{\sfc_1\sfc_2}{}^{\sfa}\,T^{\sfc_1\sfc_2\sfb}
 -2\,\epsilon_{\alpha\beta}\,Z_{\sfc}\,T^{\sfa\sfb\sfc} \,,
\\
 T^{\sfa}_\alpha\circ T^{\sfb_1\sfb_2\sfb_3}
 &= -3\,f_{\sfc}{}_{\alpha}^{\sfa[\sfb_1}\,T^{\sfb_2\sfb_3]\sfc} 
  - f_{\sfc\alpha}{}^\gamma\,T_\gamma^{\sfa\sfc \sfb_1\sfb_2\sfb_3} 
  - \tfrac{1}{2}\,f_{\sfc_1\sfc_2}{}^{\sfa}\,T_\alpha^{\sfc_1\sfc_2 \sfb_1\sfb_2\sfb_3}
\\
 &\quad +2\,Z_{\sfc}\,T^{\sfa\sfb_1\sfb_2\sfb_3\sfc}_\alpha \,,
\\
 T^{\sfa}_\alpha\circ T_\beta^{\sfb_1\cdots \sfb_5}
 &= -5\, f_{\sfc}{}_{\alpha}^{\sfa[\sfb_1}\,T_\beta^{\sfb_2\cdots \sfb_5]\sfc} \,,
\\
 T^{\sfa_1\sfa_2\sfa_3} \circ T_{\sfb} 
 &= -f_{\sfb}{}^{\sfc \sfa_1\sfa_2\sfa_3}\,T_{\sfc} 
 - 6\,\epsilon^{\gamma\delta}\,f_{[\sfb|}{}_{\gamma}^{[\sfa_1\sfa_2}\,\delta_{|\sfc]}^{\sfa_3]}\,T_\delta^{\sfc}
\\
 &\quad + 3\,f_{\sfb\sfc}{}^{[\sfa_1}\, T^{\sfa_2\sfa_3]\sfc}
 + 3\, f_{\sfc_1\sfc_2}{}^{[\sfa_1}\,\delta_{\sfb}^{\sfa_2}\,T^{\sfa_3]\sfc_1\sfc_2}
 +16\,Z_{\sfc}\,\delta_{\sfb}^{[\sfa_1}\,T^{\sfa_2\sfa_3\sfc]} \,,
\\
 T^{\sfa_1\sfa_2\sfa_3} \circ T_\beta^{\sfb} 
 &= -f_{\sfc}{}^{\sfa_1\sfa_2\sfa_3 \sfb}\,T_\beta^{\sfc}
  + 3\, f_{\sfc}{}_{\beta}^{[\sfa_1\sfa_2}\,T^{\sfa_3] \sfb\sfc}
  + \tfrac{3}{2}\, f_{\sfc_1\sfc_2}{}^{[\sfa_1}\,T_\beta^{\sfa_2\sfa_3] \sfb \sfc_1\sfc_2}
\\
 &\quad -4\,Z_{\sfc}\,T_\beta^{\sfa_1\sfa_2\sfa_3\sfb\sfc}\,,
\\
 T^{\sfa_1\sfa_2\sfa_3} \circ T^{\sfb_1\sfb_2\sfb_3} 
 &= -3\, f_{\sfc}{}^{\sfa_1\sfa_2\sfa_3 [\sfb_1}\, T^{\sfb_2\sfb_3] \sfc}
   + 3\, \epsilon^{\gamma\delta}\,f_{\sfc}{}_{\gamma}^{[\sfa_1\sfa_2}\,T_\delta^{\sfa_3]\sfb_1\sfb_2\sfb_3 \sfc} \,,
\\
 T^{\sfa_1\sfa_2\sfa_3} \circ T_\beta^{\sfb_1\cdots \sfb_5} 
 &= -5\,f_{\sfc}{}^{\sfa_1\sfa_2\sfa_3 [\sfb_1}\,T_\beta^{\sfb_2\cdots \sfb_5] \sfc} \,,
\\
 T_\alpha^{\sfa_1\cdots \sfa_5} \circ T_{\sfb} 
 &= - 10\,f_{[\sfb}{}^{[\sfa_1\cdots \sfa_4}\, \delta_{\sfc]}^{\sfa_5]}\,T_\alpha^{\sfc}
 - 30\,f_{\sfc}{}_{\alpha}^{[\sfa_1\sfa_2}\, \delta_{\sfb}^{\sfa_3}\, T^{\sfa_4\sfa_5]\sfc}
\\
 &\quad
 + 10\,f_{\sfb}{}_{\alpha}^{[\sfa_1\sfa_2}\,T^{\sfa_3\sfa_4\sfa_5]}
 + 5\,f_{\sfc\alpha}{}^{\gamma}\, \delta^{[\sfa_1}_{\sfb}\, T_\gamma^{\sfa_2\cdots \sfa_5]\sfc}
 - f_{\sfb\alpha}{}^{\gamma}\,T_\gamma^{\sfa_1\cdots \sfa_5}
\\
 &\quad
 + 5\,f_{\sfb\sfc}{}^{[\sfa_1}\, T_\alpha^{\sfa_2\cdots \sfa_5]\sfc}
 + 10\,f_{\sfc_1\sfc_2}{}^{[\sfa_1}\, \delta^{\sfa_2}_{\sfb}\,T_\alpha^{\sfa_3\sfa_4\sfa_5]\sfc_1\sfc_2}
 +36\,Z_{\sfc}\,\delta_{\sfb}^{[\sfa_1}\,T_\alpha^{\sfa_2\cdots\sfa_5\sfc]}\,,
\\
 T_\alpha^{\sfa_1\cdots \sfa_5} \circ T_\beta^{\sfb} 
 &= -5\,\epsilon_{\alpha\beta}\,f_{\sfc}{}^{[\sfa_1\cdots \sfa_4}\,T^{\sfa_5] \sfb\sfc}
 + 10\,f_{\sfc}{}_{\alpha}^{[\sfa_1\sfa_2}\,T_\beta^{\sfa_3\sfa_4\sfa_5] \sfb\sfc} \,,
\\
 T_\alpha^{\sfa_1\cdots \sfa_5} \circ T^{\sfb_1\sfb_2\sfb_3} 
 &= 5\,f_{\sfc}{}^{[\sfa_1\cdots \sfa_4}\, T_\alpha^{\sfa_5] \sfb_1\sfb_2\sfb_3\sfc} \,,
\\
 T_\alpha^{\sfa_1\cdots \sfa_5} \circ T_\beta^{\sfb_1\cdots \sfb_5} &= 0 \,.
\end{split}
\end{align}
We can obtain the lower-dimensional EDA in a similar manner by a truncation. 

In the type IIB picture, the level-0 part of the embedding tensor is
\begin{align}
 \widehat{\Theta}_{\sfa}^{(0)} = f_{\sfa\sfb}{}^{\sfc}\,K^{\sfb}{}_{\sfc} + f_{\sfa\beta}{}^\gamma\,R^\beta{}_\gamma -Z_a\,K\,,
\end{align}
and the Leibniz identity \eqref{eq:cocycle-0} requires
\begin{align}
 f_{[\sfa\sfb}{}^{\sfd}\,f_{\sfc]\sfd}{}^{\sfe} = 0\,,\qquad 
 f_{\sfa\gamma}{}^\beta\,f_{\sfb\beta}{}^\delta - f_{\sfb\gamma}{}^\beta\,f_{\sfa\beta}{}^\delta
 + f_{\sfa\sfb}{}^{\sfc}\,f_{\sfc\gamma}{}^\delta = 0\,,\qquad f_{\sfa\sfb}{}^{\sfc}\,Z_{\sfc} =0\,.
\end{align}
Then, defining
\begin{align}
\begin{alignedat}{2}
 f^2(x) &\equiv \tfrac{1}{2!}\,x^{\sfa}\,f_{\sfa}{}^{\sfb_1\sfb_2}_\beta\,R_{\sfb_1\sfb_2}^\beta\,,&\qquad
 f^4(x) &\equiv \tfrac{1}{4!}\,x^{\sfa}\,f_{\sfa}{}^{\sfb_1\cdots\sfb_2}\,R_{\sfb_1\cdots\sfb_4}\,, 
\\
 f^6(x) &\equiv \tfrac{1}{6!}\,x^{\sfa}\,f_{\sfa}{}^{\sfb_1\cdots \sfb_6}_\beta\,R_{\sfb_1\cdots \sfb_6}^\beta\,,&\qquad
 f^{7,1}(x) &\equiv \tfrac{1}{7!}\,x^{\sfa}\,f_{\sfa}{}^{\sfb_1\cdots\sfb_7,\sfb'}\,R_{\sfb_1\cdots\sfb_7,\sfb'}\,,
\end{alignedat}
\end{align}
we can express the cocycle conditions as
\begin{align}
 0&=\rmd f^2(x) \,,
\\
 0&=\rmd f^4(x) -[f^2(x),\,f^2(y)] = 0\,,
\\
 0&=\rmd f^6(x) -[f^2(x),\,f^4(y)] - [f^4(x),\,f^2(y)] \,,
\\
 0&=\rmd f^{7,1}(x) -[f^2(x),\,f^6(y)] - [f^4(x),\,f^4(y)] - [f^6(x),\,f^2(y)] \,,
\end{align}
where
\begin{align}
 \rmd f^{*}(x,y) \equiv x\cdot f^{*}(y) - y\cdot f^{*}(x) - f^{*}([x,y]) \,,
\qquad
 x\cdot f^{*}(y) \equiv x^{\sfa}\, \bigl[f^{*}(y),\,\widehat{\Theta}_{\sfa}^{(0)}\bigr]\,.
\end{align}
More explicitly, we find
\begin{align}
 0&= 4\,f_{[\sfa|\sfd}{}^{[\sfc_1|}\,f_{|\sfb]}{}^{\sfd|\sfc_2]}_\gamma - f_{\sfa\sfb}{}^{\sfd}\,f_{\sfd}{}^{\sfc_1\sfc_2}_\gamma -2\,f_{[\sfa|\gamma}{}^\delta\,f_{|\sfb]}{}^{\sfc_1\sfc_2}_\delta 
 -4\,Z_{[\sfa}\,f_{\sfb]}{}^{\sfc_1\sfc_2}_\gamma\,,
\\
 0&= 8\,f_{[\sfa|\sfd}{}^{[\sfc_1|}\,f_{|\sfb]}{}^{\sfd|\sfc_2\sfc_3\sfc_4]} - f_{\sfa\sfb}{}^{\sfd}\,f_{\sfd}{}^{\sfc_1\cdots\sfc_4} - 6\,\epsilon^{\gamma\delta}\,f_{\sfa}{}^{[\sfc_1\sfc_2}_\gamma\,f_{\sfb}{}^{\sfc_3\sfc_4]}_\delta 
 -8\,Z_{[\sfa}\,f_{\sfb]}{}^{\sfc_1\cdots\sfc_4} \,,
\\
\begin{split}
 0&= 12\,f_{[\sfa|\sfd}{}^{[\sfc_1|}\,f_{|\sfb]}{}^{\sfd|\sfc_2\cdots\sfc_6]}_\gamma - f_{\sfa\sfb}{}^{\sfd}\,f_{\sfd}{}^{\sfc_1\cdots\sfc_6}_\gamma -2\,f_{[\sfa|\gamma}{}^\delta\,f_{|\sfb]}{}^{\sfc_1\cdots\sfc_6}_\delta
\\
 &\quad +30\,f_{[\sfa|}{}^{[\sfc_1\sfc_2}_\alpha \,f_{|\sfb]}{}^{\sfc_3\cdots\sfc_6]}
 - 12\,Z_{[\sfa}\,f_{\sfb]}{}^{\sfc_1\cdots\sfc_6}_\gamma \,,
\end{split}
\\
\begin{split}
 0&= 14\,f_{[\sfa|\sfd}{}^{[\sfc_1|}\,f_{|\sfb]}{}^{\sfd|\sfc_2\cdots \sfc_7],\sfc'} + 2\,f_{\sfd[\sfa}{}^{\sfc'}\,f_{\sfb]}{}^{\sfc_1\cdots \sfc_7,\sfd}
  - f_{\sfa\sfb}{}^{\sfd}\,f_{\sfd}{}^{\sfc_1\cdots\sfc_7,\sfc'} 
\\
 &\quad - 42\,\epsilon^{\alpha\beta}\,f_{[\sfa|}{}^{[\sfc_1\sfc_2}_\alpha\,f_{|\sfb]}{}^{\sfc_3\cdots\sfc_7]\sfc'}_\beta 
 + 35\,f_{[\sfa}{}^{[\sfc_1\cdots\sfc_4}\,f_{\sfb]}{}^{\sfc_5\sfc_6\sfc_7]\sfc'} 
 - 16\,Z_{[\sfa}\,f_{\sfb]}{}^{\sfc_1\cdots \sfc_7,\sfc'}\,.
\end{split}
\end{align}
By introducing
\begin{align}
\begin{alignedat}{2}
 \bm{\rr}^2 &\equiv \tfrac{1}{2!}\,\rr^{a_1a_2}_{\alpha}\,R^\alpha_{a_1a_2}\,,&\qquad
 \bm{\rr}^4 &\equiv \tfrac{1}{4!}\,\rr^{a_1\cdots a_4}\,R_{a_1\cdots a_4}\,,
\\
 \bm{\rr}^6 &\equiv \tfrac{1}{6!}\,\rr_\alpha^{a_1\cdots a_6}\,R^\alpha_{a_1\cdots a_6}\,,&\qquad
 \bm{\rr}^{7,1} &\equiv \tfrac{1}{7!}\,\rr^{a_1\cdots a_7,a'}\,R_{a_1\cdots a_7,a'}\,,
\end{alignedat}
\end{align}
and defining $\cR\equiv \Exp{\bm{\rr}^2}\Exp{\bm{\rr}^4}\Exp{\bm{\rr}^6}\Exp{\bm{\rr}^{7,1}}$, the coboundary ansatzes are decomposed as
\begin{align}
 \mathfrak{f}^2(x) &= \rmd \bm{\rr}^2(x)\,,
\\
 \mathfrak{f}^4(x) &= \rmd \bm{\rr}^4(x)+\tfrac{1}{2}\,[\bm{\rr}^2,\,f^2(x)]\,,
\\
 \mathfrak{f}^6(x) &= \rmd \bm{\rr}^6(x) + [\bm{\rr}^2,\,f^4(x)]-\tfrac{1}{3}\,[\bm{\rr}^2,\,[\bm{\rr}^2,\,f^2(x)]]\,,
\\
\begin{split}
 \mathfrak{f}^{7,1}(x) &= \rmd \bm{\rr}^{7,1}(x) + [\bm{\rr}^2,\,f^6(x)]
 +\tfrac{1}{2}\,[\bm{\rr}^4,\,f^4(x)]
 -\tfrac{1}{2}\,[\bm{\rr}^2,\,[\bm{\rr}^2,\,f^4(x)]]
\\
 &\quad -\tfrac{1}{4}\,[\bm{\rr}^4,\,[\bm{\rr}^2,\,f^2(x)]]
 +\tfrac{1}{8}\,[\bm{\rr}^2,\,[\bm{\rr}^2,\,[\bm{\rr}^2,\,f^2(x)]]]\,,
\end{split}
\end{align}
where $\rmd \bm{\rr}^{*}(x) \equiv x^{\sfa}\, [\bm{\rr}^{*},\,\widehat{\Theta}_{\sfa}^{(0)}]$\,.
In components, we have
\begin{align}
 \mathfrak{f}_{\sfa}{}^{\sfb_1\sfb_2}_\beta &= 2\,f_{\sfa\sfc}{}^{[\sfb_1|}\,\rr^{\sfc|\sfb_2]}_\beta - f_{\sfa\beta}{}^\gamma\,\rr_{\gamma}^{\sfb_1\sfb_2} -2\,Z_{\sfa}\,\rr^{\sfb_1\sfb_2}_\beta\,,
\\
 \mathfrak{f}_{\sfa}{}^{\sfb_1\cdots \sfb_4} &= 4\,f_{\sfa\sfc}{}^{[\sfb_1|}\,\rr^{\sfc|\sfb_2\sfb_2\sfb_4]} -4\,Z_{\sfa}\,\rr^{\sfb_1\cdots\sfb_4}
 -3\,\epsilon^{\alpha\beta}\,f_{\sfa}{}^{[\sfb_1\sfb_2}_{\alpha}\,\rr^{\sfb_3\sfb_4]}_\beta \,,
\\
\begin{split}
 \mathfrak{f}_{\sfa}{}^{\sfb_1\cdots\sfb_6}_\beta &= 6\,f_{\sfa\sfc}{}^{[\sfb_1|}\,\rr^{\sfc|\sfb_2\cdots \sfb_6]}_\beta - f_{\sfa\beta}{}^\gamma\,\rr_{\gamma}^{\sfb_1\cdots\sfb_6} -6\,Z_{\sfa}\,\rr^{\sfb_1\cdots\sfb_6}_\beta
\\
 &\quad -15\,f_{\sfa}{}^{[\sfb_1\cdots\sfb_4}\,\rr^{\sfb_5\sfb_6]}_\beta
 -30\,\epsilon^{\gamma\delta}\,f_{\sfa}{}^{[\sfb_1\sfb_2}_{\gamma}\,\rr_\delta^{\sfb_3\sfb_4}\,\rr^{\sfb_5\sfb_6]}_\alpha\,,
\end{split}
\\
\begin{split}
 \mathfrak{f}_{\sfa}{}^{\sfb_1\cdots\sfb_7,\sfb'} &=7\,f_{\sfa\sfc}{}^{[\sfb_1|}\,\rr^{\sfc|\sfb_2\cdots \sfb_7],\sfb'} + f_{\sfa\sfc}{}^{\sfb'}\,\rr^{\sfb_1\cdots\sfb_7,\sfc} -8\,Z_{\sfa}\,\rr^{\sfb_1\cdots\sfb_7,\sfb'}
 +21\,\epsilon^{\gamma\delta}\,\rr^{[\sfb_1\sfb_2}_\gamma\,f_{\sfa}{}^{\sfb_3\cdots\sfb_7]\sfb'}_\delta
\\
 &\quad -\tfrac{35}{2}\,\rr^{[\sfb_1\cdots\sfb_4}\,f_{\sfa}{}^{\sfb_5\sfb_6\sfb_7]\sfb'}
 +\tfrac{105}{2}\,\epsilon^{\alpha\beta}\,\rr^{\sfb'[\sfb_1}\,f_{\sfa}{}^{\sfb_2\cdots \sfb_5}\,\rr_\beta^{\sfb_6\sfb_7]}
\\
 &\quad +\tfrac{105}{4}\,\epsilon^{\gamma\delta}\,\rr^{[\sfb_1\cdots\sfb_4}\,\bigl(\rr_\gamma^{\sfb_5\sfb_6}\,f_{\sfa}{}_\delta^{\sfb_7]\sfb'} + \rr_\gamma^{\sfb_5|\sfb'|}\,f_{\sfa}{}_\delta^{\sfb_6\sfb_7]}\bigr)
\\
 &\quad + \tfrac{315}{4}\,\epsilon^{\alpha\beta}\,\epsilon^{\gamma\delta}\,\rr^{[\sfb_1\sfb_2}_\alpha\,\rr^{|\sfb'|\sfb_3}_\beta\,\rr^{\sfb_4\sfb_5}_\gamma\,f_{\sfa}{}_\delta^{\sfb_6\sfb_7]}\,.
\end{split}
\end{align}

For the fundamental identities, assuming $n\leq 7$ for simplicity, we obtain
\begin{align}
 &f_{\sfc}{}^{\sfd\sfa}_\alpha\,f_{\sfd}{}^{\sfb_1\sfb_2}_\beta - 2\,f_{\sfd}{}^{\sfa[\sfb_1}_\alpha\,f_{\sfc}{}^{\sfb_2]\sfd}_\beta
 = f_{\sfd\alpha}{}^\gamma\,\epsilon_{\gamma\beta}\,f_{\sfc}{}^{\sfa\sfb_{12}\sfd}
 +\tfrac{1}{2}\,\epsilon_{\alpha\beta}\,f_{\sfd_{12}}{}^{\sfa}\,f_{\sfc}{}^{\sfb_{12}\sfd_{12}}
 + 2\,\epsilon_{\alpha\beta}\,f_{\sfc}{}^{\sfa\sfb_{12}\sfd}\,Z_{\sfd}\,,
\\
 &f_{\sfc}{}^{\sfd\sfa_{123}}\,f_{\sfe}{}^{\sfb_{1\cdots 4}} - 4\,f_{\sfd}{}^{\sfa_{123}[\sfb_1}\,f_{\sfc}{}^{\sfb_{234}]\sfd}
  = -3\,\epsilon^{\gamma\delta}\,f_{\sfd}{}_\gamma^{[\sfa_{12}}\,f_{\sfc}{}_\delta^{\sfa_3]\sfb_{1\cdots 4}\sfd}
 - 3\,f_{\sfd_{12}}{}^{[\sfa_1}\,\delta^{\sfa_2}_{\sfe}\,f_{\sfc}{}^{\sfa_3]\sfb_{1\cdots 4}\sfd_{12},\sfe}
\nn\\
 &\quad +3\,f_{\sfd_{12}}{}^{[\sfa_1}\,f_{\sfc}{}^{\sfa_2\sfa_3]\sfb_{1\cdots 4}[\sfd_1,\sfd_2]} 
 +16\,Z_{\sfd}\,f_{\sfc}{}^{\sfb_{1\cdots 4}[\sfa_{123},\sfd]}\,,
\\
 &f_{\sfc}{}_\alpha^{\sfd\sfa_{1\cdots 5}}\,f_{\sfe}{}^{\sfb_{1\cdots 6}}_\beta - 6\,f_{\sfd}{}_\alpha^{\sfa_{1\cdots 5}[\sfb_1}\,f_{\sfc}{}_\beta^{\sfb_{2\cdots 6}]\sfd}
 = -10\,\epsilon_{\alpha\beta}\,f_{\sfd}{}^{[\sfa_{1\cdots 4}}\,\delta^{\sfa_5]\sfd}_{\sfe_1\sfe_2}\,f_{\sfc}{}^{\sfb_{1\cdots 6}[\sfe_1,\sfe_2]}\,.
\end{align}
From $Z_{AB}{}^C\,X_C=0$, we find additional identities, such as
\begin{align}
 Z^{\sfa}_\alpha{}_{\sfb}{}^C\,X_C = \delta_{\sfa}^{\sfb}\,\bigl(f_{\sfd}{}_\gamma^{\sfe\sfc}\,f_{\sfc\alpha}{}^\gamma + f_{\sfd}{}_{\alpha}^{\sfe\sfc}\,Z_{\sfc}\bigr)\,K^{\sfd}{}_{\sfe}=0\quad\Leftrightarrow\quad
 f_{\sfa}{}_\gamma^{\sfb\sfc}\,f_{\sfc\alpha}{}^\gamma + f_{\sfa}{}_{\alpha}^{\sfb\sfc}\,Z_{\sfc}=0\,.
\end{align}
Again, they are not the whole set of the Leibniz identities. 

Now, let us construct the generalized frame fields. 
Using the level-0 generators, we define
\begin{align}
 \mathbb{E}_A{}^I \equiv \bigl(\Exp{-h_\alpha{}^{\beta}\,R^\alpha{}_\beta} \Exp{-(h_R)_{\sfa}{}^{\sfb}\,\tilde{K}^{\sfa}{}_{\sfb}}\bigr)_A{}^I\,.
\end{align}
Here, $h_R$ is related to the right-invariant vector fields as $e_{\sfa}^m=(\Exp{-(h_R)})_{\sfa}{}^m$ and we also define $\lambda_{\alpha}{}^{\dot{\beta}}\equiv (\Exp{-h})_{\alpha}{}^{\dot{\beta}}$\,. 
We further multiply $\Exp{(\tilde{K}+t_0)\Delta}$ and 
\begin{align}
 \bm{\Pi}\equiv \Exp{-\frac{1}{2!}\,\pi^{\sfa_1\sfa_2}_\alpha\,R_{\sfa_1\sfa_2}^\alpha}
 \Exp{-\frac{1}{4!}\,\pi^{\sfa_1\cdots \sfa_4}\,R_{\sfa_1\cdots \sfa_4}}
 \Exp{-\frac{1}{6!}\,\pi^{\sfa_1\cdots \sfa_6}_\alpha\,R_{\sfa_1\cdots \sfa_6}^\alpha}
 \Exp{-\frac{1}{7!}\,\pi^{\sfa_1\cdots \sfa_7,\sfa'}\,R_{\sfa_1\cdots \sfa_7,\sfa'}},
\end{align}
to obtain the generalized frame fields $E_A{}^I$\,. 
For example, when $n\leq 7$\,, we find
\begin{align}
 E_{\sfa} &= e_{\sfa}\,, \qquad
 E_\alpha^\sfa = \pi_\alpha^{\sfa\sfb}\,e_{\sfb} + \Exp{-2\Delta}\lambda_\alpha{}^{\dot{\alpha}}\,r^\sfa \,,
\\
 E^{\sfa_1\sfa_2\sfa_3} &= \bigl(\pi^{\sfa_1\sfa_2\sfa_3\sfb} + \tfrac{3}{2}\,\epsilon^{\gamma\delta}\,\pi_\gamma^{[\sfa_1|\sfb|}\,\pi_\delta^{\sfa_2\sfa_3]}\bigr)\,e_{\sfb}
 + 3\Exp{-2\Delta}\epsilon^{\gamma\delta}\,\lambda_\gamma{}^{\dot{\alpha}}\,\pi_\delta^{[\sfa_1\sfa_2}\,r^{\sfa_3]}
 + \Exp{-4\Delta}r^{\sfa_1\sfa_2\sfa_3} \,,
\\
\begin{split}
 E_\alpha^{\sfa_1\cdots\sfa_5} &=
 \bigl(\pi_\alpha^{\sfa_1\cdots\sfa_5 \sfb} - 10\,\pi^{\sfb [\sfa_1\sfa_2\sfa_3}\, \pi_\alpha^{\sfa_4\sfa_5]} 
  - 5\,\epsilon^{\gamma\delta}\,\pi_\gamma^{\sfb[\sfa_1}\,\pi_\delta^{\sfa_2\sfa_3}\,\pi_\alpha^{\sfa_4\sfa_5]} \bigr)\,e_{\sfb} 
\\
 &\quad -5\Exp{-2\Delta}\lambda_\alpha{}^{\dot{\alpha}}\,\pi^{[\sfa_1\cdots\sfa_4}\,r^{\sfa_5]} 
 + 15 \Exp{-2\Delta}\epsilon^{\gamma\delta}\,\lambda_\gamma{}^{\dot{\alpha}}\,\pi_\delta^{[\sfa_1\sfa_2}\,\pi_\alpha^{\sfa_3\sfa_4}\, r^{\sfa_5]} 
\\
 &\quad 
  + 10\Exp{-4\Delta}\pi_\alpha^{[\sfa_1\sfa_2}\,r^{\sfa_3\sfa_4\sfa_5]} 
  + \Exp{-6\Delta}\lambda_\alpha{}^{\dot{\alpha}}\,r^{\sfa_1\cdots \sfa_5}\,,
\end{split}
\\
\begin{split}
 E^{\sfa_{1\cdots 6},\sfa'}&=\bigl(6\,\epsilon^{\gamma\delta}\,\pi_\gamma^{\sfb[\sfa_1\cdots\sfa_5}\,\pi_\delta^{\sfa_6]\sfa'}
 +30\,\epsilon^{\gamma\delta}\,\pi_\gamma^{\sfa'[\sfa_1}\,\pi_\delta^{\sfa_2\sfa_3}\,\pi^{\sfa_4\sfa_5\sfa_6]\sfb}
 +10\,\pi^{\sfb [\sfa_1\sfa_2\sfa_3}\, \pi^{\sfa_4\sfa_5\sfa_6]\sfa'}
\\
 &\quad\ 
 -\tfrac{15}{2}\, \epsilon^{\alpha\beta}\,\epsilon^{\gamma\delta}\,\pi_\alpha^{\sfa'[\sfa_1}\,\pi_\beta^{\sfa_2\sfa_3}\,\pi_\gamma^{\sfa_4\sfa_5}\, \pi_\delta^{\sfa_6]\sfb}\bigr)\,e_{\sfb} 
\\
 &\quad 
  + \Exp{-2\Delta}\lambda_{\beta}{}^{\dot{\alpha}}\,\bigl[- \epsilon^{\beta\gamma}\,\pi_{\gamma}^{\sfa_{1\cdots 6}}\,r^{\sfa'}
  + 30\,\bigl(\epsilon^{\beta\gamma}\,\pi^{[\sfa_{1\cdots 4}} 
  + \epsilon^{\gamma\delta}\,\epsilon^{\beta\epsilon}\, \pi_\delta^{[\sfa_{12}}\,\pi_\epsilon^{\sfa_{34}}\bigr)\,\pi_{\gamma}^{|\sfa'|\sfa_5}\,r^{\sfa_6]}\bigr] 
\\
 &\quad 
  -20\Exp{-4\Delta}\pi^{\sfa'[\sfa_{123}}\, r^{\sfa_{456}]}
  +30\Exp{-4\Delta}\epsilon^{\gamma\delta}\,\pi_\gamma^{\sfa'[\sfa_1}\, \pi_\delta^{\sfa_{23}} \, r^{\sfa_{456}]}
\\
 &\quad 
 + 6\Exp{-6\Delta}\lambda_{\gamma}{}^{\dot{\alpha}}\,\epsilon^{\gamma\delta}\,\pi_\delta^{\sfa'[\sfa_1}\,r^{\sfa_{2 \cdots 6}]} 
 +\Exp{-8\Delta}r^{\sfa_{1 \cdots 6}} \otimes r^{\sfa'} \,,
\end{split}
\end{align}
where $r^{\sfa_1\cdots \sfa_p}\equiv r^{\sfa_1}\wedge\cdots\wedge r^{\sfa_p}$\,. 

In the general case, the differential identity gives
\begin{align}
 D_{\sfa}\bm{\pi}^2 &= f_{\sfa}{}^2 + [\bm{\pi}^2,\,\widehat{\Theta}^{(0)}] \,,
\\
 D_{\sfa}\bm{\pi}^4 &= f_{\sfa}{}^4 + [\bm{\pi}^4,\,\widehat{\Theta}^{(0)}] + \tfrac{1}{2}\,[\bm{\pi}^2,\,f^2_{\sfa}] \,,
\\
 D_{\sfa}\bm{\pi}^6 &= f_{\sfa}{}^6 + [\bm{\pi}^6,\,\widehat{\Theta}^{(0)}] 
 + [\bm{\pi}^2,\, f^4_{\sfa}] 
 + \tfrac{1}{3}\,[\bm{\pi}^2,\,[\bm{\pi}^2,\,f^2_{\sfa}]] \,,
\\
\begin{split}
 D_{\sfa}\bm{\pi}^{7,1} &= f_{\sfa}{}^{7,1} + [\bm{\pi}^{7,1},\,\widehat{\Theta}^{(0)}] 
 + [\bm{\pi}^2,\, f^6_{\sfa}] 
 + \tfrac{1}{2}\,[\bm{\pi}^4,\,f^4_{\sfa}] 
\\
 &\quad + \tfrac{1}{2}\,[\bm{\pi}^2,\,[\bm{\pi}^2,\,f^4_{\sfa}]]
 + \tfrac{1}{4}\,[\bm{\pi}^4,\,[\bm{\pi}^2,\,f^2_{\sfa}]]
 + \tfrac{1}{8}\,[\bm{\pi}^2,\,[\bm{\pi}^2,\,[\bm{\pi}^2,\,f^2_{\sfa}]] \,,
\end{split}
\\
 D_{\sfa} \Delta &= Z_{\sfa}\,,\quad (\lambda\,D_{\sfa}\lambda^{-1})_\alpha{}^\beta = f_{\sfa\alpha}{}^\beta\,,
\end{align}
In components, we find
\begin{align}
 D_{\sfa}\pi^{\sfb_1\sfb_2}_\beta &= f_{\sfa}{}_\beta^{\sfb_1\sfb_2} + 2\,f_{\sfa\sfc}{}^{[\sfb_1|}\,\pi_\beta^{\sfc|\sfb_2]} - f_{\sfa\beta}{}^{\gamma}\,\pi_\gamma^{\sfb_1\sfb_2} -2\,Z_{\sfa}\,\pi_\beta^{\sfb_1\sfb_2}\,,
\\
 D_{\sfa}\pi^{\sfb_1\cdots\sfb_4} &= f_{\sfa}{}^{\sfb_1\cdots\sfb_4} + 4\,f_{\sfa\sfc}{}^{[\sfb_1|}\,\pi^{\sfc|\sfb_2\sfb_3\sfb_4]} -4\,Z_{\sfa}\,\pi^{\sfb_1\cdots\sfb_4} -3\,\epsilon^{\alpha\beta}\,f_{\sfa}{}^{[\sfb_1\sfb_2}_\alpha\,\pi^{\sfb_3\sfb_4]}_\beta \,,
\\
\begin{split}
 D_{\sfa}\pi_\beta^{\sfb_1\cdots\sfb_6} &= f_{\sfa}{}_\beta^{\sfb_1\cdots\sfb_6} + 6\,f_{\sfa\sfc}{}^{[\sfb_1|}\,\pi_\beta^{\sfc|\sfb_2\cdots \sfb_6]} - f_{\sfa\beta}{}^{\gamma}\,\pi_\gamma^{\sfb_1\cdots\sfb_6} -6\,Z_{\sfa}\,\pi^{\sfb_1\cdots\sfb_6}_\beta
\\
 &\quad -15\,f_{\sfa}{}^{[\sfb_1\cdots\sfb_4}\,\pi^{\sfb_5\sfb_6]}_\beta 
 +30\,\epsilon^{\gamma\delta}\,f_{\sfa}{}_\gamma^{[\sfb_1\sfb_2}\,\pi^{\sfb_3\sfb_4}_\delta\,\pi^{\sfb_5\sfb_6]}_\alpha\,,\end{split}
\\
\begin{split}
 D_{\sfa}\pi^{\sfb_{1\cdots 7},\sfb'} &= f_{\sfa}{}^{\sfb_{1\cdots 7},\sfb'} + 7\,f_{\sfa\sfc}{}^{[\sfb_1|}\,\pi^{\sfc|\sfb_{2\cdots 7}],\sfb'}
 + f_{\sfa\sfc}{}^{\sfb'}\,\pi^{\sfb_{1\cdots 7},\sfc} -8\,Z_{\sfa}\,\pi^{\sfb_{1\cdots 7},\sfb'} 
\\
 &\quad +21\,\epsilon^{\gamma\delta}\,f_{\sfa}{}_\gamma^{\sfb'[\sfb_{1\cdots 5}}\,\pi^{\sfb_{67}]}_\delta
 +\tfrac{35}{2}\,f_{\sfa}{}^{\sfb'[\sfb_{123}}\,\pi^{\sfb_{4\cdots 7}]}
\\
 &\quad -\tfrac{105}{2}\,\epsilon^{\alpha\beta}\,f_{\sfa}{}^{[\sfb_{1\cdots 4}}\,\pi^{\sfb_{56}}_\alpha\,\pi^{\sfb_7]\sfb'}_\beta
 +\tfrac{105}{4}\,\epsilon^{\gamma\delta}\,f_{\sfa}{}_\gamma^{[\sfb_{12}}\,\pi^{\sfb_{3\cdots 6}}\,\pi_\delta^{\sfb_7]\sfb'}
\\
 &\quad -\tfrac{105}{4}\,\epsilon^{\gamma\delta}\,f_{\sfa}{}_\gamma^{\sfb'[\sfb_1}\,\pi^{\sfb_{2\cdots 5}}\,\pi^{\sfb_{67}]}_\delta
 -\tfrac{315}{4}\,\epsilon^{\alpha\beta}\,\epsilon^{\gamma\delta}\,\pi^{[\sfb_{12}}_\alpha\,\pi^{\sfb_3|\sfb'|}_\beta\,\pi_\gamma^{\sfb_{45}}\,f_{\sfa}{}_\delta^{\sfb_{67]}}\,.
\end{split}
\end{align}
If we define the Nambu--Lie structures as
\begin{align}
\begin{alignedat}{2}
 \pi^{m_1m_2}_{\alpha} &\equiv \pi^{\sfb_1\sfb_2}_{\alpha}\,e_{\sfb_1}^{m_1}\,e_{\sfb_2}^{m_2}\,,&\qquad
 \pi^{m_1\cdots m_4} &\equiv \pi^{\sfb_1\cdots \sfb_4}\,e_{\sfb_1}^{m_1}\cdots e_{\sfb_4}^{m_4}\,,
\\
 \pi^{m_1\cdots m_6}_{\alpha} &\equiv \pi^{\sfb_1\cdots \sfb_6}_{\alpha}\,e_{\sfb_1}^{m_1}\cdots e_{\sfb_6}^{m_6}\,,&\qquad
 \pi^{m_1\cdots m_7,m'} &\equiv \pi^{\sfb_1\cdots \sfb_7,\sfb'}\,e_{\sfb_1}^{m_1}\cdots e_{\sfb_7}^{m_7}\,e_{\sfb'}^{m'}\,,
\end{alignedat}
\end{align}
we find
\begin{align}
 &\Lie_{v_{\sfa}} \pi^{m_1m_2}_{\alpha} = \Exp{-2\Delta} f_{\sfa}{}^{\sfb_1\sfb_2}_{\alpha}\,v_{\sfb_1}^{m_1}\,v_{\sfb_2}^{m_2}\,,\quad 
 \Lie_{\sfa} \Delta = Z_{\sfa}\,,\quad 
 (\lambda\,\Lie_{\sfa}\lambda^{-1})_\alpha{}^\beta = f_{\sfa\alpha}{}^\beta\,,
\\
 &\Lie_{v_{\sfa}} \pi^{m_1\cdots m_4} 
 +3\,\epsilon^{\alpha\beta}\,\pi^{[m_1m_2}_{\alpha}\,\Lie_{v_{\sfa}}\pi^{m_3m_4]}_{\beta}
 = \Exp{-4\Delta} f_{\sfa}{}^{\sfb_1\sfb_2}\,v_{\sfb_1}^{m_1}\,v_{\sfb_2}^{m_2}\,,
\\
 &\Lie_{v_{\sfa}} \pi^{m_{1\cdots 6}}_{\alpha} 
 +15\,\bigl(\delta_{\alpha}^{\gamma}\,\pi^{[m_{1\cdots 4}} 
 - \epsilon^{\beta\gamma}\,\pi_{\alpha}^{[m_{12}}\,\pi^{m_{34}}_{\beta} \bigr)\,\Lie_{v_{\sfa}}\pi^{m_{56}]}_{\gamma}
 =\Exp{-6\Delta} f_{\sfa}{}^{\sfb_{1\cdots 6}}\,v_{\sfb_1}^{m_1}\cdots v_{\sfb_6}^{m_6}\,,
\\
\begin{split}
 &\Lie_{v_{\sfa}} \pi^{m_{1\cdots 7},m'} 
 -21\,\epsilon^{\alpha\beta}\,\pi^{m'[m_{1\cdots 5}}\,\Lie_{v_{\sfa}}\pi^{m_{67}]}
 - \tfrac{35}{2}\,\pi^{[m_{1\cdots 4}}\,\Lie_{v_{\sfa}}\pi^{m_{567}]m'}
\\
 &\tfrac{105}{2}\,\epsilon^{\alpha\beta}\,\pi^{[m_{1\cdots 4}}\,\pi^{m_{56}}_{\alpha}\,\Lie_{v_{\sfa}}\pi^{m_7]m'}_{\beta}
 -\tfrac{105}{2}\,\epsilon^{\alpha\beta}\,\pi^{[m_{1\cdots 4}}\,\pi^{m_{5}|m'|}_{\alpha}\,\Lie_{v_{\sfa}}\pi^{m_{67}}_{\beta}
\\
 &-\tfrac{105}{4}\,\epsilon^{\alpha\beta}\,\epsilon^{\gamma\delta}\,\pi^{[m_{12}}_{\alpha}\,\pi^{m_3|m'|}_{\beta}\,\pi^{m_{45}}_{\gamma}\,\Lie_{v_{\sfa}}\pi^{m_{67}}_{\delta} = \Exp{-8\Delta} f_{\sfa}{}^{\sfb_{1\cdots 7},\sfb'}\,v_{\sfb_1}^{m_1}\cdots v_{\sfb_7}^{m_7}\,v_{\sfb'}^{m'}\,.
\end{split}
\end{align}

In the case of the coboundary EDA, we can solve for the Nambu--Lie structure by means of the classical $r$-matrices. 
Here, use the formula \eqref{eq:U-formula} for simplicity.
Then we find
\begin{align}
 U_I{}^J = \bigl(\Exp{\Exp{-2\Delta}\rho^2} \Exp{\Exp{-4\Delta}\rho^4} \Exp{\Exp{-6\Delta}\rho^6} \Exp{\Exp{-8\Delta}\rho^{7,1}}\bigr)_I{}^J\,.
\end{align}
One can obtain the Nambu--Lie structure $\bm{\Pi}_I{}^J\equiv \mathbb{E}_I{}^A\,\bm{\Pi}_A{}^B\,\mathbb{E}_B{}^J$ by computing
\begin{align}
 \bm{\Pi}_I{}^J = \cR_I{}^K\, (\Omega_\Delta^{-1})_K{}^J \,,\qquad
 \cR_I{}^J \equiv \mathbb{E}_I{}^A\,\cR_A{}^B\,\mathbb{E}_B{}^J\,.
\label{eq:pi-formula}
\end{align}
For example, we have
\begin{align}
 \pi^{m_1m_2}_\alpha
 &= \rr^{\sfa_1\sfa_2}_\beta\,\bigl(\lambda_\alpha{}^\beta\,\Exp{-2\Delta}v_{\sfa_1}^{m_1}\,v_{\sfa_2}^{m_2} - \delta_\alpha^\beta\,e_{\sfa_1}^{m_1}\,e_{\sfa_2}^{m_2}\bigr)\,,
\\
 \pi^{m_1\cdots m_4} 
 &= \rr^{\sfa_1\cdots \sfa_4} \,\bigl(\Exp{-4\Delta}v_{\sfa_1}^{m_1}\cdots v_{\sfa_4}^{m_4} - e_{\sfa_1}^{m_1}\cdots e_{\sfa_4}^{m_4}\bigr)
 -3\Exp{-2\Delta}\epsilon^{\gamma\delta}\,\pi^{[m_1m_2}\,\rr^{m_3m_4]}\,,
\end{align}
which are similar to the relations \eqref{eq:M-pi-sol1}--\eqref{eq:M-pi-sol3} found in the M-theory picture. 

In the fully general situation, for example, we obtain the generalized CYBE for $\pi_\alpha^{\sfa_1\sfa_2}$ as
\begin{align}
\begin{split}
 \text{CYBE}^{\sfa}_\alpha{}_\beta^{\sfb_1\sfb_2}
 &= -\bigl[2 \, f_{\sfc\sfd}{}^{[\sfb_1}\, \rr_\beta^{\sfb_2]\sfd}\,\rr_\alpha^{\sfc\sfa} 
 -f_{\sfc\beta}{}^\gamma\,\rr_\alpha^{\sfa\sfc}\, \rr_\gamma^{\sfb_1\sfb_2}
 - 2 Z_{\sfc}\, \rr_\alpha^{\sfa\sfc}\, \rr_\beta^{\sfb_1\sfb_2}
\\
 &\qquad + \bigl(f_{\sfc\sfd}{}^{\sfa} - 4\,Z_{[\sfc}\, \delta_{\sfd]}^{\sfa}\bigr) \bigl(\tfrac{1}{2}\,\epsilon_{\alpha\beta}\,\rr^{\sfc\sfd \sfb_1\sfb_2}
 - \rr_{(\alpha}^{\sfc[\sfb_1}\, \rr_{\beta)}^{\sfb_2]\sfd}
 - \tfrac{1}{2}\, \rr_{[\alpha}^{\sfc\sfd}\, \rr_{\beta]}^{\sfb_1\sfb_2}\bigr)
\\
 &\qquad + f_{\sfc\alpha}{}^\gamma\,\bigl(\epsilon_{\gamma\beta}\, \rr^{\sfa\sfc\sfb_1\sfb_2} 
 - 2\,\rr_{(\gamma}^{\sfa[\sfb_1}\, \rr_{\beta)}^{\sfb_2]\sfc}
 - \rr_{[\gamma}^{\sfa\sfc}\,\rr_{\beta]}^{\sfb_1\sfb_2}\bigr)\bigr]=0\,.
\end{split}
\end{align}
By considering CYBE \eqref{eq:gen-YB} for non-negative generators and $\text{CYBE}^{a}_\alpha{}^0=0$, we find additional conditions, such as
\begin{align}
 \rr_\alpha^{\sfa\sfb}\,Z_{\sfb} =0\,,\quad
 \rr_\alpha^{\sfb\sfc}\,f_{\sfb\sfc}{}^{\sfa} = 0\,,\quad
 \rr_{\alpha}^{\sfa\sfb}\,f_{\sfb\beta}{}^\gamma = 0\,.
\label{eq:unimodular}
\end{align}
In particular, when $(r_\alpha^{\sfa\sfb}) =(r_1^{\sfa\sfb},\,r_2^{\sfa\sfb}) =(r^{\sfa\sfb},0)$ and $f_{\sfa1}{}^2=0$ are satisfied, we find
\begin{align}
 \text{CYBE}^{\sfa}_1{}_1^{\sfb_1\sfb_2}
 &= 3\,f_{\sfc\sfd}{}^{[\sfa}\,\rr^{|\sfc|\sfb_1}\, \rr^{\sfb_2]\sfd} 
 + 3\,\rr^{[\sfa\sfb_1}\, \rr^{\sfb_2]\sfc}\,(f_{\sfc1}{}^1+2\,Z_{\sfc}\bigr) =0\,,
\end{align}
which reduces to the standard classical Yang--Baxter equation when the dilaton flux satisfies $f_{\sfa1}{}^1=-2\,Z_{\sfa}$\,.\footnote{From $f_{\sfa\beta}{}^\beta=0$\,, the dilaton flux satisfies $f_{\sfa1}{}^1 = -f_{\sfa 2}{}^2$\,.}
The second requirement of Eq.~\eqref{eq:unimodular} shows that the classical matrix should be unimodular in the terminology of \cite{1608.03570}. 

We can straightforwardly compute other components of the CYBE, but this is a very hard task. 
Here, assuming $f_{\sfa\beta}{}^\gamma=0$ and $Z_{\sfa}=0$, we show the main part of the CYBE for multi-vectors. 
For simplicity, again we consider the case of a single multi-vector. 
Since $\rr_\alpha^{\sfa_1\sfa_2}$ is already considered, let us consider $\cR=\Exp{\rr^4}$ and $\cR=\Exp{\rr^6}$\,. 
In each case, we find
\begin{align}
 \text{CYBE}^{\sfa_{123}}{}^{\sfb_{1\cdots 4}}
 &= -\bigl[ 4\,f_{\sfc\sfd}{}^{[\sfb_1}\,\rr^{|\sfc|\sfb_{234}]}\,\rr^{\sfd\sfa_{123}}
\nn\\
 &\qquad +3\,f_{\sfc\sfd}{}^{[\sfa_1}\,\bigl(4\,\delta_{\sfe_{1\cdots 4}}^{\sfa_{23}]\sfc\sfd}\,\delta_{\sff}^{[\sfe_1}\,\rr^{\sfe_{234}][\sfb_1}\,\rr^{\sfb_{234}]\sff}
 -\tfrac{1}{2}\,\rr^{\sfa_{23}]\sfc\sfd}\,\rr^{\sfb_{1\cdots 4}}\bigr)\bigr] =0\,,
\\
 \text{CYBE}^{\sfa_{1\cdots5}}_\alpha{}^{\sfb_{1\cdots 6}}_\beta
 &= -\bigl[ 6\,f_{\sfc\sfd}{}^{[\sfb_1}\,\rr_\alpha^{|\sfc|\sfb_{234}]}\,\rr_\beta^{\sfd\sfa_{123}}
 -45\,f_{\sfc\sfd}{}^{[\sfa_1}\, \delta_{\sfe_1\cdots\sfe_6}^{\sfa_{2\cdots 5}]\sfc\sfd}\,\delta_{\sff}^{[\sfe_1}\,\rr^{\sfe_{2\cdots6}][\sfb_1}_\alpha\,\rr^{\sfb_{2\cdots6}]\sff}_\beta
\nn\\
 &\qquad -\tfrac{5}{4}\,f_{\sfc\sfd}{}^{[\sfa_1}\,\bigl( \rr_\beta^{\sfa_{2\cdots5}]\sfc\sfd}\,\rr_\alpha^{\sfb_{1\cdots 6}}
 -5\,\rr_\alpha^{\sfa_{2\cdots5}]\sfc\sfd}\,\rr_\beta^{\sfb_{1\cdots 6}}\bigr)\bigr] =0\,.
\end{align}

\section{Summary and discussions}
\label{sec:discussion}

In this paper, we presented the general structure of the ExDA by using the generalized Lie derivative in ExFT. 
As is known in the Poisson--Lie $T$-duality, if we choose a maximally isotropic subalgebra $\mathfrak{g}$ of the ExDA, we can systematically construct the generalized frame fields $E_A{}^I$\,. 
We have generally shown that, for any choice of the subalgebra $\mathfrak{g}$\,, the constructed generalized frame fields $E_A{}^I$ satisfy the algebra
\begin{align}
 \gLie_{E_A} E_B{}^I = - X_{AB}{}^C\,E_C{}^I\,,
\end{align}
which means that the target space can be called a generalized parallelizable space. 
Here, we have considered DFT and the $E_{n(n)}$ EFT ($n\leq 8$) as particular examples of ExFT, but our presentation does not depend on the details of the duality group $\cG$\,. 
Thus the generalized parallelizability will be realized even for other ExFTs, such as the heterotic DFT \cite{1103.2136}. 
It is also interesting to consider the extension to the $E_{11}$ EFT (see for example \cite{1907.02080}). 

Among the Leibniz identities, the particularly interesting ones are the cocycle conditions. 
By using the duality algebra, we found a general definition of the coboundary operator $\delta_n$\,. 
In particular, the operator $\delta_0$ is non-trivial and the coboundary ansatz \eqref{eq:cR-def} shows that the embedding tensor $\widehat{\Theta}_a$ is twisted by the classical $r$-matrices. 
Moreover, we have provided a general formula \eqref{eq:pi-formula} of the Nambu--Lie structure $\pi$ for a general coboundary ExDA. 

As a particular class of the non-Abelian duality, we discussed the Yang--Baxter deformations. 
The explicit form of the generalized CYBE needs to be clarified in future studies, but once a solution of the CYBE is found,\footnote{When we consider concrete examples, we can easily check the CYBE by computing $\mathfrak{X}_{AB}{}^C = \cX_{AB}{}^C$ (recall Eq.~\eqref{eq:gen-YB}). The only difficulty is to identity the full set of independent conditions on a general $\rr^{\Pa}$ from $\mathfrak{X}_{AB}{}^C = \cX_{AB}{}^C$.} we can easily perform the Yang--Baxter deformation,
\begin{align}
 \cM_{IJ} \to \cM'_{IJ}=\bigl(\Omega_\Delta\,\cM\,\Omega_\Delta^{\rmT}\bigr)_{IJ}\,.
\label{eq:YB-deformation}
\end{align}
Here, the matrix $(\Omega_\Delta)_I{}^J$ is made of the multi-vector fields $\Exp{-(p+q)\Delta}\rho^{i_1\cdots i_p,i_{p+1}\cdots i_{p+q}}$ where
\begin{align}
 \rho^{i_1\cdots i_p,i_{p+1}\cdots i_{p+q}} = \rr^{a_1\cdots a_p, a_{p+1}\cdots a_{p+q}}\,v_{a_1}^{i_1}\cdots v_{a_{p+q}}^{i_{p+q}}\,,
\end{align}
and $\rr$ is the generalized classical $r$-matrix. 
Recently, the Yang--Baxter deformations have been understood as the local $\beta$-transformations \cite{1702.02861,1705.02063,1705.07116,1710.06784,1803.05903,1811.09056,1906.09053} which are characterized by the bi-vector $\rho^{mn}=\rr^{ab}\,v_a^m\,v_b^n$\,. 
As a natural extension, the tri-vector deformation in 11D supergravity has been proposed in \cite{1906.09052,2002.01915}, and our general formula \eqref{eq:YB-deformation} supports the conjectural multi-vector deformations. 
In this paper, we consider that the physical space is a group manifold, but the usual Yang--Baxter deformation can be performed in more general curved spaces. 
It is important future work to extend the generalized Yang--Baxter deformation to a more general setup. 
It is also important to study the non-Abelian duality for non-coboundary ExDAs. 

In this paper, we have parameterized the generators of $\mathfrak{g}$ as $T_a$ and introduced the corresponding physical coordinates $x^a\,(=\delta^a_i\,x^i)$\,. 
However, we can change this parameterization. 
For example, in the M-theory picture, we may find a maximally isotropic subalgebra $\mathfrak{g}$ that is spanned by a mixture of $T_a$ and $T^{a_1a_2}$\,. 
In this case, we construct the group element $g$ by exponentiating these generators with the corresponding coordinates $x^a$ and $y_{a_1a_2}$\,. 
In fact, even in this case, we can systematically construct the generalized frame fields by following the same procedure. 
By construction, ExFT has the formal duality symmetry,
\begin{align}
 x^I \to (\Lambda^{-1})_J{}^I\,x^J\,,\qquad \cM_{IJ} \to \Lambda_I{}^K\,\Lambda_J{}^L\, \cM_{KL}\qquad (\Lambda\in \cG)\,,
\label{eq:formal-duality}
\end{align}
where $\Lambda_I{}^J$ is a constant matrix. 
Then, for an arbitrary Leibniz algebra which is generated by the generalized Lie derivative and contains a maximally isotropic algebra $\mathfrak{g}$, we can perform the formal duality transformation \eqref{eq:formal-duality} such that $\mathfrak{g}$ is spanned by $T'_a$\,. 
After the duality rotation, the Leibniz algebra will have the form of an ExDA. 
In other words, if a Leibniz algebra generated by the generalized Lie derivative contains a maximally isotropic algebra, it will be related to an ExDA through a formal duality. 
In this sense, EDA is very universal.

In this paper, we have defined the ExDA by using the generalized Lie derivative in ExFT. 
In this case, the structure constants have the form \eqref{eq:EDA-structure}, which has been studied in the maximal gauged supergravities. 
If we consider less supersymmetric cases, the structure constants will have a more general form. 
It is an interesting to study whether we can construct certain generalized parallelizable spaces for this generalized algebras. 
For example, in the case of the usual Drinfel'd double, a deformation of the structure constants $X_{AB}{}^C$,
\begin{align}
 X_{AB}{}^C \to X'_{AB}{}^C \equiv X_{AB}{}^C + 2\,X_{[A}\,\delta_{B]}^C \,,
\end{align}
was studied in \cite{1407.7106,1705.05082}. 
Here, the vector $X_A$ is supposed to be a null vector $\eta^{AB}\,X_A\,X_B = 0$\,. 
To be more explicit, by parameterizing $(X_A)=(\alpha_a,\,\beta^a)$ and defining $f'_{ab}{}^c\equiv f_{ab}{}^c + 2\,\alpha_{[a}\,\delta_{b]}^c$ and $f'_c{}^{ab} \equiv f_c{}^{ab} + 2\,\beta^{[a}\,\delta_b^{c]}$\,, this deformed Lie algebra is given by
\begin{align}
\begin{split}
 T_a\circ T_b &= f'_{ab}{}^c \, T_c \,,\quad
 T^a\circ T^b = f'_c{}^{ab}\, T^c \,,
\\
 T_a\circ T^b &= \bigl(f'_a{}^{bc} +\beta^c\,\delta_a^b -2\beta^b\,\delta_a^c \bigr)\, T_c - \bigl(f'_{ac}{}^b -\alpha_c\,\delta_a^b +2\,\alpha_a\,\delta_c^b \bigr)\, T^c = - T^b\circ T_a \,.
\end{split}
\end{align}
In \cite{1407.7106,1705.05082}, the standard procedures of the Poisson--Lie $T$-duality was straightforwardly extended and the generalized duality, called the Jacobi--Lie $T$-duality, was discussed. 
Obviously, we find $X'_{[ABC]}\neq X'_{ABC}$ and we cannot realize this algebra by means of the generalized Lie derivative in DFT. 
Deformations of the generalized Lie derivative in ExFT have been studied in \cite{1604.08602,1612.05230,1708.02589} and it is interesting to extend the ExDA by using such deformed Lie derivatives.

\subsection*{Acknowledgments}

This work is supported by JSPS Grant-in-Aids for Scientific Research (C) 18K13540 and (B) 18H01214. 

\newpage

\appendix

\section{Conventions}
\label{app:En-algebra}

We (anti-)symmetrize the indices as
\begin{align}
 A_{(a_1\cdots a_p)} = \tfrac{1}{p!}\,\bigl[A_{a_1\cdots a_p} + \text{(permutations)}\bigr]\,,\quad
 A_{[a_1\cdots a_p]} = \tfrac{1}{p!}\,\bigl[A_{a_1\cdots a_p} \pm \text{(permutations)}\bigr]\,.
\end{align}
We define the antisymmetrized Kronecker delta as $\delta^{a_1\cdots a_p}_{b_1\cdots b_p}\equiv \delta^{[a_1}_{[b_1}\cdots \delta^{a_p]}_{b_1]}$\,. 

We sometimes use a short-hand notation $A_{a_{1\cdots p}}\equiv A_{a_1\cdots p}$ and may simply denote it as $A_p$\,. 
In this Appendix, we use a further condensed notation, such as
\begin{align}
 A_{\bar{a}_p \bar{b}_q} \equiv \tfrac{1}{\sqrt{p!\,q!}}\,A_{a_{1\cdots p} b_{1\cdots q}}\,.
\end{align}
Namely, in order to reproduce the standard notation, each barred index $\bar{a}_p$ is replaced by the antisymmetrized indices $a_{1\cdots p}$ and we additionally multiply the factor $\tfrac{1}{\sqrt{p!}}$ to the expression. 
In this notation, the antisymmetrized Kronecker delta $\delta^{a_1\cdots a_p}_{b_1\cdots b_p}$ appears with a factor $p!$\,.
Namely, for example, we have
\begin{align}
 \delta^{\bar{a}_p}_{\bar{b}_p} \equiv \tfrac{1}{\sqrt{p!\,p!}}\, p!\,\delta^{a_{1\cdots p}}_{b_{1\cdots p}}=\delta^{a_{1\cdots p}}_{b_{1\cdots p}}\,,\qquad
 \delta^{c\bar{a}_p}_{\bar{b}_{p+1}} \equiv \tfrac{1}{\sqrt{p!\,(p+1)!}}\, (p+1)!\,\delta^{ca_{1\cdots p}}_{b_{1\cdots (p+1)}}\,.
\end{align}
This notation significantly simplifies various expression. 

For $E_{n(n)}$ group, we define various representations as described in Table \ref{tab:reps}.
\begin{table}[b]
\begin{center}
{\small\tabcolsep=1.2mm \renewcommand\arraystretch{0.8}
 \begin{tabular}{|c||*9{c|}|c|}\hline
 $n$ & $R_1$ & $R_2$ & $R_3$ & $R_4$ & $R_5$ & $R_6$ & $R_7$ & $R_8$ & $R_9$ & $R_{{\tt SC}}$ \\\hline\hline
 2 & \begin{tabular}{c}${\bf 2}$\\ ${\bf 1}$\end{tabular} & ${\bf 2}$ & ${\bf 1}$ & ${\bf 1}$ & ${\bf 2}$ & \begin{tabular}{c}${\bf 2}$ \\ ${\bf 1}$\end{tabular} & \begin{tabular}{c}${\bf 3}$ \\ ${\bf 1}$\end{tabular} & \begin{tabular}{c}${\bf 3}$ \\ ${\bf 2}$\end{tabular} & \begin{tabular}{c}${\bf 4}$ \\ $2\times {\bf 2}$\end{tabular} & ${\bf 2}$ \\\hline
 3 & $({\bf \overline{3}},{\bf 2})$ & $({\bf 3},{\bf 1})$ & $({\bf 1},{\bf 2})$ & $({\bf \overline{3}},{\bf 1})$ & $({\bf 3},{\bf 2})$ & \begin{tabular}{c}$({\bf 8},{\bf 1})$ \\ $({\bf 1},{\bf 3})$\end{tabular} & \begin{tabular}{c}$({\bf 6},{\bf 2})$ \\ $({\bf \overline{3}},{\bf 2})$\end{tabular} & \begin{tabular}{c}$({\bf 15},{\bf 1})$ \\ $({\bf 3},{\bf 3})$ \\ $2\times({\bf 3},{\bf 1})$\end{tabular} & & $({\bf 3},{\bf 1})$ \\\hline
 4 & ${\bf \overline{10}}$ & ${\bf 5}$ & ${\bf \overline{5}}$ & ${\bf 10}$ & ${\bf 24}$ & \begin{tabular}{c}${\bf \overline{40}}$ \\ ${\bf \overline{15}}$\end{tabular} & \begin{tabular}{c}${\bf 70}$ \\ ${\bf 45}$ \\ ${\bf 5}$\end{tabular} & & & ${\bf 5}$ \\\hline
 5 & ${\bf 16}$ & ${\bf 10}$ & ${\bf \overline{16}}$ & ${\bf 45}$ & ${\bf 144}$ & \begin{tabular}{c}${\bf 320}$ \\ ${\bf 126}$ \\ ${\bf 10}$\end{tabular} & & & & ${\bf 10}$ \\\hline
 6 & ${\bf 27}$ & ${\bf \overline{27}}$ & ${\bf 78}$ & ${\bf 351}$ & \begin{tabular}{c}${\bf \overline{1728}}$ \\ ${\bf \overline{27}}$\end{tabular} & & & & & ${\bf \overline{27}}$ \\\hline
 7 & ${\bf 56}$ & ${\bf 133}$ & ${\bf 912}$ & \begin{tabular}{c}${\bf 8645}$ \\ ${\bf 133}$\end{tabular} & & & & & & \begin{tabular}{c}${\bf 133}$ \\ ${\bf 1}$\end{tabular} \\\hline
 8 & ${\bf 248}$ & \begin{tabular}{c}${\bf 3875}$ \\ ${\bf 1}$\end{tabular} & \begin{tabular}{c}${\bf 147250}$ \\ ${\bf 3875}$ \\ ${\bf 248}$\end{tabular} & & & & & & & \begin{tabular}{c}${\bf 3875}$ \\ ${\bf 248}$ \\ ${\bf 1}$\end{tabular} \\\hline
\end{tabular}
}
\caption{Representations of $E_{n(n)}$ group. The $R_p$ representation is the $U$-duality multiplet of $p$-form fields in $(11-n)$ dimensions \cite{0705.0752,0705.1304}. The adjoint representation $R_{\tt adj}$ of $E_{n(n)}$ coincides with $R_{9-n}$ and the embedding tensor $\Theta_A{}^{\bm{\alpha}}$ transforms in the $R_{10-n}$ representation.\label{tab:reps}}
\end{center}\end{table}

\newpage

\subsection{$E_{n(n)}$ algebra in the M-theory picture}

In the M-theory picture, the non-vanishing commutators for the $\mathfrak{e}_{n(n)}$ algebra $(n\leq 8)$ are given as follows:
\begin{align}
 [K^a{}_b,\, K^c{}_d]
 &= \delta_b^c\,K^a{}_d - \delta_d^a\, K^c{}_b \,,
\\
 [K^a{}_b,\, R^{\bar{c}_3} ]
  &= \delta_{b\bar{d}_2}^{\bar{c}_3} \,R^{a\bar{d}_2}\,,
\qquad
 [K^a{}_b,\, R_{\bar{c}_3} ]
  = - \delta_{\bar{c}_3}^{a\bar{d}_2} \, R_{b\bar{d}_2}\,, 
\\
 [K^a{}_b,\, R^{\bar{c}_6} ]
  &= \delta_{b\bar{d}_5}^{\bar{c}_6} \, R^{a\bar{d}_5}\,, 
\qquad
 [K^a{}_b,\, R_{\bar{c}_6} ]
  = - \delta^{a\bar{d}_5}_{\bar{c}_6}\, R_{b\bar{d}_5}\,,
\\
 [K^a{}_b,\,R^{\bar{c}_8,c'}]
 &= \delta^{\bar{c}_8}_{b\bar{d}_7}\, R^{a\bar{d}_7,c'} + \delta_b^{c'}\,R^{\bar{c}_8,a}\,, 
\qquad
 [K^a{}_b,\,R_{\bar{c}_8,c'}]
  = - \delta_{\bar{c}_8}^{a\bar{d}_7}\, R_{b\bar{d}_7,c'} - \delta_{c'}^a\,R_{\bar{c}_8,b}\,,
\\
 [R^{\bar{a}_3},\, R^{\bar{b}_3}]
 &= R^{\bar{a}_3 \bar{b}_3}\,,
\quad
 [R^{\bar{a}_3},\, R^{\bar{b}_6}]
  = \delta^{\bar{b}_6}_{\bar{c}_5c'}\, R^{\bar{a}_3 \bar{c}_5,c'}\,,
\\
 [ R^{\bar{a}_3},\,R_{\bar{b}_3} ]
  &= - \delta_{\bar{e}_2 c}^{\bar{a}_3}\,\delta_{\bar{b}_3}^{\bar{e}_2 d}\, K^c{}_d + \tfrac{1}{3}\,\delta_{\bar{b}_3}^{\bar{a}_3}\, K \,, 
\qquad
 [ R^{\bar{a}_3},\, R_{\bar{b}_6} ]
  = - \delta_{\bar{b}_6}^{\bar{a}_3 \bar{c}_3}\,R_{\bar{c}_3} \,, 
\\
 [R^{\bar{a}_3},\, R_{\bar{b}_8,b}]
 &= - \delta^{\bar{a}_3 \bar{c}_5}_{\bar{b}_8}\, R_{\bar{c}_5 b} \,,
\qquad
 [R^{\bar{a}_6},\, R_{\bar{b}_3}]
  = \delta^{\bar{a}_6}_{\bar{b}_3 \bar{c}_3}\,R^{\bar{c}_3} \,, 
\\
 [R^{\bar{a}_6},\,R_{\bar{b}_6} ]
  &= - \delta^{\bar{a}_6}_{\bar{e}_5 c}\, \delta_{\bar{b}_6}^{\bar{e}_5 d}\, K^c{}_d + \tfrac{2}{3}\,\delta_{\bar{b}_6}^{\bar{a}_6}\, K \,,
\qquad
 [R^{\bar{a}_6},\, R_{\bar{b}_8,b}]
  = - \delta^{\bar{a}_6 \bar{c}_2}_{\bar{b}_8}\, R_{\bar{c}_2b} \,,
\\
 [R^{\bar{a}_8,a},\, R_{\bar{b}_3}]
 &= \delta^{\bar{a}_8}_{\bar{b}_3 \bar{c}_5}\, R^{\bar{c}_5 a} \,,
\qquad
 [R^{\bar{a}_8,a}, R_{\bar{b}_6}]
  = \delta^{\bar{a}_8}_{\bar{b}_6 \bar{c}_2}\, R^{\bar{c}_2 a} \,,
\\
 [R^{\bar{a}_8,a}, R_{\bar{b}_8,b}]
 &= - \delta^{\bar{a}_8}_{\bar{b}_8}\, K^a{}_b \,,
\quad
 [R_{\bar{a}_3},\,R_{\bar{b}_3}] = R_{\bar{a}_3\bar{b}_3}\,,
\quad
 [R_{\bar{a}_3},\, R_{\bar{b}_6}]
  = \delta_{\bar{b}_6}^{\bar{c}_5 c'}\, R_{\bar{a}_3\bar{c}_5,c'}\,,
\end{align}
where $K\equiv K^a{}_a$\,. 

The matrix representations $(t_{\bm{\alpha}})_A{}^B$ in the $R_1$ representation are as follows:
\begin{align}
 K^c{}_d &\equiv \tilde{K}^c{}_d - \beta_n\,\delta_d^c\,t_0 \,,
\\
 \tilde{K}^c{}_d &\equiv {\arraycolsep=1mm \begin{pmatrix}
 \delta_a^c\,\delta_d^b & 0 & 0 & 0 & 0 & 0 & 0 \\
 0 & K_2 & 0 & 0 & 0 & 0 & 0\\
 0 & 0 & K_5 & 0 & 0 & 0 & 0 \\
 0 & 0 & 0 & K_{7,1} & 0 & 0 & 0 \\
 0 & 0 & 0 & 0 & K_{8,3} & 0 & 0 \\
 0 & 0 & 0 & 0 & 0 & K_{8,6} & 0 \\
 0 & 0 & 0 & 0 & 0 & 0 & K_{8,8,1}
 \end{pmatrix} } 
\\
 &\left[\begin{array}{l}
 K_p \equiv - \delta_{\bar{d}\bar{e}_{p-1}}^{\bar{a}_p}\,\delta_{\bar{b}_p}^{\bar{c}\bar{e}_{p-1}}\,,\quad
 K_{s,t} \equiv -\delta^{\bar{a}_s}_{d \bar{e}_{s-1}} \delta^{c \bar{e}_{s-1}}_{\bar{b}_s}\delta^{\bar{a}'_t}_{\bar{b}'_t} - \delta^{\bar{a}_s}_{\bar{b}_s} \delta^{\bar{a}'_t}_{d\bar{e}_{t-1}}\delta^{c\bar{e}_{t-1}}_{\bar{b}'_t}\,,
\\
 K_{8,8,1} \equiv - \delta^{\bar{a}_8}_{d\bar{e}_7} \delta^{c\bar{e}_7}_{\bar{b}_8}\delta^{\bar{a}'_8}_{\bar{b}'_8}\delta^{a\dprime}_{b\dprime} - \delta^{\bar{a}_8}_{\bar{b}_8} \delta^{\bar{a}'_8}_{d\bar{e}_7}\delta^{c\bar{e}_7}_{\bar{b}'_8}\delta^{a\dprime}_{b\dprime} - \delta^{\bar{a}_8}_{\bar{b}_8} \delta^{\bar{a}'_8}_{\bar{b}_8} \delta_{d}^{a\dprime} \delta^{c}_{b\dprime} \end{array}\right]\,, 
\\
 R_{\bar{c}_3} &\equiv {\footnotesize 
 {\arraycolsep=0.5mm \begin{pmatrix}
 0 & 0 & 0 & 0 & 0 & 0 & 0 \\
 \delta^{b \bar{a}_2}_{\bar{c}_3} & 0 & 0 & 0 & 0 & 0 & 0 \\
 0 & -\delta_{\bar{b}_2\bar{c}_3}^{\bar{a}_5} & 0 & 0 & 0 & 0 & 0 \\
 0 & 0 & \!\!\! -\delta_{\bar{b}_5 \bar{d}_2}^{\bar{a}_7} \delta_{\bar{c}_3}^{\bar{d}_2 a'} + \frac{1}{4} \delta_{\bar{b}_5 \bar{c}_3}^{\bar{a}_7 a'} \!\!\! & 0 & 0 & 0 & 0 \\
 0 & 0 & 0 & \!\!\! -\delta_{\bar{b}_7d}^{\bar{a}_8} \delta_{\bar{c}_3}^{d\bar{e}_2} \delta_{\bar{e}_2b'}^{\bar{a}'_3} + \frac{1}{4}\delta_{\bar{b}_7 b'}^{\bar{a}_8} \delta_{\bar{c}_3}^{\bar{a}'_3} \!\!\! & 0 & 0 & 0 \\
 0 & 0 & 0 & 0 & -\delta_{\bar{b}_8}^{\bar{a}_8} \delta_{\bar{b}'_3 \bar{c}_3}^{\bar{a}'_6} & 0 & 0 \\
 0 & 0 & 0 & 0 & 0 & -\delta_{\bar{b}_8}^{\bar{a}_8} \delta_{\bar{b}'_6 \bar{d}_2}^{\bar{a}'_8} \delta_{\bar{c}_3}^{a\dprime \bar{d}_2} & 0 
 \end{pmatrix} } } , 
\\
 R^{\bar{c}_3} &\equiv {\footnotesize 
 {\arraycolsep=0.5mm \begin{pmatrix}
 0 & -\delta_{a\bar{b}_2}^{\bar{c}_3} & 0 & 0 & 0 & 0 & 0 \\
 0 & 0 & \delta^{\bar{a}_2\bar{c}_3}_{\bar{b}_5} & 0 & 0 & 0 & 0 \\
 0 & 0 & 0 & \delta^{\bar{a}_5 \bar{d}_2}_{\bar{b}_7} \delta^{\bar{c}_3}_{\bar{d}_2b'} -\frac{1}{4} \delta^{\bar{a}_5 \bar{c}_3}_{\bar{b}_7 b'} & 0 & 0 & 0 \\
 0 & 0 & 0 & 0& \delta^{\bar{a}_7 d}_{\bar{b}_8} \delta^{\bar{c}_3}_{d \bar{e}_2} \delta^{\bar{e}_2a'}_{\bar{b}_3'}-\frac{1}{4}\delta^{\bar{a}_7 a'}_{\bar{b}_8} \delta^{\bar{c}_3}_{\bar{b}'_3} & 0 & 0 \\
 0 & 0 & 0 & 0& 0 & \delta^{\bar{a}_8}_{\bar{b}_8} \delta^{\bar{a}'_3\bar{c}_3}_{\bar{b}'_6} & 0 \\
 0 & 0 & 0 & 0& 0 & 0 & \delta^{\bar{a}_8}_{\bar{b}_8} \delta^{\bar{a}'_6 \bar{d}_2}_{\bar{b}'_8} \delta^{\bar{c}_3}_{b\dprime \bar{d}_2} \\
 0 & 0 & 0 & 0& 0 & 0 & 0 
 \end{pmatrix} } } , 
\\
 R_{\bar{c}_6} &\equiv {\footnotesize {\arraycolsep=1mm \begin{pmatrix}
 0 & 0 & 0 & 0 & 0 & 0 & 0 \\
 0 & 0 & 0 & 0 & 0 & 0 & 0 \\
 \delta^{b\bar{a}_5}_{\bar{c}_6} & 0 & 0 & 0 & 0 & 0 & 0 \\
 0 & -\delta_{\bar{b}_2}^{a'd} \delta_{\bar{c}_6 d}^{\bar{a}_7} - \frac{1}{2} \delta_{\bar{c}_6 \bar{b}_2}^{\bar{a}_7 a'} & 0 & 0 & 0 & 0 & 0 \\
 0 & 0 & -\delta_{\bar{b}_5 \bar{d}_3}^{\bar{a}_8} \delta_{\bar{c}_6}^{\bar{a}'_3\bar{d}_3} & 0 & 0 & 0 & 0 \\
 0 & 0 & 0 & -\delta_{d\bar{b}_7}^{\bar{a}_8} \delta_{\bar{c}_6b'}^{\bar{a}'_6 d} - \frac{1}{2} \delta_{\bar{b}_7b'}^{\bar{a}_8} \delta_{\bar{c}_6}^{\bar{a}'_6} & 0 & 0 & 0 \\
 0 & 0 & 0 & 0 & \delta_{\bar{b}_8}^{\bar{a}_8} \delta_{\bar{c}_6 \bar{d}_2}^{\bar{a}'_8} \delta_{\bar{b}'_3}^{a\dprime \bar{d}_2} & 0 & 0 \end{pmatrix} } } ,
\\
 R^{\bar{c}_6} &\equiv {\footnotesize \arraycolsep=1mm \begin{pmatrix}
 0 & 0 & -\delta_{a\bar{b}_5}^{\bar{c}_6} & 0 & 0 & 0 & 0 \\
 0 & 0 & 0 & \delta^{\bar{a}_2}_{b'd} \delta^{\bar{c}_6d}_{\bar{b}_7} + \frac{1}{2} \delta^{\bar{c}_6\bar{a}_2}_{\bar{b}_7b'} & 0 & 0 & 0 \\
 0 & 0 & 0 & 0 & \delta^{\bar{a}_5\bar{d}_3}_{\bar{b}_8} \delta^{\bar{c}_6}_{\bar{b}'_2 \bar{d}_3} & 0 & 0 \\
 0 & 0 & 0 & 0& 0 & \delta^{d \bar{a}_7}_{\bar{b}_8} \delta^{\bar{c}_6 a'}_{\bar{b}'_6 d} + \frac{1}{2} \delta^{\bar{a}_7a'}_{\bar{b}_8} \delta^{\bar{c}_6}_{\bar{b}'_6} & 0 \\
 0 & 0 & 0 & 0& 0 & 0 & -\delta^{\bar{a}_8}_{\bar{b}_8} \delta^{\bar{c}_6\bar{d}_2}_{\bar{b}'_8} \delta^{\bar{a}'_3}_{b\dprime\bar{d}_2} \\
 0 & 0 & 0 & 0& 0 & 0 & 0 \\
 0 & 0 & 0 & 0& 0 & 0 & 0 \end{pmatrix}} , 
\\
 R_{\bar{c}_8,c'} &\equiv {\footnotesize \arraycolsep=1mm \begin{pmatrix}
 0 & 0 & 0 & 0 & 0 & 0 & 0 \\
 0 & 0 & 0 & 0 & 0 & 0 & 0 \\
 0 & 0 & 0 & 0 & 0 & 0 & 0 \\
 \delta^{b\bar{a}_7}_{\bar{c}_8}\delta^{a'}_{c'}-\frac{1}{4} \delta_{\bar{c}_8}^{\bar{a}_7a'}\delta_{c'}^b & 0 & 0 & 0 & 0 & 0 & 0 \\
 0 & \delta_{\bar{c}_8}^{\bar{a}_8} \delta^{\bar{a}'_3}_{c' \bar{b}_2} & 0 & 0 & 0 & 0 & 0 \\
 0 & 0 & \delta_{\bar{c}_8}^{\bar{a}_8} \delta_{c' \bar{b}_5}^{\bar{a}'_6} & 0 & 0 & 0 & 0 \\
 0 & 0 & 0 & \delta_{\bar{c}_8}^{\bar{a}_8} \delta_{c'\bar{b}_7}^{\bar{a}'_8}\delta_{b'}^{a\dprime} - \frac{1}{4}\delta_{\bar{c}_8}^{\bar{a}_8}\delta_{\bar{b}_7b'}^{\bar{a}'_8}\delta_{c'}^{a\dprime} & 0 & 0 & 0 \end{pmatrix}} ,
\\
 R^{\bar{c}_8,c'} &\equiv {\footnotesize \arraycolsep=0.5mm \begin{pmatrix}
 0 & 0 & 0 & -\delta_{a\bar{b}_7}^{\bar{c}_8}\delta_{b'}^{c'}+\frac{1}{4} \delta^{\bar{c}_8}_{\bar{b}_7b'}\delta^{c'}_{a} & 0 & 0 & 0 \\
 0 & 0 & 0 & 0 & - \delta^{\bar{c}_8}_{\bar{b}_8} \delta_{\bar{b}'_3}^{c'\bar{a}_2} & 0 & 0 \\
 0 & 0 & 0 & 0 & 0 & -\delta^{\bar{c}_8}_{\bar{b}_8} \delta^{c' \bar{a}_5}_{\bar{b}'_6} & 0 \\
 0 & 0 & 0 & 0& 0 & 0 & -\delta^{\bar{c}_8}_{\bar{b}_8}\delta^{c'\bar{a}_7}_{\bar{b}'_8}\delta^{a'}_{b\dprime} + \frac{1}{4}\delta^{\bar{c}_8}_{\bar{b}_8}\delta^{\bar{a}_7a'}_{\bar{b}'_8}\delta^{c'}_{b\dprime} \\
 0 & 0 & 0 & 0& 0 & 0 & 0 \\
 0 & 0 & 0 & 0& 0 & 0 & 0 \\
 0 & 0 & 0 & 0& 0 & 0 & 0 \end{pmatrix}} .
\end{align}

\newpage

\subsection{$E_{n(n)}$ algebra in the type IIB picture}
\label{app:IIB-algebra}

In the type IIB picture, the non-vanishing commutators for the $\mathfrak{e}_{n(n)}$ algebra $(n\leq 8)$ are given as follows:
\begin{align}
 [K^\sfa{}_\sfb,\, K^\sfc{}_\sfd ]
 &= \delta_\sfb^\sfc\,K^\sfa{}_\sfd - \delta_\sfd^\sfa\, K^\sfc{}_\sfb \,,
\\
 [K^\sfa{}_\sfb,\, R^{\bar{\sfc}_2}_\alpha ]
 &= \delta_{\sfb\sfd}^{\bar{\sfc}_2} \,R^{\sfa\sfd}_\alpha\,,
\qquad
 [K^\sfa{}_\sfb,\, R_{\bar{\sfc}_2}^\alpha ]
  = - \delta^{\sfa\sfd}_{\bar{\sfc}_2} \,R_{\sfb\sfd}^\alpha\,,
\\
 [K^\sfa{}_\sfb,\, R^{\bar{\sfc}_4} ]
 &= \delta_{\sfb\bar{\sfd}_3}^{\bar{\sfc}_4} \, R^{\sfa\bar{\sfd}_3}\,,
\qquad
 [K^\sfa{}_\sfb,\, R_{\bar{\sfc}_4} ]
  = - \delta^{\sfa\bar{\sfd}_3}_{\bar{\sfc}_4} \, R_{\sfb\bar{\sfd}_3}\,,
\\
 [K^\sfa{}_\sfb,\, R^{\bar{\sfc}_6}_\alpha ]
 &= \delta_{\sfb\bar{\sfd}_5}^{\bar{\sfc}_6} \, R^{\sfa\bar{\sfd}_5}_\alpha\,,
\qquad
 [K^\sfa{}_\sfb,\, R_{\bar{\sfc}_6}^\alpha ]
  = - \delta^{\sfa\bar{\sfd}_5}_{\bar{\sfc}_6} \, R_{\sfb\bar{\sfd}_5}^\alpha\,,
\\
 [K^\sfa{}_\sfb,\, R^{\bar{\sfc}_7,\sfc'} ]
 &= \delta_{\sfb\bar{\sfd}_6}^{\bar{\sfc}_7} \, R^{\sfa\bar{\sfd}_6,\sfc'} + \delta_\sfb^{\sfc'}\,R^{\bar{\sfc}_7,\sfa}\,,
\qquad
 [K^\sfa{}_\sfb,\, R_{\bar{\sfc}_7,\sfc'} ]
  = - \delta^{\sfa\bar{\sfd}_6}_{\bar{\sfc}_7} \, R_{\sfb\bar{\sfd}_6,\sfc'} - \delta^\sfa_{\sfc'}\,R_{\bar{\sfc}_7,\sfb}\,,
\\
 [R^\alpha{}_\beta,\, R^\gamma{}_\delta]
 &= \delta^\gamma_\beta\,R^\alpha{}_\delta - \delta^\alpha_\delta\,R^\gamma{}_\beta \,,
\\
 [R^\alpha{}_\beta,\, R^{\bar{\sfc}_2}_\gamma]
 &= -\bigl(\delta^\alpha_\gamma\,\delta^\delta_\beta -\tfrac{1}{2}\,\delta^\alpha_\beta\,\delta^\delta_\gamma\bigr)\,R^{\bar{\sfc}_2}_\delta \,, \qquad
 [R^\alpha{}_\beta,\, R_{\bar{\sfc}_2}^\gamma]
  = \bigl(\delta^\alpha_\delta\,\delta^\gamma_\beta -\tfrac{1}{2}\,\delta^\alpha_\beta\,\delta_\delta^\gamma\bigr)\,R_{\bar{\sfc}_2}^\delta \,,
\\
 [R^\alpha{}_\beta,\, R^{\bar{\sfc}_6}_\gamma]
 &= -\bigl(\delta^\alpha_\gamma\,\delta^\delta_\beta -\tfrac{1}{2}\,\delta^\alpha_\beta\,\delta^\delta_\gamma\bigr)\,R^{\bar{\sfc}_6}_\delta \,, \qquad
 [R^\alpha{}_\beta,\, R_{\bar{\sfc}_6}^\gamma]
  = \bigl(\delta^\alpha_\delta\,\delta^\gamma_\beta -\tfrac{1}{2}\,\delta^\alpha_\beta\,\delta_\delta^\gamma\bigr)\,R_{\bar{\sfc}_6}^\delta \,,
\\
 [R^{\bar{\sfa}_2}_\alpha,\, R^{\bar{\sfb}_2}_\beta ]
 &= \epsilon_{\alpha\beta}\,R^{\bar{\sfa}_2\bar{\sfb}_2} \,,
\quad
 [R^{\bar{\sfa}_2}_\alpha,\, R^{\bar{\sfb}_4} ]
  = - R_\alpha^{\bar{\sfa}_2\bar{\sfb}_4} \,,
\quad
 [R^{\bar{\sfa}_2}_\alpha,\, R^{\bar{\sfb}_6}_\beta ]
  = \epsilon_{\alpha\beta}\, \delta^{\bar{\sfb}_6}_{\bar{\sfc}_5\sff}\,R^{\bar{\sfa}_2\bar{\sfc}_4,\sff} \,,
\\
 [R^{\bar{\sfa}_2}_\alpha,\, R_{\bar{\sfb}_2}^\beta ]
 &= -\delta_\alpha^\beta\,\delta^{\bar{\sfa}_2}_{\sfc\sfe}\,\delta_{\bar{\sfb}_2}^{\sfd\sfe}\, K^\sfc{}_\sfd
 + \tfrac{1}{4}\,\delta_\alpha^\beta\,\delta^{\bar{\sfa}_2}_{\bar{\sfb}_2}\, K 
 + \delta^{\bar{\sfa}_2}_{\bar{\sfb}_2}\,R^\beta{}_{\alpha}\,,
\\
 [R^{\bar{\sfa}_2}_\alpha,\, R_{\bar{\sfb}_4} ]
 &= - \epsilon_{\alpha\beta}\,\delta^{\bar{\sfa}_2\bar{\sfc}_2}_{\bar{\sfb}_4}\,R^\beta_{\bar{\sfc}_2} \,,
\qquad
 [R^{\bar{\sfa}_2}_\alpha,\, R_{\bar{\sfb}_6}^\beta ]
  = \delta_\alpha^\beta\, \delta_{\bar{\sfb}_6}^{\bar{\sfa}_2\bar{\sfc}_4}\,R_{\bar{\sfc}_4} \,,
\\ 
 [R^{\bar{\sfa}_2}_\alpha,\, R_{\bar{\sfb}_7,\sfb'} ]
  &= - \epsilon_{\alpha\beta}\, \delta_{\bar{\sfb}_7}^{\bar{\sfa}_2\bar{\sfc}_5}\,R^\beta_{\bar{\sfc}_5\sfb'} \,,
\\
 [R^{\bar{\sfa}_4},\, R^{\bar{\sfb}_4} ]
 &= -\delta^{\bar{\sfb}_4}_{\bar{\sfc}_3\sfd}\,R^{\bar{\sfa}_4\bar{\sfc}_3,\sfd} \,,
\qquad
 [R^{\bar{\sfa}_4},\, R_{\bar{\sfb}_2}^\alpha]
  = \epsilon^{\alpha\beta}\,\delta_{\bar{\sfb}_2\bar{\sfc}_2}^{\bar{\sfa}_4}\,R_\beta^{\bar{\sfc}_2} \,,
\\
 [R^{\bar{\sfa}_4},\, R_{\bar{\sfb}_4} ]
 &= -\delta^{\bar{\sfa}_4}_{\sfc\bar{\sfe}_3}\,\delta_{\bar{\sfb}_4}^{\sfd\bar{\sfe}_3}\, K^\sfc{}_\sfd
 + \tfrac{2}{4}\, \delta^{\bar{\sfa}_4}_{\bar{\sfb}_4}\, K 
\\
 [R^{\bar{\sfa}_4},\, R_{\bar{\sfb}_6}^\alpha ]
 &= -\delta^{\bar{\sfa}_4\bar{\sfc}_2}_{\bar{\sfb}_6}\, R^\alpha_{\bar{\sfc}_2}\,,
\qquad
 [R^{\bar{\sfa}_4},\, R_{\bar{\sfb}_7,\sfb'} ]
  = \delta^{\bar{\sfa}_4\bar{\sfc}_3}_{\bar{\sfb}_7}\, R_{\bar{\sfc}_3\sfb'}
\\
 [R^{\bar{\sfa}_6}_\alpha,\, R_{\bar{\sfb}_2}^\beta ]
 &= - \delta^\beta_\alpha\, \delta^{\bar{\sfa}_6}_{\bar{\sfb}_2\bar{\sfc}_4}\,R^{\bar{\sfc}_4} \,,
\qquad
 [R^{\bar{\sfa}_6}_\alpha,\, R_{\bar{\sfb}_4} ]
  = \delta_{\bar{\sfb}_4\bar{\sfc}_2}^{\bar{\sfa}_6}\, R_\alpha^{\bar{\sfc}_2}\,,
\\
 [R^{\bar{\sfa}_6}_\alpha,\, R_{\bar{\sfb}_6}^\beta ]
 &= -\delta_\alpha^\beta\,\delta^{\bar{\sfa}_6}_{\sfc\bar{\sfe}_5}\,\delta_{\bar{\sfb}_6}^{\sfd\bar{\sfe}_5}\, K^\sfc{}_\sfd
 + \tfrac{3}{4}\,\delta_\alpha^\beta\,\delta^{\bar{\sfa}_6}_{\bar{\sfb}_6} \, K 
 + \delta^{\bar{\sfa}_6}_{\bar{\sfb}_6}\,R^\beta{}_{\alpha}\,,
\\
 [R^{\bar{\sfa}_6}_\alpha,\, R_{\bar{\sfb}_7,\sfb} ]
 &= \epsilon_{\alpha\beta}\,\delta^{\bar{\sfa}_6\sfc}_{\bar{\sfb}_7}\, R^\beta_{\sfc\sfb}\,,
\\
 [R^{\bar{\sfa}_7,\sfa'},\, R_{\bar{\sfb}_2}^\alpha ]
 &= \epsilon^{\alpha\beta}\, \delta^{\bar{\sfa}_7}_{\bar{\sfb}_2\bar{\sfc}_5}\,R_\beta^{\bar{\sfc}_5\sfa'} \,,
\qquad
 [R^{\bar{\sfa}_7,\sfa'},\, R_{\bar{\sfb}_4} ]
  = - \delta_{\bar{\sfb}_4\bar{\sfc}_3}^{\bar{\sfa}_7}\, R^{\bar{\sfc}_3\sfa'}
\\
 [R^{\bar{\sfa}_7,\sfa} ,\, R_{\bar{\sfb}_6}^\alpha]
 &= -\epsilon_{\alpha\beta}\,\delta_{\bar{\sfb}_6\sfc}^{\bar{\sfa}_7}\, R_\beta^{\sfc\sfa}\,,
\qquad
 [R^{\bar{\sfa}_7,\sfa} ,\, R_{\bar{\sfb}_7,\sfb}]
  = - \delta_{\bar{\sfb}_7}^{\bar{\sfa}_7}\, K^\sfa{}_\sfb\,,
\\
 [R_{\bar{\sfa}_2}^\alpha,\, R_{\bar{\sfb}_2}^\beta ]
 &= \epsilon^{\alpha\beta}\,R_{\bar{\sfa}_2\bar{\sfb}_2} \,,
\quad
 [R_{\bar{\sfa}_2}^\alpha,\, R_{\bar{\sfb}_4} ]
  = - R_{\bar{\sfa}_2\bar{\sfb}_4}^\alpha \,,
\quad
 [R_{\bar{\sfa}_2}^\alpha,\, R_{\bar{\sfb}_6}^\beta ]
  = \epsilon^{\alpha\beta}\, \delta_{\bar{\sfb}_6}^{\bar{\sfc}_5\sfd}\,R_{\bar{\sfa}_2\bar{\sfc}_5,\sfd} \,,
\\
 [R_{\bar{\sfa}_4},\, R_{\bar{\sfb}_4} ]
 &= - \delta_{\bar{\sfb}_4}^{\bar{\sfc}_3\sff}\,R_{\bar{\sfa}_4\bar{\sfc}_3,\sff} \,,
\end{align}
where $K\equiv K^{\sfa}{}_{\sfa}$ and $\epsilon^{12}=\epsilon_{12}=1$\,.

\newpage

The matrix representation in the $R_1$ representation are as follows:
\begin{align}
 K^{\sfc}{}_{\sfd} &\equiv \tilde{K}^{\sfc}{}_{\sfd} - \beta_n\, \delta_{\sfd}^{\sfc}\,t_0 \,,
\\
 K^{\sfc}{}_{\sfd} &\equiv 
 {\small{\arraycolsep=1mm \begin{pmatrix}
 \delta_{\sfa}^{\sfc}\,\delta_{\sfd}^{\sfb} & 0 & 0 & 0 & 0 & 0 & 0 & 0 & 0 & 0 \\
 0 & \delta_\alpha^\beta\,K_1 & 0 & 0 & 0 & 0 & 0 & 0 & 0 & 0 \\
 0 & 0 & K_3 & 0 & 0 & 0 & 0 & 0 & 0 & 0 \\
 0 & 0 & 0 & \delta_\alpha^\beta\,K_5 & 0 & 0 & 0 & 0 & 0 & 0 \\
 0 & 0 & 0 & 0 & K_{6,1} & 0 & 0 & 0 & 0 & 0 \\
 0 & 0 & 0 & 0 & 0 & \delta^{(\alpha_{1}}_{(\beta_{1}}\,\delta^{\alpha_{2})}_{\beta_{2})}\, K_7 & 0 & 0 & 0 & 0 \\
 0 & 0 & 0 & 0 & 0 & 0 & \delta_\alpha^\beta\,K_{7,2} & 0 & 0 & 0 \\
 0 & 0 & 0 & 0 & 0 & 0 & 0 & K_{7,4} & 0 & 0 \\
 0 & 0 & 0 & 0 & 0 & 0 & 0 & 0 & \delta_\alpha^\beta\,K_{7,6} & 0 \\
 0 & 0 & 0 & 0 & 0 & 0 & 0 & 0 & 0 & K_{7,7,1} 
\end{pmatrix}}} , 
\\
 &\left[\begin{array}{l}
 K_{p} \equiv -\delta_{\sfd\bar{\sfe}_{p-1}}^{\bar{\sfa}_p}\,\delta_{\bar{\sfb}_p}^{\sfc \bar{\sfe}_{p-1}} \,, \quad
 K_{s,t} \equiv -\delta_{\bar{\sfb}_s}^{\sfc\bar{\sfe}_{s-1}}\,\delta_{\sfd\bar{\sfe}_{s-1}}^{\bar{\sfa}_s}\,\delta_{\bar{\sfb}'_t}^{\bar{\sfa}'_t} -\delta_{\bar{\sfb}_s}^{\bar{\sfa}_s}\,\delta_{\sfd\bar{d}_{t-1}}^{\bar{\sfa}'_t}\,\delta_{\bar{\sfb}'_t}^{\sfc\bar{d}_{t-1}}\,,
\\
 K_{7,7,1}\equiv -\delta_{\bar{\sfa}_7}^{\sfc\bar{\sfe}_6} \delta_{\sfd\bar{\sfe}_6}^{\bar{\sfb}_7}\delta_{\bar{\sfa}'_7}^{\bar{\sfb}'_7}\delta_{\sfa\dprime}^{\sfb\dprime} - \delta_{\bar{\sfa}_7}^{\bar{\sfb}_7} \delta_{\bar{\sfa}'_7}^{\sfc\bar{\sfe}_6}\delta_{\sfd\bar{\sfe}_6}^{\bar{\sfb}'_7}\delta_{\sfa\dprime}^{\sfb\dprime} - \delta_{\bar{\sfa}_7}^{\bar{\sfb}_7} \delta_{\bar{\sfa}'_7}^{\bar{\sfb}'_7} \delta^{\sfc}_{\sfa\dprime} \delta_{\sfd}^{\sfb\dprime}
 \end{array}\right]\,,
\\
 R^\gamma{}_\delta &\equiv 
 {\small{\arraycolsep=1mm \begin{pmatrix}
 0 & 0 & 0 & 0 & 0 & 0 & 0 & 0 & 0 & 0 \\
 0 & R_1\,\delta_{\sfb}^{\sfa} & 0 & 0 & 0 & 0 & 0 & 0 & 0 & 0 \\
 0 & 0 & 0 & 0 & 0 & 0 & 0 & 0 & 0 & 0 \\
 0 & 0 & 0 & R_1 \,\delta^{\bar{\sfa}_5}_{\bar{\sfb}_5} & 0 & 0 & 0 & 0 & 0 & 0 \\
 0 & 0 & 0 & 0 & 0 & 0 & 0 & 0 & 0 & 0 \\
 0 & 0 & 0 & 0 & 0 & R_2\,\delta^{\bar{\sfa}_7}_{\bar{\sfb}_7} & 0 & 0 & 0 & 0 \\
 0 & 0 & 0 & 0 & 0 & 0 & R_1 \,\delta^{\bar{\sfa}_7}_{\bar{\sfb}_7}\,\delta^{\bar{\sfa}'_2}_{\bar{\sfb}'_2} & 0 & 0 & 0 \\
 0 & 0 & 0 & 0 & 0 & 0 & 0 & 0 & 0 & 0 \\
 0 & 0 & 0 & 0 & 0 & 0 & 0 & 0 & R_1\,\delta^{\bar{\sfa}_7}_{\bar{\sfb}_7}\,\delta^{\bar{\sfa}'_6}_{\bar{\sfb}'_6} & 0 \\
 0 & 0 & 0 & 0 & 0 & 0 & 0 & 0 & 0 & 0 
\end{pmatrix}}} ,
\\
 &\Bigl[R_1 \equiv \delta_\alpha^\gamma\,\delta^\beta_\delta-\tfrac{1}{2}\delta_\alpha^\beta\delta^\gamma_\delta\,,\qquad 
 R_2 \equiv \delta_{(\alpha_1}^\gamma\,\delta_{\alpha_2)}^\epsilon\,\delta^{(\beta_1}_\delta\,\delta^{\beta_2)}_\epsilon - \tfrac{1}{2}\,\delta_{(\alpha_1}^{(\beta_1}\,\delta_{\alpha_2)}^{\beta_2)}\,\delta^\gamma_\delta\Bigr]\,,
\\
 R^\gamma_{\bar{\sfc}_2}
 &\equiv {\footnotesize{\arraycolsep=0.2mm 
 \begin{pmatrix}
 0 & 0 & 0 & 0 & 0 & 0 & 0 & 0 & 0 & 0 \\
 -\delta^\gamma_\alpha\delta^{\sfa\sfb}_{\bar{\sfc}_2} & 0 & 0 & 0 & 0 & 0 & 0 & 0 & 0 & 0 \\
 0 & \!\!\! -\epsilon^{\beta\gamma}\delta^{\bar{\sfa}_3}_{\sfb\bar{\sfc}_2} \!\!\! & 0 & 0 & 0 & 0 & 0 & 0 & 0 & 0 \\
 0 & 0 & \!\!\! -\delta^\gamma_\alpha\delta^{\bar{\sfa}_5}_{\bar{\sfb}_3\bar{\sfc}_2} \!\!\!\!\!\! & 0 & 0 & 0 & 0 & 0 & 0 & 0 \\
 0 & 0 & 0 & \!\!\! \epsilon^{\beta\gamma} \biggl[\genfrac{}{}{0pt}{1}{-\delta^{\bar{\sfa}_6}_{\bar{\sfb}_5\sfd} \delta^{\sfa'\sfd}_{\bar{\sfc}_2}}{-c_2\,\delta_{\bar{\sfb}_5\bar{\sfc}_2}^{\bar{\sfa}_6\sfa'}}\biggr] \!\!\!\!\!\! & 0 & 0 & 0 & 0 & 0 & 0 \\
 0 & 0 & 0 & \!\!\!\!\!\! -\delta_{(\alpha_1}^{\beta}\delta_{\alpha_2)}^{\gamma} \delta_{\bar{\sfb}_5\bar{\sfc}_2}^{\bar{\sfa}_7} \!\!\!\!\!\! & 0 & 0 & 0 & 0 & 0 & 0 \\
 0 & 0 & 0 & 0 & \!\!\!\!\!\! \delta^\gamma_\alpha \biggl[\genfrac{}{}{0pt}{1}{-\delta^{\bar{\sfa}_7}_{\bar{\sfb}_6\sfd} \delta^{\sfd\sfe}_{\bar{\sfc}_2} \delta^{\bar{\sfa}'_2}_{\sfe\sfb'}}{-c_2\,\delta^{\bar{\sfa}_7}_{\bar{\sfb}_6\sfb'} \delta^{\bar{\sfa}'_2}_{\bar{\sfc}_2}}\biggr] & -\delta^{(\beta_1}_\alpha \epsilon^{\beta_2)\gamma} \delta^{\bar{\sfa}_7}_{\bar{\sfb}_7}\delta^{\bar{\sfa}'_2}_{\bar{\sfc}_2} \!\!\!\!\!\! & 0 & 0 & 0 & 0 \\
 0 & 0 & 0 & 0 & 0 & 0 & \!\!\!\!\!\! -\epsilon^{\beta\gamma}\delta_{\bar{\sfb}_7}^{\bar{\sfa}_7}\delta_{\bar{\sfb}'_2\bar{\sfc}_2}^{\bar{\sfa}'_4} \!\!\! & 0 & 0 & 0 \\
 0 & 0 & 0 & 0 & 0 & 0 & 0 & \!\!\!\!\!\! -\delta^\gamma_\alpha\delta_{\bar{\sfb}_7}^{\bar{\sfa}_7}\delta_{\bar{\sfb}'_4\bar{\sfc}_2}^{\bar{\sfa}'_6} \!\!\! & 0 & 0 \\
 0 & 0 & 0 & 0 & 0 & 0 & 0 & 0 & \!\!\! -\epsilon^{\beta\gamma}\delta_{\bar{\sfb}_7}^{\bar{\sfa}_7}\delta_{\bar{\sfb}'_6\sfd}^{\bar{\sfa}'_7}\delta_{\bar{\sfc}_2}^{\sfa\dprime\sfd} & 0 
\end{pmatrix}}}
\nn\\
 &\quad \bigl[c_2 \equiv \tfrac{2}{7}+\tfrac{1}{7\sqrt{2}}\bigr]\,,
\\
 R_\gamma^{\bar{\sfc}_2}
 &\equiv {\footnotesize{\arraycolsep=0.2mm 
 \begin{pmatrix}
 0 & \delta_\gamma^\beta\delta_{\sfb\sfa}^{\bar{\sfc}_2} & 0 & 0 & 0 & 0 & 0 & 0 & 0 & 0 \\
 0 & 0 & \!\!\!\epsilon_{\alpha\gamma}\delta_{\bar{\sfb}_3}^{\sfa\bar{\sfc}_2}\!\!\! & 0 & 0 & 0 & 0 & 0 & 0 & 0 \\
 0 & 0 & 0 & \delta_\gamma^\beta\delta_{\bar{\sfb}_5}^{\bar{\sfa}_3\bar{\sfc}_2} & 0 & 0 & 0 & 0 & 0 & 0 \\
 0 & 0 & 0 & 0 & \!\!\!\!\!\! \epsilon_{\alpha\gamma} \biggl[\genfrac{}{}{0pt}{1}{\delta_{\bar{\sfb}_6}^{\bar{\sfa}_5\sfd} \delta_{\sfb'\sfd}^{\bar{\sfc}_2}}{+c_2\,\delta^{\bar{\sfa}_5\bar{\sfc}_2}_{\bar{\sfb}_6\sfb'}}\biggr] & \delta^{(\beta_1}_{(\alpha} \delta^{\beta_2)}_{\gamma)} \delta^{\bar{\sfa}_5\bar{\sfc}_2}_{\bar{\sfb}_7} \!\!\! & 0 & 0 & 0 & 0 \\
 0 & 0 & 0 & 0 & 0 & 0 & \!\!\!\!\!\!\delta_\gamma^\beta \biggl[\genfrac{}{}{0pt}{1}{\delta_{\bar{\sfb}_7}^{\bar{\sfa}_6\sfd} \delta_{\sfd\sfe}^{\bar{\sfc}_2} \delta_{\bar{\sfb}'_2}^{\sfe\sfa'}}{+c_2\,\delta_{\bar{\sfb}_7}^{\bar{\sfa}_6\sfa'} \delta_{\bar{\sfb}'_2}^{\bar{\sfc}_2}}\biggr]\!\!\!\!\!\! & 0 & 0 & 0 \\
 0 & 0 & 0 & 0 & 0 & 0 & \!\!\!\!\!\! \delta_{(\alpha_1}^\beta \epsilon_{\alpha_2)\gamma} \delta_{\bar{\sfb}_7}^{\bar{\sfa}_7}\delta_{\bar{\sfb}'_2}^{\bar{\sfc}_2}\!\!\!\!\!\! & 0 & 0 & 0 \\
 0 & 0 & 0 & 0 & 0 & 0 & 0 & \epsilon_{\alpha\gamma}\delta^{\bar{\sfa}_7}_{\bar{\sfb}_7}\delta^{\bar{\sfa}'_2\bar{\sfc}_2}_{\bar{\sfb}'_4} & 0 & 0 \\
 0 & 0 & 0 & 0 & 0 & 0 & 0 & 0 & \!\!\!\!\!\! \delta_\gamma^\beta\delta^{\bar{\sfa}_7}_{\bar{\sfb}_7}\delta^{\bar{\sfa}'_4\bar{\sfc}_2}_{\bar{\sfb}'_6} \!\!\!\!\!\! & 0 \\
 0 & 0 & 0 & 0 & 0 & 0 & 0 & 0 & 0 & \epsilon_{\alpha\gamma}\delta^{\bar{\sfa}_7}_{\bar{\sfb}_7}\delta^{\bar{\sfa}'_6\sfd}_{\bar{\sfb}'_7}\delta^{\bar{\sfc}_2}_{\sfb\dprime\sfd} \\
 0 & 0 & 0 & 0 & 0 & 0 & 0 & 0 & 0 & 0 
\end{pmatrix}}} ,
\\
 R_{\bar{\sfc}_4} &\equiv {\footnotesize {\arraycolsep=1mm 
 \begin{pmatrix}
 0 & 0 & 0 & 0 & 0 & 0 & 0 & 0 & 0 & 0 \\
 0 & 0 & 0 & 0 & 0 & 0 & 0 & 0 & 0 & 0 \\
 -\delta_{\bar{\sfc}_4}^{\bar{\sfa}_3\sfb} & 0 & 0 & 0 & 0 & 0 & 0 & 0 & 0 & 0 \\
 0 & \delta^\beta_\alpha \delta_{\sfb\bar{\sfc}_4}^{\bar{\sfa}_5} & 0 & 0 & 0 & 0 & 0 & 0 & 0 & 0 \\
 0 & 0 & -\biggl[\genfrac{}{}{0pt}{1}{ \delta_{\bar{\sfb}_3\bar{\sfd}_3}^{\bar{\sfa}_6}\delta_{\bar{\sfc}_4}^{\sfa'\bar{\sfd}_3}}{+c_4\,\delta_{\bar{\sfb}_3\bar{\sfc}_4}^{\bar{\sfa}_6\sfa'}}\biggr] & 0 & 0 & 0 & 0 & 0 & 0 & 0 \\
 0 & 0 & 0 & 0 & 0 & 0 & 0 & 0 & 0 & 0 \\
 0 & 0 & 0 & \!\!\!\!\!\! -\delta^\beta_\alpha\delta_{\bar{\sfb}_5\bar{\sfd}_2}^{\bar{\sfa}_7}\delta_{\bar{\sfc}_4}^{\bar{\sfd}_2\bar{\sfa}'_2} \!\!\!\!\!\! & 0 & 0 & 0 & 0 & 0 & 0 \\
 0 & 0 & 0 & 0 & \!\!\! -\biggl[\genfrac{}{}{0pt}{1}{ \delta_{\bar{\sfb}_6\sfd}^{\bar{\sfa}_7}\delta_{\bar{\sfe}_3\sfb'}^{\bar{\sfa}'_4}\delta_{\bar{\sfc}_4}^{\sfd\bar{\sfe}_3}}{+c_4\,\delta_{\bar{\sfb}_6\sfb'}^{\bar{\sfa}_7}\delta_{\bar{\sfc}_4}^{\bar{\sfa}'_4}}\biggr] \!\!\! & 0 & 0 & 0 & 0 & 0 \\
 0 & 0 & 0 & 0 & 0 & 0 & \delta^\beta_\alpha\delta^{\bar{\sfa}_7}_{\bar{\sfb}_7}\delta^{\bar{\sfa}'_6}_{\bar{\sfb}'_2\bar{\sfc}_4} & 0 & 0 & 0 \\
 0 & 0 & 0 & 0 & 0 & 0 & 0 & -\delta^{\bar{\sfa}_7}_{\bar{\sfb}_7}\delta^{\sfa'\bar{\sfd}_3}_{\bar{\sfb}'_4}\delta^{\bar{\sfa}'_7}_{\bar{\sfd}_3\bar{\sfc}_4} & 0 & 0 
\end{pmatrix}}}
\nn\\
 &\quad \bigl[c_4\equiv \tfrac{4}{7} + \tfrac{2}{7\sqrt{2}}\bigr]\,,
\\
 R^{\bar{\sfc}_4} &\equiv {\footnotesize {\arraycolsep=1mm 
 \begin{pmatrix}
 0 & 0 & \delta^{\bar{\sfc}_4}_{\bar{\sfb}_3\sfa} & 0 & 0 & 0 & 0 & 0 & 0 & 0 \\
 0 & 0 & 0 & -\delta_\alpha^\beta \delta^{\sfa\bar{\sfc}_4}_{\bar{\sfb}_5} & 0 & 0 & 0 & 0 & 0 & 0 \\
 0 & 0 & 0 & 0 & \!\!\! \biggl[\genfrac{}{}{0pt}{1}{ \delta^{\bar{\sfa}_3\bar{\sfd}_3}_{\bar{\sfb}_6}\delta^{\bar{\sfc}_4}_{\sfb'\bar{\sfd}_3}}{+c_4\,\delta^{\bar{\sfa}_3\bar{\sfc}_4}_{\bar{\sfb}_6\sfb'}}\biggr] & 0 & 0 & 0 & 0 & 0 \\
 0 & 0 & 0 & 0 & 0 & 0 & \delta_\alpha^\beta\delta^{\bar{\sfa}_5\bar{\sfd}_2}_{\bar{\sfb}_7}\delta^{\bar{\sfc}_4}_{\bar{\sfd}_2\bar{\sfb}'_2} & 0 & 0 & 0 \\
 0 & 0 & 0 & 0 & 0 & 0 & 0 & \!\!\! \biggl[\genfrac{}{}{0pt}{1}{ \delta^{\bar{\sfa}_6\sfd}_{\bar{\sfb}_7}\delta^{\bar{\sfe}_3\sfa'}_{\bar{\sfb}'_4}\delta^{\bar{\sfc}_4}_{\sfd\bar{\sfe}_3}}{+c_4\,\delta^{\bar{\sfa}_6\sfa'}_{\bar{\sfb}_7}\delta^{\bar{\sfc}_4}_{\bar{\sfb}'_4}}\biggr] \!\!\! & 0 & 0 \\
 0 & 0 & 0 & 0 & 0 & 0 & 0 & 0 & 0 & 0 \\
 0 & 0 & 0 & 0 & 0 & 0 & 0 & 0 & \!\!\! -\delta_\alpha^\beta\delta_{\bar{\sfb}_7}^{\bar{\sfa}_7}\delta_{\bar{\sfb}'_6}^{\bar{\sfa}'_2\bar{\sfc}_4} \!\!\! & 0 \\
 0 & 0 & 0 & 0 & 0 & 0 & 0 & 0 & 0 & \delta_{\bar{\sfb}_7}^{\bar{\sfa}_7}\delta_{\sfb'\bar{\sfd}_3}^{\bar{\sfa}'_4}\delta_{\bar{\sfb}'_7}^{\bar{\sfd}_3\bar{\sfc}_4} \\
 0 & 0 & 0 & 0 & 0 & 0 & 0 & 0 & 0 & 0 \\
 0 & 0 & 0 & 0 & 0 & 0 & 0 & 0 & 0 & 0 
\end{pmatrix}}} ,
\\
 R^\gamma_{\bar{\sfc}_6} &\equiv {\footnotesize {\arraycolsep=0.4mm 
 \begin{pmatrix}
 0 & 0 & 0 & 0 & 0 & 0 & 0 & 0 & 0 & 0 \\[-1mm]
 0 & 0 & 0 & 0 & 0 & 0 & 0 & 0 & 0 & 0 \\[-1mm]
 0 & 0 & 0 & 0 & 0 & 0 & 0 & 0 & 0 & 0 \\[-1mm]
 -\delta^\gamma_\alpha\delta_{\bar{\sfc}_6}^{\bar{\sfa}_5\sfb} & 0 & 0 & 0 & 0 & 0 & 0 & 0 & 0 & 0 \\[-0.5mm]
 0 & \!\!\! \epsilon^{\beta\gamma}\biggl[\genfrac{}{}{0pt}{1}{\delta_{\bar{\sfc}_6}^{\bar{\sfa}_6}\delta^{\sfa'}_\sfb}{-c_6\,\delta_{\sfb\bar{\sfc}_6}^{\bar{\sfa}_6\sfa'}}\biggr] \!\!\! & 0 & 0 & 0 & 0 & 0 & 0 & 0 & 0 \\[-0.5mm]
 0 & \!\!\! \delta_{(\alpha_1}^{(\beta} \delta_{\alpha_2)}^{\gamma)}\delta_{\sfb\bar{\sfc}_6}^{\bar{\sfa}_7} \!\!\! & 0 & 0 & 0 & 0 & 0 & 0 & 0 & 0 \\[-0.5mm]
 0 & 0 & \delta^\gamma_\alpha\delta_{\bar{\sfc}_6\sfd}^{\bar{\sfa}_7}\delta_{\bar{\sfb}_3}^{\bar{\sfa}'_2\sfd} & 0 & 0 & 0 & 0 & 0 & 0 & 0 \\[-0.5mm]
 0 & 0 & 0 & \!\!\! -\epsilon^{\beta\gamma}\delta_{\bar{\sfc}_6\sfd}^{\bar{\sfa}_7}\delta_{\bar{\sfb}_5}^{\bar{\sfa}'_4\sfd} \!\!\! & 0 & 0 & 0 & 0 & 0 & 0 \\[-0.5mm]
 0 & 0 & 0 & 0 & \!\!\! \delta^\gamma_\alpha\biggl[\genfrac{}{}{0pt}{1}{-\delta_{\bar{\sfb}_6}^{\bar{\sfa}'_6}\delta_{\sfb'\bar{\sfc}_6}^{\bar{\sfa}_7}}{+c_6\,\delta_{\bar{\sfb}_6\sfb'}^{\bar{\sfa}_7}\delta_{\bar{\sfc}_6}^{\bar{\sfa}'_6}}\biggr] & -\delta_\alpha^{(\beta_1} \epsilon^{\beta_2)\gamma} \delta_{\bar{\sfb}_7}^{\bar{\sfa}_7}\delta_{\bar{\sfc}_6}^{\bar{\sfa}'_6} \!\!\! & 0 & 0 & 0 & 0 \\
 0 & 0 & 0 & 0 & 0 & 0 & \!\!\!\!\!\! -\epsilon^{\beta\gamma}\delta_{\bar{\sfb}_7}^{\bar{\sfa}_7}\delta^{\bar{\sfa}'_7}_{\bar{\sfc}_6\sfd}\delta_{\bar{\sfb}'_2}^{\sfa\dprime\sfd} & 0 & 0 & 0 
\end{pmatrix}}} ,
\\
 R_\gamma^{\bar{\sfc}_6} &\equiv {\footnotesize {\arraycolsep=0.4mm 
 \begin{pmatrix}
 0 & 0 & 0 & \delta_\gamma^\beta \delta^{\bar{\sfc}_6}_{\bar{\sfb}_5\sfa} & 0 & 0 & 0 & 0 & 0 & 0 \\[-0.5mm]
 0 & 0 & 0 & 0 & \!\!\! \epsilon_{\alpha\gamma}\biggl[\genfrac{}{}{0pt}{1}{-\delta^{\bar{\sfc}_6}_{\bar{\sfb}_6}\delta_{\sfb'}^\sfa}{+c_6\,\delta^{\sfa\bar{\sfc}_6}_{\bar{\sfb}_6\sfb'}}\biggr] & -\delta_{(\alpha}^{(\beta_1}\delta^{\beta_2)}_{\gamma)}\delta^{\sfa\bar{\sfc}_6}_{\bar{\sfb}_7} \!\!\! & 0 & 0 & 0 & 0 \\[-0.5mm]
 0 & 0 & 0 & 0 & 0 & 0 & \!\!\!\!\!\! -\delta_\gamma^\beta\delta^{\bar{\sfc}_6\sfd}_{\bar{\sfb}_7}\delta^{\bar{\sfa}_3}_{\bar{\sfb}'_2\sfd} \!\!\! & 0 & 0 & 0 \\[-0.5mm]
 0 & 0 & 0 & 0 & 0 & 0 & 0 & \epsilon_{\alpha\gamma}\delta^{\bar{\sfc}_6\sfd}_{\bar{\sfb}_7}\delta^{\bar{\sfa}_5}_{\bar{\sfb}'_4\sfd} & 0 & 0 \\[-0.5mm]
 0 & 0 & 0 & 0 & 0 & 0 & 0 & 0 & \!\!\! \delta_\gamma^\beta\biggl[\genfrac{}{}{0pt}{1}{\delta^{\bar{\sfa}_6}_{\bar{\sfb}'_6}\delta^{\sfa'\bar{\sfc}_6}_{\bar{\sfb}_7}}{-c_6\,\delta^{\bar{\sfa}_6\sfa'}_{\bar{\sfb}_7}\delta^{\bar{\sfc}_6}_{\bar{\sfb}'_6}}\biggr] \!\!\! & 0 \\[-0.5mm]
 0 & 0 & 0 & 0 & 0 & 0 & 0 & 0 & \!\!\! \delta^\beta_{(\alpha_1} \epsilon_{\alpha_2)\gamma} \delta^{\bar{\sfa}_7}_{\bar{\sfb}_7}\delta^{\bar{\sfc}_6}_{\bar{\sfb}'_6} \!\!\! & 0 \\[-0.5mm]
 0 & 0 & 0 & 0 & 0 & 0 & 0 & 0 & 0 & \epsilon_{\alpha\gamma}\delta^{\bar{\sfa}_7}_{\bar{\sfb}_7}\delta_{\bar{\sfb}'_7}^{\bar{\sfc}_6\sfd}\delta^{\bar{\sfa}'_2}_{\sfb\dprime\sfd} \\[-1mm]
 0 & 0 & 0 & 0 & 0 & 0 & 0 & 0 & 0 & 0 \\[-1mm]
 0 & 0 & 0 & 0 & 0 & 0 & 0 & 0 & 0 & 0 \\[-1mm]
 0 & 0 & 0 & 0 & 0 & 0 & 0 & 0 & 0 & 0 
\end{pmatrix}}} ,
\\
 R_{\bar{\sfc}_7,\sfc'} &\equiv {\footnotesize {\arraycolsep=1mm 
 \begin{pmatrix}
 0 & 0 & 0 & 0 & 0 & 0 & 0 & 0 & 0 & 0 \\[-1mm]
 0 & 0 & 0 & 0 & 0 & 0 & 0 & 0 & 0 & 0 \\[-1mm]
 0 & 0 & 0 & 0 & 0 & 0 & 0 & 0 & 0 & 0 \\[-1mm]
 0 & 0 & 0 & 0 & 0 & 0 & 0 & 0 & 0 & 0 \\[-1mm]
 \Bigl[\genfrac{}{}{0pt}{1}{\delta_{\bar{\sfc}_7}^{\sfb\bar{\sfa}_6}\delta_{\sfc'}^{\sfa'}}{-c_{7,1}\,\delta_{\bar{\sfc}_7}^{\bar{\sfa}_6\sfa'}\delta^\sfb_{\sfc'}}\Bigr] & 0 & 0 & 0 & 0 & 0 & 0 & 0 & 0 & 0 \\[-1mm]
 0 & 0 & 0 & 0 & 0 & 0 & 0 & 0 & 0 & 0 \\[-1mm]
 0 & -\delta^\beta_\alpha \delta_{\bar{\sfc}_7}^{\bar{\sfa}_7} \delta_{\sfb\sfc'}^{\bar{\sfa}'_2} & 0 & 0 & 0 & 0 & 0 & 0 & 0 & 0 \\[-1mm]
 0 & 0 & - \delta_{\bar{\sfc}_7}^{\bar{\sfa}_7}\delta_{\bar{\sfb}_3\sfc'}^{\bar{\sfa}'_4} & 0 & 0 & 0 & 0 & 0 & 0 & 0 \\[-1mm]
 0 & 0 & 0 & -\delta^\beta_\alpha \delta_{\bar{\sfc}_7}^{\bar{\sfa}_7} \delta_{\bar{\sfb}_5\sfc'}^{\bar{\sfa}'_6} & 0 & 0 & 0 & 0 & 0 & 0 \\[-1mm]
 0 & 0 & 0 & 0 & \biggl[\genfrac{}{}{0pt}{1}{\delta_{\bar{\sfc}_7}^{\bar{\sfa}_7}\delta_{\bar{\sfb}_6\sfc'}^{\bar{\sfa}'_7}\delta_{\sfb'}^{\sfa\dprime}}{-c_{7,1}\,\delta_{\bar{\sfc}_7}^{\bar{\sfa}_7}\delta_{\bar{\sfb}_6\sfb'}^{\bar{\sfa}'_7}\delta^{\sfa\dprime}_{\sfc'}}
\biggr] & 0 & 0 & 0 & 0 & 0 
\end{pmatrix}}} ,
\\
 R^{\bar{\sfc}_7,\sfc'} &\equiv {\footnotesize {\arraycolsep=1mm 
 \begin{pmatrix}
 0 & 0 & 0 & 0 & \biggl[\genfrac{}{}{0pt}{1}{-\delta^{\bar{\sfc}_7}_{\sfa\bar{\sfb}_6}\delta^{\sfc'}_{\sfb'}}{+c_{7,1}\,\delta^{\bar{\sfc}_7}_{\bar{\sfb}_6\sfb'}\delta_\sfa^{\sfc'}}\biggr] & 0 & 0 & 0 & 0 & 0 \\[-1mm]
 0 & 0 & 0 & 0 & 0 & 0 & \delta_\alpha^\beta \delta^{\bar{\sfc}_7}_{\bar{\sfb}_7}\delta^{\sfa\sfc'}_{\bar{\sfb}'_2} & 0 & 0 & 0 \\[-1mm]
 0 & 0 & 0 & 0 & 0 & 0 & 0 & \delta^{\bar{\sfc}_7}_{\bar{\sfb}_7}\delta^{\bar{\sfa}_3\sfc'}_{\bar{\sfb}'_4} & 0 & 0 \\[-1mm]
 0 & 0 & 0 & 0 & 0 & 0 & 0 & 0 & \delta_\alpha^\beta \delta^{\bar{\sfc}_7}_{\bar{\sfb}_7}\delta^{\bar{\sfa}_5\sfc'}_{\bar{\sfb}'_6} & 0 \\[-1mm]
 0 & 0 & 0 & 0 & 0 & 0 & 0 & 0 & 0 & \biggl[\genfrac{}{}{0pt}{1}{-\delta^{\bar{\sfc}_7}_{\bar{\sfb}_7}\delta^{\bar{\sfa}_6\sfc'}_{\bar{\sfb}'_7}\delta^{\sfa'}_{\sfb\dprime}}{+c_{7,1}\,\delta^{\bar{\sfc}_7}_{\bar{\sfb}_7}\delta^{\bar{\sfa}_6\sfa'}_{\bar{\sfb}'_7}\delta_{\sfb\dprime}^{\sfc'}}\biggr] \\[-1mm]
 0 & 0 & 0 & 0 & 0 & 0 & 0 & 0 & 0 & 0 \\[-1mm]
 0 & 0 & 0 & 0 & 0 & 0 & 0 & 0 & 0 & 0 \\[-1mm]
 0 & 0 & 0 & 0 & 0 & 0 & 0 & 0 & 0 & 0 \\[-1mm]
 0 & 0 & 0 & 0 & 0 & 0 & 0 & 0 & 0 & 0 \\[-1mm]
 0 & 0 & 0 & 0 & 0 & 0 & 0 & 0 & 0 & 0 
\end{pmatrix}}} 
\nn\\
 &\quad \Bigl[ c_6 \equiv \tfrac{1}{7}-\tfrac{3}{7\sqrt{2}}\qquad c_{7,1}\equiv \tfrac{1}{7}+\tfrac{4}{7\sqrt{2}}\Bigr]\,.
\end{align}

\subsection{Explicit matrix form of $\chi$}
\label{app:chi}

We here determine the explicit form of the matrices $\chi_{\bm{\alpha}B}$ and $\chi^{A\bm{\beta}}$ by requiring
\begin{align}
 (t_{\bm{\alpha}})_A{}^B = f_{\bm{\alpha}\bm{\beta}}{}^{\bm{\gamma}}\,\chi_{\bm{\gamma}A}\,\chi^{B\bm{\beta}}\,. 
\end{align}

In the M-theory picture, the adjoint index $\bm{\alpha}$ and the vector index $B$ are decomposed as
\begin{align}
\begin{split}
 (\chi_{\bm{\alpha}B})
 &= \bigl(\chi_{\bm{\alpha}b},\, \tfrac{\chi_{\bm{\alpha}}{}^{b_{12}}}{\sqrt{2!}} ,\, \tfrac{\chi_{\bm{\alpha}}{}^{b_{1\cdots 5}}}{\sqrt{5!}} ,\, \tfrac{\chi_{\bm{\alpha}}{}^{b_{1\cdots 7},b}}{\sqrt{7!}} ,\, \tfrac{\chi_{\bm{\alpha}}{}^{8,b_{123}}}{\sqrt{3!}} ,\, \tfrac{\chi_{\bm{\alpha}}{}^{8,b_{1\cdots 6}}}{\sqrt{6!}} ,\, \chi_{\bm{\alpha}}{}^{8,8,b}\bigr)\,,
\\
 (\chi_{\bm{\alpha}B})
 &= \bigl(\chi_{R_{8,a}B},\, 
 \tfrac{\chi_{R_{a_{1\cdots 6}}B}}{\sqrt{6!}},\, 
 \tfrac{\chi_{R_{a_{123}}B}}{\sqrt{3!}},\, 
 \chi_{K^a{}_cB},\, 
 \tfrac{\chi_{R^{a_{123}}B}}{\sqrt{3!}},\, 
 \tfrac{\chi_{R^{a_{1\cdots 6}}B}}{\sqrt{6!}},\, 
 \chi_{R^{8,a}B} \bigr)\,.
\end{split}
\end{align}
Under this decomposition, components of $(\chi_{\bm{\alpha}B})$ and $(\chi^{A\bm{\beta}})$ are determined as follows:
\begin{align}
 (\chi_{\bm{\alpha}B}) &= \begin{pmatrix}
 0 & 0 & 0 & 0 & 0 & 0 & \delta_a^b \\
 0 & 0 & 0 & 0 & 0 & \delta_{a_{1\cdots 6}}^{b_{1\cdots 6}} & 0 \\
 0 & 0 & 0 & 0 & \delta_{a_{123}}^{b_{123}} & 0 & 0 \\
 0 & 0 & 0 & \frac{\epsilon^{ab_{1\cdots 7}}\delta_c^b - \frac{1}{4} \,\epsilon^{b_{1\cdots 7}b}\delta_c^a}{\sqrt{7!}} & 0 & 0 & 0 \\
 0 & 0 & \frac{-\epsilon^{a_{123}b_{1\cdots 5}}}{\sqrt{3!\,5!}} & 0 & 0 & 0 & 0 \\
 0 & \frac{\epsilon^{a_{1\cdots 6}b_{12}}}{\sqrt{6!\,2!}} & 0 & 0 & 0 & 0 & 0 \\
 \delta^a_b & 0 & 0 & 0 & 0 & 0 & 0
\end{pmatrix}.
\\
 (\chi^{A\bm{\beta}}) &= \begin{pmatrix}
 0 & 0 & 0 & 0 & 0 & 0 & \delta^a_b \\
 0 & 0 & 0 & 0 & 0 & \frac{\epsilon_{a_{12}b_{1\cdots 6}}}{\sqrt{2!\,6!}} & 0 \\
 0 & 0 & 0 & 0 & \frac{\epsilon_{a_{1\cdots 5}b_{123}}}{\sqrt{5!\,3!}} & 0 & 0 \\
 0 & 0 & 0 & \frac{\epsilon_{ba_{1\cdots 7}}\delta^d_{a'} + \frac{1}{12} \,\epsilon_{a_{1\cdots 7}a'}\delta^d_b}{\sqrt{7!}} & 0 & 0 & 0 \\
 0 & 0 & \delta_{a_{123}}^{b_{123}} & 0 & 0 & 0 & 0 \\
 0 & \delta_{a_{1\cdots 6}}^{b_{1\cdots 6}} & 0 & 0 & 0 & 0 & 0 \\
 \delta^b_a & 0 & 0 & 0 & 0 & 0 & 0
\end{pmatrix},
\end{align}
where $\epsilon^{1\cdots 8}=\epsilon_{1\cdots8}=1$\,.

In the type IIB picture, we consider the following decomposition:
\begin{align}
\begin{split}
 (\chi_{\bm{\alpha}B})
 &= \bigl(\chi_{\bm{\alpha}\sfb},\, \chi_{\bm{\alpha}}{}^{\sfb}_\beta,\, \tfrac{\chi_{\bm{\alpha}}{}^{\sfb_{123}}}{\sqrt{3!}},\, \tfrac{\chi_{\bm{\alpha}}{}^{\sfb_{1\cdots 5}}_\beta}{\sqrt{5!}},\, \tfrac{\chi_{\bm{\alpha}}{}^{\sfb_{1\cdots 6},\sfb}}{\sqrt{6!}},\, \chi_{\bm{\alpha}}{}^7_{\beta_{(12)}} ,\,
 \tfrac{\chi_{\bm{\alpha}}{}^{7,\sfb_{12}}_\beta}{\sqrt{2!}},\, \tfrac{\chi_{\bm{\alpha}}{}^{7,\sfb_{1\cdots 4}}}{\sqrt{4!}},\, \tfrac{\chi_{\bm{\alpha}}{}^{7,\sfb_{1\cdots 6}}_\beta}{\sqrt{6!}},\, \chi_{\bm{\alpha}}{}^{7,7,\sfb}\bigr)\,,
\\
 (\chi_{\bm{\alpha}B})
 &= \bigl( \chi_{R_{7,\sfa}B} ,\, 
 \tfrac{\chi_{R_{\sfa_{1\cdots 6}}^\alpha B}}{\sqrt{6!}},\, 
 \tfrac{\chi_{R_{\sfa_{1\cdots 4}} B}}{\sqrt{4!}},\, 
 \tfrac{\chi_{R_{\sfa_{12}}^\alpha B}}{\sqrt{2!}},\, 
 \chi_{R^{\alpha_1}{}_{\alpha_2} B} ,\, 
 \chi_{K^{\sfc}{}_{\sfd} B} ,\, 
 \tfrac{\chi_{R^{\sfa_{12}}_\alpha B}}{\sqrt{2!}},\, 
 \tfrac{\chi_{R^{\sfa_{1\cdots 4}} B}}{\sqrt{4!}},\, 
 \tfrac{\chi_{R^{\sfa_{1\cdots 6}}_\alpha B}}{\sqrt{6!}},\, 
 \chi_{R^{7,\sfa}B} \bigr) \,.
\end{split}
\end{align}
Then, the matrices $(\chi_{\bm{\alpha}B})$ and $(\chi^{A\bm{\beta}})$ are determined as
\begin{align}
 (\chi_{\bm{\alpha}B})
 &={\tiny \arraycolsep=0mm \begin{pmatrix}
 0 & 0 & 0 & 0 & 0 & 0 & 0 & 0 & 0 & \delta_{\sfa}^{\sfb} \\
 0 & 0 & 0 & 0 & 0 & 0 & 0 & 0 & \delta^\alpha_\beta\,\delta_{\sfa_{1\cdots6}}^{\sfb_{1\cdots6}} & 0 \\
 0 & 0 & 0 & 0 & 0 & 0 & 0 & \delta_{\sfa_{1\cdots4}}^{\sfb_{1\cdots4}} & 0 & 0 \\
 0 & 0 & 0 & 0 & 0 & 0 & \delta^\alpha_\beta\,\delta_{\sfa_{12}}^{\sfb_{12}} & 0 & 0 & 0 \\
 0 & 0 & 0 & 0 & 0 & \!\!\!\delta_{(\beta_1}^{\alpha_1}\,\epsilon_{\beta_2)\alpha_2}\!\!\! & 0 & 0 & 0 & 0 \\
 0 & 0 & 0 & 0 & \!\!\!\frac{\epsilon^{\sfa\sfb_{1\cdots 6}}\delta_{\sfc}^{\sfb'} - c_{7,1}\,\epsilon^{\sfb_{1\cdots 6}\sfb}\delta_{\sfc}^{\sfa}}{\sqrt{6!}}\!\!\! & 0 & 0 & 0 & 0 & 0 \\
 0 & 0 & 0 & \!\!\!\frac{-\epsilon_{\alpha\beta}\,\epsilon^{\sfa_{12}\sfb_{1\cdots5}}}{\sqrt{2!\,5!}}\!\!\! & 0 & 0 & 0 & 0 & 0 & 0 \\
 0 & 0 & \frac{\epsilon^{\sfa_{1\cdots 4}\sfb_{123}}}{\sqrt{4!\,3!}} & 0 & 0 & 0 & 0 & 0 & 0 & 0 \\
 0 & \frac{\epsilon_{\alpha\beta}\,\epsilon^{\sfa_{1\cdots 6}\sfb}}{\sqrt{6!}} & 0 & 0 & 0 & 0 & 0 & 0 & 0 & 0 \\
 \delta^{\sfa}_{\sfb} & 0 & 0 & 0 & 0 & 0 & 0 & 0 & 0 & 0 \\
\end{pmatrix}}.
\\
 (\chi^{A\bm{\beta}})
 &={\tiny \arraycolsep=0mm \begin{pmatrix}
 0 & 0 & 0 & 0 & 0 & 0 & 0 & 0 & 0 & \delta^{\sfa}_{\sfb} \\
 0 & 0 & 0 & 0 & 0 & 0 & 0 & 0 & -\frac{\epsilon^{\alpha\beta}\,\epsilon_{\sfa\sfb_{1\cdots 6}}}{\sqrt{6!}} & 0 \\
 0 & 0 & 0 & 0 & 0 & 0 & 0 & \frac{\epsilon_{\sfa_{123}\sfb_{1\cdots 4}}}{\sqrt{3!\,4!}} & 0 & 0 \\
 0 & 0 & 0 & 0 & 0 & 0 & \!\!\!\frac{\epsilon^{\alpha\beta}\,\epsilon_{\sfa_{1\cdots5}\sfa_{12}}}{\sqrt{5!\,2!}}\!\!\! & 0 & 0 & 0 \\
 0 & 0 & 0 & 0 & 0 & \!\!\!\frac{\epsilon_{\sfa_{1\cdots 6}\sfb}\delta^{\sfd}_{\sfa'} - \frac{1+\frac{1}{2\sqrt{2}}}{7} \,\epsilon_{\sfa_{1\cdots 6}\sfa'}\delta^{\sfd}_{\sfb}}{\sqrt{6!}}\!\!\! & 0 & 0 & 0 & 0 \\
 0 & 0 & 0 & 0 & \!\!\!\delta^{(\alpha_1}_{\beta_1}\,\epsilon^{\alpha_2)\beta_2}\!\!\! & 0 & 0 & 0 & 0 & 0 \\
 0 & 0 & 0 & \delta^\alpha_\beta\,\delta_{\sfa_{12}}^{\sfb_{12}} & 0 & 0 & 0 & 0 & 0 & 0 \\
 0 & 0 & \delta_{\sfa_{1\cdots4}}^{\sfb_{1\cdots4}} & 0 & 0 & 0 & 0 & 0 & 0 & 0 \\
 0 & \delta^\alpha_\beta\,\delta_{\sfa_{1\cdots6}}^{\sfb_{1\cdots6}} & 0 & 0 & 0 & 0 & 0 & 0 & 0 & 0 \\
 \delta_{\sfa}^{\sfb} & 0 & 0 & 0 & 0 & 0 & 0 & 0 & 0 & 0 \\
\end{pmatrix}},
\end{align}
where $\epsilon^{1\cdots 7}=\epsilon_{1\cdots7}=1$\,.

In both the M-theory and the type IIB pictures, we can clearly see that the matrix vanishes when $n\leq 7$\,.


\begin{thebibliography}{99}
\bibitem{hep-th:9502122} 
  C.~Klimcik and P.~Severa,
  ``Dual nonAbelian duality and the Drinfeld double,''
  Phys.\ Lett.\ B {\bf 351}, 455 (1995)
  [hep-th/9502122].



\bibitem{hep-th:9509095} 
  C.~Klimcik,
  ``Poisson-Lie T duality,''
  Nucl.\ Phys.\ Proc.\ Suppl.\  {\bf 46}, 116 (1996)
  [hep-th/9509095].



\bibitem{1911.06320} 
  Y.~Sakatani,
  ``$U$-duality extension of Drinfel’d double,''
  PTEP {\bf 2020}, no. 2, 023B08 (2020)
  [arXiv:1911.06320 [hep-th]].



\bibitem{1911.07833} 
  E.~Malek and D.~C.~Thompson,
  ``Poisson-Lie U-duality in Exceptional Field Theory,''
  JHEP {\bf 2004}, 058 (2020)
  [arXiv:1911.07833 [hep-th]].



\bibitem{2006.12452} 
  C.~D.~A.~Blair, D.~C.~Thompson and S.~Zhidkova,
  ``Exploring Exceptional Drinfeld Geometries,''
  arXiv:2006.12452 [hep-th].



\bibitem{2001.09983} 
  Y.~Sakatani and S.~Uehara,
  ``Non-Abelian $U$-duality for membranes,''
  PTEP {\bf 2020}, no. 7, 073B01 (2020)
  [arXiv:2001.09983 [hep-th]].



\bibitem{2003.06164} 
  L.~Hlavaty,
  ``Classification of 6D Leibniz algebras,''
  PTEP {\bf 2020}, no. 7, 071B01 (2020)
  [arXiv:2003.06164 [hep-th]].



\bibitem{2007.01213} 
  E.~T.~Musaev,
  ``On non-abelian U-duality of 11D backgrounds,''
  arXiv:2007.01213 [hep-th].



\bibitem{2007.08510} 
  E.~Malek, Y.~Sakatani and D.~C.~Thompson,
  ``E$_{6(6)}$ Exceptional Drinfel'd Algebras,''
  arXiv:2007.08510 [hep-th].



\bibitem{hep-th:0104081} 
  P.~C.~West,
  ``E$_{11}$ and M theory,''
  Class.\ Quant.\ Grav.\  {\bf 18}, 4443 (2001)
  [hep-th/0104081].



\bibitem{hep-th:0107181} 
  I.~Schnakenburg and P.~C.~West,
  ``Kac-Moody symmetries of 2B supergravity,''
  Phys.\ Lett.\ B {\bf 517}, 421 (2001)
  [hep-th/0107181].



\bibitem{hep-th:0307098} 
  P.~C.~West,
  ``E$_{11}$, SL(32) and central charges,''
  Phys.\ Lett.\ B {\bf 575}, 333 (2003)
  [hep-th/0307098].



\bibitem{hep-th:0402140} 
  P.~C.~West,
  ``The IIA, IIB and eleven-dimensional theories and their common E$_{11}$ origin,''
  Nucl.\ Phys.\ B {\bf 693}, 76 (2004)
  [hep-th/0402140].



\bibitem{hep-th:9302036} 
  W.~Siegel,
  ``Two vierbein formalism for string inspired axionic gravity,''
  Phys.\ Rev.\ D {\bf 47}, 5453 (1993)
  [hep-th/9302036].



\bibitem{hep-th:9305073} 
  W.~Siegel,
  ``Superspace duality in low-energy superstrings,''
  Phys.\ Rev.\ D {\bf 48}, 2826 (1993)
  [hep-th/9305073].



\bibitem{hep-th:9308133} 
  W.~Siegel,
  ``Manifest duality in low-energy superstrings,''
  hep-th/9308133.



\bibitem{0904.4664} 
  C.~Hull and B.~Zwiebach,
  ``Double Field Theory,''
  JHEP {\bf 0909}, 099 (2009)
  [arXiv:0904.4664 [hep-th]].



\bibitem{1006.4823} 
  O.~Hohm, C.~Hull and B.~Zwiebach,
  ``Generalized metric formulation of double field theory,''
  JHEP {\bf 1008}, 008 (2010)
  [arXiv:1006.4823 [hep-th]].



\bibitem{1008.1763} 
  D.~S.~Berman and M.~J.~Perry,
  ``Generalized Geometry and M theory,''
  JHEP {\bf 1106}, 074 (2011)
  [arXiv:1008.1763 [hep-th]].



\bibitem{1111.0459} 
  D.~S.~Berman, H.~Godazgar, M.~J.~Perry and P.~West,
  ``Duality Invariant Actions and Generalised Geometry,''
  JHEP {\bf 1202}, 108 (2012)
  [arXiv:1111.0459 [hep-th]].



\bibitem{1206.7045} 
  P.~West,
  ``E11, generalised space-time and equations of motion in four dimensions,''
  JHEP {\bf 1212}, 068 (2012)
  [arXiv:1206.7045 [hep-th]].



\bibitem{1208.5884} 
  D.~S.~Berman, M.~Cederwall, A.~Kleinschmidt and D.~C.~Thompson,
  ``The gauge structure of generalised diffeomorphisms,''
  JHEP {\bf 1301}, 064 (2013)
  [arXiv:1208.5884 [hep-th]].



\bibitem{1308.1673} 
  O.~Hohm and H.~Samtleben,
  ``Exceptional Form of D=11 Supergravity,''
  Phys.\ Rev.\ Lett.\  {\bf 111}, 231601 (2013)
  [arXiv:1308.1673 [hep-th]].



\bibitem{1312.0614} 
  O.~Hohm and H.~Samtleben,
  ``Exceptional Field Theory I: $E_{6(6)}$ covariant Form of M-Theory and Type IIB,''
  Phys.\ Rev.\ D {\bf 89}, no. 6, 066016 (2014)
  [arXiv:1312.0614 [hep-th]].



\bibitem{1312.4542} 
  O.~Hohm and H.~Samtleben,
  ``Exceptional field theory. II. E$_{7(7)}$,''
  Phys.\ Rev.\ D {\bf 89}, 066017 (2014)
  [arXiv:1312.4542 [hep-th]].



\bibitem{1406.3348} 
  O.~Hohm and H.~Samtleben,
  ``Exceptional field theory. III. E$_{8(8)}$,''
  Phys.\ Rev.\ D {\bf 90}, 066002 (2014)
  [arXiv:1406.3348 [hep-th]].



\bibitem{1810.11446} 
  S.~Demulder, F.~Hassler and D.~C.~Thompson,
  ``Doubled aspects of generalised dualities and integrable deformations,''
  JHEP {\bf 1902}, 189 (2019)
  [arXiv:1810.11446 [hep-th]].



\bibitem{Tseytlin:1990nb} 
  A.~A.~Tseytlin,
  ``Duality Symmetric Formulation of String World Sheet Dynamics,''
  Phys.\ Lett.\ B {\bf 242}, 163 (1990).



\bibitem{Tseytlin:1990va} 
  A.~A.~Tseytlin,
  ``Duality symmetric closed string theory and interacting chiral scalars,''
  Nucl.\ Phys.\ B {\bf 350}, 395 (1991).



\bibitem{1111.1828} 
  N.~B.~Copland,
  ``A Double Sigma Model for Double Field Theory,''
  JHEP {\bf 1204}, 044 (2012)
  [arXiv:1111.1828 [hep-th]].



\bibitem{hep-th:9512040} 
  C.~Klimcik and P.~Severa,
  ``Poisson-Lie T duality and loop groups of Drinfeld doubles,''
  Phys.\ Lett.\ B {\bf 372}, 65 (1996)
  [hep-th/9512040].



\bibitem{hep-th:9605212} 
  C.~Klimcik and P.~Severa,
  ``NonAbelian momentum winding exchange,''
  Phys.\ Lett.\ B {\bf 383}, 281 (1996)
  [hep-th/9605212].



\bibitem{1508.05832} 
  C.~Klimcik,
  ``$\eta$ and $\lambda$ deformations as E -models,''
  Nucl.\ Phys.\ B {\bf 900}, 259 (2015)
  [arXiv:1508.05832 [hep-th]].



\bibitem{1208.1232} 
  M.~Hatsuda and K.~Kamimura,
  ``SL(5) duality from canonical M2-brane,''
  JHEP {\bf 1211}, 001 (2012)
  [arXiv:1208.1232 [hep-th]].



\bibitem{1509.02915} 
  M.~J.~Duff, J.~X.~Lu, R.~Percacci, C.~N.~Pope, H.~Samtleben and E.~Sezgin,
  ``Membrane Duality Revisited,''
  Nucl.\ Phys.\ B {\bf 901}, 1 (2015)
  [arXiv:1509.02915 [hep-th]].



\bibitem{0807.4527} 
  M.~Grana, R.~Minasian, M.~Petrini and D.~Waldram,
  ``T-duality, Generalized Geometry and Non-Geometric Backgrounds,''
  JHEP {\bf 0904}, 075 (2009)
  [arXiv:0807.4527 [hep-th]].



\bibitem{1401.3360} 
  K.~Lee, C.~Strickland‐Constable and D.~Waldram,
  ``Spheres, generalised parallelisability and consistent truncations,''
  Fortsch.\ Phys.\  {\bf 65}, no. 10-11, 1700048 (2017)
  [arXiv:1401.3360 [hep-th]].



\bibitem{1410.8145} 
  O.~Hohm and H.~Samtleben,
  ``Consistent Kaluza-Klein Truncations via Exceptional Field Theory,''
  JHEP {\bf 1501}, 131 (2015)
  [arXiv:1410.8145 [hep-th]].



\bibitem{hep-th:9804056} 
  F.~Cordaro, P.~Fre, L.~Gualtieri, P.~Termonia and M.~Trigiante,
  ``N=8 gaugings revisited: An Exhaustive classification,''
  Nucl.\ Phys.\ B {\bf 532}, 245 (1998)
  [hep-th/9804056].



\bibitem{hep-th:0010076} 
  H.~Nicolai and H.~Samtleben,
  ``Maximal gauged supergravity in three-dimensions,''
  Phys.\ Rev.\ Lett.\  {\bf 86}, 1686 (2001)
  [hep-th/0010076].



\bibitem{hep-th:0212239} 
  B.~de Wit, H.~Samtleben and M.~Trigiante,
  ``On Lagrangians and gaugings of maximal supergravities,''
  Nucl.\ Phys.\ B {\bf 655}, 93 (2003)
  [hep-th/0212239].



\bibitem{hep-th:0311224} 
  B.~de Wit, H.~Samtleben and M.~Trigiante,
  ``Maximal supergravity from IIB flux compactifications,''
  Phys.\ Lett.\ B {\bf 583}, 338 (2004)
  [hep-th/0311224].



\bibitem{hep-th:0412173} 
  B.~de Wit, H.~Samtleben and M.~Trigiante,
  ``The Maximal D=5 supergravities,''
  Nucl.\ Phys.\ B {\bf 716}, 215 (2005)
  [hep-th/0412173].



\bibitem{hep-th:0501243} 
  B.~de Wit and H.~Samtleben,
  ``Gauged maximal supergravities and hierarchies of nonAbelian vector-tensor systems,''
  Fortsch.\ Phys.\  {\bf 53}, 442 (2005)
  [hep-th/0501243].



\bibitem{hep-th:0507289} 
  B.~de Wit, H.~Samtleben and M.~Trigiante,
  ``Magnetic charges in local field theory,''
  JHEP {\bf 0509}, 016 (2005)
  [hep-th/0507289].



\bibitem{0705.2101} 
  B.~de Wit, H.~Samtleben and M.~Trigiante,
  ``The Maximal D=4 supergravities,''
  JHEP {\bf 0706}, 049 (2007)
  [arXiv:0705.2101 [hep-th]].



\bibitem{0801.1294} 
  B.~de Wit, H.~Nicolai and H.~Samtleben,
  ``Gauged Supergravities, Tensor Hierarchies, and M-Theory,''
  JHEP {\bf 0802}, 044 (2008)
  [arXiv:0801.1294 [hep-th]].



\bibitem{0809.5180} 
  A.~Le Diffon and H.~Samtleben,
  ``Supergravities without an Action: Gauging the Trombone,''
  Nucl.\ Phys.\ B {\bf 811}, 1 (2009)
  [arXiv:0809.5180 [hep-th]].



\bibitem{1112.3989} 
  A.~Coimbra, C.~Strickland-Constable and D.~Waldram,
  ``$E_{d(d)} \times \mathbb{R}^+$ generalised geometry, connections and M theory,''
  JHEP {\bf 1402}, 054 (2014)
  [arXiv:1112.3989 [hep-th]].



\bibitem{1212.1586} 
  A.~Coimbra, C.~Strickland-Constable and D.~Waldram,
  ``Supergravity as Generalised Geometry II: $E_{d(d)} \times \mathbb{R}^+$ and M theory,''
  JHEP {\bf 1403}, 019 (2014)
  [arXiv:1212.1586 [hep-th]].



\bibitem{2006.09777} 
  D.~S.~Berman and C.~D.~A.~Blair,
  ``The Geometry, Branes and Applications of Exceptional Field Theory,''
  arXiv:2006.09777 [hep-th].



\bibitem{hep-th:0212291} 
  P.~C.~West,
  ``Very extended E(8) and A(8) at low levels, gravity and supergravity,''
  Class.\ Quant.\ Grav.\  {\bf 20}, 2393 (2003)
  [hep-th/0212291].



\bibitem{1303.2035} 
  H.~Godazgar, M.~Godazgar and M.~J.~Perry,
  ``E8 duality and dual gravity,''
  JHEP {\bf 1306}, 044 (2013)
  [arXiv:1303.2035 [hep-th]].



\bibitem{1405.7894} 
  A.~G.~Tumanov and P.~West,
  ``Generalised vielbeins and non-linear realisations,''
  JHEP {\bf 1410}, 009 (2014)
  [arXiv:1405.7894 [hep-th]].



\bibitem{1909.01335} 
  J.~J.~Fernández-Melgarejo, Y.~Sakatani and S.~Uehara,
  ``Exotic branes and mixed-symmetry potentials II: Duality rules and exceptional $p$-form gauge fields,''
  PTEP {\bf 2020}, no. 5, 053B03 (2020)
  [arXiv:1909.01335 [hep-th]].



\bibitem{1610.01620} 
  W.~D.~Linch and W.~Siegel,
  ``F-brane Dynamics,''
  arXiv:1610.01620 [hep-th].



\bibitem{1708.06342} 
  Y.~Sakatani and S.~Uehara,
  ``$\eta$-symbols in exceptional field theory,''
  PTEP {\bf 2017}, no. 11, 113B01 (2017)
  [arXiv:1708.06342 [hep-th]].



\bibitem{1410.8148} 
  J.~A.~Rosabal,
  ``On the exceptional generalised Lie derivative for $d\geq7$,''
  JHEP {\bf 1509}, 153 (2015)
  [arXiv:1410.8148 [hep-th]].



\bibitem{hep-th:9710163} 
  K.~Sfetsos,
  ``Canonical equivalence of nonisometric sigma models and Poisson-Lie T duality,''
  Nucl.\ Phys.\ B {\bf 517}, 549 (1998)
  [hep-th/9710163].



\bibitem{1511.02491}
  Y.~Kosmann-Schwarzbach,
  ``Multiplicativity, from Lie groups to generalized geometry,''
  arXiv:1511.02491 [math.SG].



\bibitem{math/9812064}
  I.~Vaisman,
  ``Nambu-Lie Groups,''
  Journal of Lie Theory 10.1, 181-194 (2000)
  [math/9812064 [math.DG]].



\bibitem{1903.12175} 
  Y.~Sakatani,
  ``Type II DFT solutions from Poisson-Lie T-duality/plurality,''
  PTEP, 073B04 (2019)
  [arXiv:1903.12175 [hep-th]].



\bibitem{1304.1472} 
  D.~Geissbuhler, D.~Marques, C.~Nunez and V.~Penas,
  ``Exploring Double Field Theory,''
  JHEP {\bf 1306}, 101 (2013)
  [arXiv:1304.1472 [hep-th]].



\bibitem{1701.07819} 
  Y.~Sakatani and S.~Uehara,
  ``Connecting M-theory and type IIB parameterizations in Exceptional Field Theory,''
  PTEP {\bf 2017}, no. 4, 043B05 (2017)
  [arXiv:1701.07819 [hep-th]].



\bibitem{hep-th:0210095} 
  C.~Klimcik,
  ``Yang-Baxter sigma models and dS/AdS T duality,''
  JHEP {\bf 0212}, 051 (2002)
  [hep-th/0210095].



\bibitem{1608.03570} 
  R.~Borsato and L.~Wulff,
  ``Target space supergeometry of $\eta$ and $\lambda$-deformed strings,''
  JHEP {\bf 1610}, 045 (2016)
  [arXiv:1608.03570 [hep-th]].



\bibitem{1103.2136} 
  O.~Hohm and S.~K.~Kwak,
  ``Double Field Theory Formulation of Heterotic Strings,''
  JHEP {\bf 1106}, 096 (2011)
  [arXiv:1103.2136 [hep-th]].



\bibitem{1907.02080} 
  G.~Bossard, A.~Kleinschmidt and E.~Sezgin,
  ``On supersymmetric E$_{11}$ exceptional field theory,''
  JHEP {\bf 1910}, 165 (2019)
  [arXiv:1907.02080 [hep-th]].



\bibitem{1702.02861} 
  T.~Araujo, I.~Bakhmatov, E.~Ó.~Colgáin, J.~Sakamoto, M.~M.~Sheikh-Jabbari and K.~Yoshida,
  ``Yang-Baxter $\sigma$-models, conformal twists, and noncommutative Yang-Mills theory,''
  Phys.\ Rev.\ D {\bf 95}, no. 10, 105006 (2017)
  [arXiv:1702.02861 [hep-th]].



\bibitem{1705.02063} 
  T.~Araujo, I.~Bakhmatov, E.~Ó.~Colgáin, J.~Sakamoto, M.~M.~Sheikh-Jabbari and K.~Yoshida,
  ``Conformal twists, Yang–Baxter $\sigma$-models \& holographic noncommutativity,''
  J.\ Phys.\ A {\bf 51}, no. 23, 235401 (2018)
  [arXiv:1705.02063 [hep-th]].



\bibitem{1705.07116} 
  J.~Sakamoto, Y.~Sakatani and K.~Yoshida,
  ``Homogeneous Yang-Baxter deformations as generalized diffeomorphisms,''
  J.\ Phys.\ A {\bf 50}, no. 41, 415401 (2017)
  [arXiv:1705.07116 [hep-th]].



\bibitem{1710.06784} 
  I.~Bakhmatov, Ö.~Kelekci, E.~Ó Colgáin and M.~M.~Sheikh-Jabbari,
  ``Classical Yang-Baxter Equation from Supergravity,''
  Phys.\ Rev.\ D {\bf 98}, no. 2, 021901 (2018)
  [arXiv:1710.06784 [hep-th]].



\bibitem{1803.05903} 
  J.~Sakamoto and Y.~Sakatani,
  ``Local $\beta$-deformations and Yang-Baxter sigma model,''
  JHEP {\bf 1806}, 147 (2018)
  [arXiv:1803.05903 [hep-th]].



\bibitem{1811.09056} 
  I.~Bakhmatov and E.~T.~Musaev,
  ``Classical Yang-Baxter equation from $\beta$-supergravity,''
  JHEP {\bf 1901}, 140 (2019)
  [arXiv:1811.09056 [hep-th]].



\bibitem{1906.09053} 
  A.~Çatal-Özer and S.~Tunalı,
  ``Yang-Baxter Deformation as an $O(d,d)$ Transformation,''
  Class.\ Quant.\ Grav.\  {\bf 37}, no. 7, 075003 (2020)
  [arXiv:1906.09053 [hep-th]].



\bibitem{1906.09052} 
  I.~Bakhmatov, N.~S.~Deger, E.~T.~Musaev, E.~Ó.~Colgáin and M.~M.~Sheikh-Jabbari,
  ``Tri-vector deformations in $d=11$ supergravity,''
  JHEP {\bf 1908}, 126 (2019)
  [arXiv:1906.09052 [hep-th]].



\bibitem{2002.01915} 
  I.~Bakhmatov, K.~Gubarev and E.~T.~Musaev,
  ``Non-abelian tri-vector deformations in $d=11$ supergravity,''
  JHEP {\bf 2005}, 113 (2020)
  [arXiv:2002.01915 [hep-th]].



\bibitem{1407.7106}
  A. Rezaei-Aghdam, M. Sephid,
  ``Classical r-matrices of real low-dimensional Jacobi–Lie bialgebras and their Jacobi–Lie groups,''
   Int.\ J.\ Geom.\ Methods Mod.\ Phys.\ {\bf 13}, 1650087 (2016)
  [arXiv:1407.7106 [math-ph]].



\bibitem{1705.05082} 
  A.~Rezaei-Aghdam and M.~Sephid,
  ``Jacobi–Lie symmetry and Jacobi–Lie T-dual sigma models on group manifolds,''
  Nucl.\ Phys.\ B {\bf 926}, 602 (2018)
  [arXiv:1705.05082 [hep-th]].



\bibitem{1604.08602} 
  F.~Ciceri, A.~Guarino and G.~Inverso,
  ``The exceptional story of massive IIA supergravity,''
  JHEP {\bf 1608}, 154 (2016)
  [arXiv:1604.08602 [hep-th]].



\bibitem{1612.05230} 
  F.~Ciceri, G.~Dibitetto, J.~J.~Fernandez-Melgarejo, A.~Guarino and G.~Inverso,
  ``Double Field Theory at SL(2) angles,''
  JHEP {\bf 1705}, 028 (2017)
  [arXiv:1612.05230 [hep-th]].



\bibitem{1708.02589} 
  G.~Inverso,
  ``Generalised Scherk-Schwarz reductions from gauged supergravity,''
  JHEP {\bf 1712}, 124 (2017)
  [arXiv:1708.02589 [hep-th]].



\bibitem{0705.0752} 
  F.~Riccioni and P.~C.~West,
  ``The E$_{11}$ origin of all maximal supergravities,''
  JHEP {\bf 0707}, 063 (2007)
  [arXiv:0705.0752 [hep-th]].



\bibitem{0705.1304} 
  E.~A.~Bergshoeff, I.~De Baetselier and T.~A.~Nutma,
  ``E$_{11}$ and the embedding tensor,''
  JHEP {\bf 0709}, 047 (2007)
  [arXiv:0705.1304 [hep-th]].
\end{thebibliography}
\end{document}